\documentclass{jfm}
\usepackage{amssymb}
\usepackage{amsfonts}
\usepackage{amsmath}
\usepackage{euscript}
\usepackage{color}
\usepackage{graphicx}
\usepackage{rotating}
\usepackage{bm}
\usepackage{curves}
\usepackage[svgnames,dvipsnames]{pstricks}
\usepackage{lettrine}
\usepackage{mathrsfs}
\usepackage{booktabs}
\usepackage{dcolumn}
\usepackage{natbib}
\usepackage{tikz}
\DeclareGraphicsExtensions{{},.pdf}
\providecommand{\tsc}[1]{{\text{\sc#1}}}

\providecommand{\figref}[1]{{\textup{(Fig.~\ref{#1})}}}
\providecommand{\subfigref}[2]{{\textup{(Fig.~\ref{#1}{#2})}}}

\providecommand{\tabrefnp}[1]{{\textup{Tab.~\ref{#1}}}}
\providecommand{\figrefnp}[1]{{\textup{Fig.~\ref{#1}}}}
\providecommand{\subfigrefnp}[2]{{\textup{Fig.~\ref{#1}{#2}}}}

\providecommand{\eqrefsab}   [2]{\textup{(\ref{#1}, \ref{#2})}}
\providecommand{\eqrefsabc}  [3]{\textup{(\ref{#1}, \ref{#2}, \ref{#3})}}

\providecommand{\figrefsab}   [2]{{\textup{(Figs.~\ref{#1}, \ref{#2})}}}
\providecommand{\figrefsatob} [2]{{\textup{(Figs.~\ref{#1}--\ref{#2})}}}
\providecommand{\parref}[1]{{\textup{(\S\ref{#1})}}}

\providecommand{\parrefsatob}[2]{\textup{(\S\ref{#1}--\S\ref{#2})}}
\providecommand{\parrefnp}[1]{{\textup{\S\ref{#1}}}}
\providecommand{\parrefsatob}[2]{{\textup{(\S\ref{#1}--\S\ref{#2})}}}
\providecommand{\const}{{\rm const}}

\providecommand{\ie}{{\em ie}}

\providecommand{\viz}{{\em viz}}
\newcommand{\abs}[1]{\left\lvert#1\right\rvert}

\newcommand{\circled}[1]{\tikz[baseline=(char.base)]{\node[shape=circle,draw,inner sep=2pt] (char) {#1};}}
\title[Wall effects on pressure fluctuations in turbulent channel flow]{Wall effects on pressure fluctuations in turbulent channel flow}
\author[G.A. Gerolymos, D. S\'en\'echal and I. Vallet]
{G.\ns A.\ns G\ls E\ls R\ls O\ls L\ls Y\ls M\ls O\ls S,\ns
D.\ns S\ls E\ls N\ls E\ls C\ls H\ls A\ls L,\ns
I.\ns V\ls A\ls L\ls L\ls E\ls T\footnote{Email address for correspondence: isabelle.vallet@upmc.fr}}
\affiliation{Universit\'e Pierre-et-Marie-Curie (UPMC), 4 place Jussieu, 75005 Paris, France}
\date{\today}
\begin{document}
\maketitle
%-----------------------------------------------------------------------------------------------------------------------------------
%
%-----------------------------------------------------------------------------------------------------------------------------------
\begin{abstract}
The purpose of the present paper is to study the influence of wall-echo on pressure fluctuations $p'$,
and on statistical correlations containing $p'$, {\em viz} redistribution $\phi_{ij}$, pressure diffusion $d_{ij}^{(p)}$, and velocity/pressure-gradient $\Pi_{ij}$.
We extend the usual analysis of turbulent correlations containing pressure fluctuations in wall-bounded \tsc{dns} computations [Kim J.: {\em J. Fluid Mech.} {\bf 205} (1989) 421--451],
separating $p'$ not only into rapid $p_{(\mathrm{r})}'$ and slow $p_{(\mathrm{s})}'$ parts [Chou P.Y.: {\em Quart. Appl. Math.} {\bf 3} (1945) 38--54],
but further into volume ($p'_{(\mathrm{r};\mathfrak{V})}$ and $p'_{(\mathrm{s};\mathfrak{V})}$)
and surface (wall-echo; $p'_{(\mathrm{r};w)}$ and $p'_{(\mathrm{s};w)}$) terms.
An algorithm, based on a Green's function approach, is developed to compute the above splittings
for various correlations containing pressure fluctuations (redistribution, pressure diffusion, velocity/pressure-gradient),
in fully developed turbulent plane channel flow. This exact analysis confirms previous results based on a method-of-images approximation [Manceau R., Wang M., Laurence D.: {\em J. Fluid Mech.} {\bf 438} (2001) 307--338]
showing that, at the wall, $p'_{(\mathfrak{V})}$ and $p'_{(w)}$ are usually of the same sign and approximately equal.
The above results are then used to study the contribution of each mechanism on the pressure correlations in low Reynolds-number plane channel flow,
and to discuss standard second-moment-closure modelling practices.
\end{abstract}
%-----------------------------------------------------------------------------------------------------------------------------------
%-----------------------------------------------------------------------------------------------------------------------------------
%
%
%
%
%
%
%
%
%
\section{Introduction}\label{WEPFTCF_s_I}
%
%
%
%
%
%
%
%
%
%-----------------------------------------------------------------------------------------------------------------------------------

Understanding the physics of turbulent fluctuations of pressure $p'$ is of major importance, not only because of their direct implication
in noise \citep{Hu_Morfey_Sandham_2002a,
                Hu_Morfey_Sandham_2006a}
and excitation of immersed solid surfaces \citep{Corcos_1964a},
but also because they appear in correlations present in the transport equations for the Reynolds-stresses and the dissipation tensor \citep{Pope_2000a,
                                                                                                                                            Jovanovic_2004a}.
Traditionally the analysis of $p'$ is based on the Poisson equation for the fluctuating pressure \citep{Chou_1945a},
which, at the incompressible flow limit ($\rho\approxeq{\rm const}=\bar\rho\;;\forall t,x$ and $\mu\approxeq{\rm const}=\bar\mu\;;\forall t,x$), in a nonrotating frame-of-reference, reads
\begin{alignat}{6}
\nabla^2p'=\nabla^2(p'_\mathrm{r}+p'_\mathrm{s})
          =\underbrace{-2\rho\dfrac{\partial u_k'  }
                                   {\partial x_\ell}
                             \dfrac{\partial \bar u_\ell}
                                   {\partial         x_k}}_{\textstyle\text{rapid}}\;
           \underbrace{-\rho\dfrac{\partial^2                 }
                                  {\partial x_\ell\partial x_k}(u_k'u_\ell'-\overline{u_k'u_\ell'})}_{\textstyle\text{slow}}
                                                                                                                                    \label{Eq_WEPFTCF_s_I_001}
\end{alignat}

The incompressible flow Poisson equation \eqref{Eq_WEPFTCF_s_I_001} suggests that \citep{Chou_1945a} solenoidal \citep{Hamba_1999a} pressure
fluctuations, associated with the fluctuating velocity field, are generated by 2 separate mechanisms \citep{Rotta_1951a,
                                                                                                            Rotta_1951b,
                                                                                                            Lumley_1978a,
                                                                                                            Piquet_1999a,
                                                                                                            Pope_2000a,
                                                                                                            Jovanovic_2004a}:
\begin{enumerate}
\item the interaction of velocity fluctuations with
      mean-velocity-gradients called meanflow/turbulence interaction terms,
      also termed rapid pressure fluctuations, because they interact immediately with an imposed mean-velocity-gradient, or linear pressure fluctuations,
      because the corresponding source-term in \eqref{Eq_WEPFTCF_s_I_001} is linear in velocity fluctuations, and
\item the turbulence/turbulence interaction, also termed slow pressure fluctuations, because they react much slower than the rapid ones, which are directly
      driven by mean-velocity-gradients, or nonlinear pressure fluctuations, because the corresponding source-term in \eqref{Eq_WEPFTCF_s_I_001} is quadratic in velocity fluctuations.
\end{enumerate}

This idea of distinguishing between pressure fluctuations associated with
mean-flow-gradients ($p'_{(\rm r)}$) and pressure fluctuations associated with turbulence/turbulence interactions only ($p'_{(\rm s)}$) can be applied in general to
all correlations which contain the fluctuating pressure \citep{Rotta_1951a,
                                                               Rotta_1951b,
                                                               Lumley_1978a,
                                                               Hanjalic_1994a,
                                                               Piquet_1999a,
                                                               Pope_2000a,
                                                               Jovanovic_2004a}.

\cite{Chou_1945a} pointed out that, because of the linearity in $p'$ of the incompressible flow Poisson equation \eqref{Eq_WEPFTCF_s_I_001},
separate solutions can be obtained for each of the 2 source-terms,
based on the freespace Green's function for the Poisson equation \citep{Katz_Plotkin_1991a_Poisson_eq_freespace_Greenfuncion}
\begin{subequations}
                                                                                                                                    \label{Eq_WEPFTCF_s_I_002}
\begin{alignat}{6}
p_{(\mathrm{r})}'(\vec{x},t)
&=\underbrace{
              \frac{1}{2\pi}\iiint_\mathfrak{V}\left(\rho\dfrac{\partial\overline{\mathfrak{u}}_\ell}
                                                           {\partial\mathfrak{x}_k              }
                                                     \dfrac{\partial\mathfrak{u}_k'             }
                                                           {\partial\mathfrak{x}_\ell           }\right)\dfrac{d\mathfrak{v}(\vec{\mathfrak{x}},t)}
                                                                                                              {\abs{\vec{\mathfrak{x}}-\vec{x}}   }
             }_{\textstyle p'_{(\mathrm{r};\mathfrak{V})}(\vec{x},t)}
                                                                                                                                    \nonumber\\
&+\underbrace{
              \frac{1}{4\pi}\iint_{\partial\mathfrak{V}}\left(\dfrac{                               1}
                                                                    {\abs{\vec{\mathfrak{x}}-\vec{x}}}
                                                              \dfrac{\partial\mathfrak{p}'_\mathrm{r}}
                                                                    {\partial\mathfrak{n}            }
                                                             -\mathfrak{p}'_\mathrm{r}
                                                              \dfrac{\partial                        }
                                                                    {\partial\mathfrak{n}            }
                                                              \left[\dfrac{1                               }
                                                                          {\abs{\vec{\mathfrak{x}}-\vec{x}}}\right]\right)\;d\mathfrak{S}(\vec{\mathfrak{x}},t)
             }_{\textstyle p'_{(\mathrm{r};w)}(\vec{x},t)}
                                                                                                                                    \label{Eq_WEPFTCF_s_I_002a}\\
p'_{(\mathrm{s})}(\vec{x},t)
&=\underbrace{
              \frac{1}{4\pi}\iiint_\mathfrak{V}\left(\rho\dfrac{\partial^2\mathfrak{u}_k'\mathfrak{u}_\ell'           }
                                                               {\partial\mathfrak{x}_\ell\partial\mathfrak{x}_k       }
                                                    -\rho\dfrac{\partial^2\overline{\mathfrak{u}_k'\mathfrak{u}_\ell'}}
                                                               {\partial\mathfrak{x}_\ell\partial\mathfrak{x}_k       }\right)\dfrac{d\mathfrak{v}(\vec{\mathfrak{x}},t)}
                                                                                                                                {\abs{\vec{\mathfrak{x}}-\vec{x}}   }
             }_{\textstyle p'_{(\mathrm{s};\mathfrak{V})}(\vec{x},t)}
                                                                                                                                    \nonumber\\
&+\underbrace{
              \frac{1}{4\pi}\iint_{\partial\mathfrak{V}}\left(\dfrac{1                              }
                                                                    {\abs{\vec{\mathfrak{x}}-\vec{x}}}
                                                              \dfrac{\partial\mathfrak{p}'_\mathrm{s}}
                                                                    {\partial\mathfrak{n}            }
                                                             -\mathfrak{p}'_\mathrm{s}
                                                              \dfrac{\partial                        }
                                                                    {\partial\mathfrak{n}            }
                                                              \left[\dfrac{1                                }
                                                                          {\abs{\vec{\mathfrak{x}}-\vec{x}}}\right]\right)\;d\mathfrak{S}(\vec{\mathfrak{x}},t)
             }_{\textstyle p'_{(\mathrm{s};w)}(\vec{x},t)}
                                                                                                                                    \label{Eq_WEPFTCF_s_I_002b}
\end{alignat}
\end{subequations}
where $p(\vec{x},t)$ and $\vec{u}(\vec{x},t)$ are the pressure and velocity at point $\vec{x}$,
and the volume $d\mathfrak{v}(\vec{\mathfrak{x}},t)$ and surface $d\mathfrak{S}(\vec{\mathfrak{x}},t)$ integrals are taken over all
other points $\vec{\mathfrak{x}}$ where the pressure and velocity are $\mathfrak{p}(\vec{\mathfrak{x}},t)$ and $\vec{\mathfrak{u}}(\vec{\mathfrak{x}},t)$.
Notice that if \eqref{Eq_WEPFTCF_s_I_002} are multiplied
by a function of $\vec{x}$, this function can be entered into the integrals which are over $\vec{\mathfrak{x}}$.
In the case of unbounded flow, where $\partial\mathfrak{V}$ is very far (at infinity), only the volume integrals remain.
On the other hand, for flow near solid boundaries, the surface integrals indicate that the unsteady pressure field reacts to the
presence of the wall (surface integral; $\mathfrak{n}$ is the normal-to-the-wall coordinate, directed outwards from the fluid volume $\mathfrak{V}$). Terms related to the surface integrals are usually
called wall-echo terms \citep{Hanjalic_1994a},
since for an isolated infinite plane solid boundary they can be related to reflection from the wall, using the method of images \citep{Shir_1973a,
                                                                                                                                       Launder_Reece_Rodi_1975a,
                                                                                                                                       Gibson_Launder_1978a,
                                                                                                                                       Piquet_1999a,
                                                                                                                                       Pope_2000a,
                                                                                                                                       Jovanovic_2004a}.
Wall-echo terms, as defined by the surface integrals in \eqref{Eq_WEPFTCF_s_I_002}, contain the entire effect of the wall on the pressure fluctuations produced by the inhomogeneous
velocity field (already influenced by the wall), and cannot distinguish between different mechanisms by which the presence of the wall affects turbulence \citep{Hunt_Graham_1978a},
whose precise identification would require further specific decomposition (projection) of the \tsc{dns}-computed velocity field \citep{Marmanis_1998a,
                                                                                                                                       Perot_1999a}.

Starting from the seminal paper of \cite{Chou_1945a} all models \citep{Rotta_1951a,
                                                                       Rotta_1951b,
                                                                       Lumley_1978a,
                                                                       Piquet_1999a,
                                                                       Pope_2000a,
                                                                       Jovanovic_2004a}
for the redistribution tensor $\phi_{ij}$, the velocity/pressure-gradient tensor $\Pi_{ij}$
or the pressure transport $\overline{p'u_i'}$ appearing in the pressure-diffusion tensor $d_{ij}^{(p)}$, are traditionally composed of 4 parts
corresponding to the splitting $p'=p'_{(\mathrm{r};\mathfrak{V})}+p'_{(\mathrm{r};w)}+p'_{(\mathrm{s};\mathfrak{V})}+p'_{(\mathrm{s};w)}$ \eqref{Eq_WEPFTCF_s_I_002}.
There is at present no possibility to separately measure $p'_{(\mathrm{r};\mathfrak{V})}$, $p'_{(\mathrm{r};w)}$, $p'_{(\mathrm{s};\mathfrak{V})}$, and $p'_{(\mathrm{s};w)}$,
and, despite advances in experimental techniques for the measurement of $p'$ \citep{Tsuji_Fransson_Alfredsson_Johansson_2007a,
                                                                                    Tsuji_Ishihara_2003a},%}}
 the simultaneous measurement of $p'$ and $\partial_{x_j} u_i'$ in the the wall-vicinity is a difficult challenge \citep{Naka_Omori_Obi_Masuda_2006a}.
Direct numerical simulation (\tsc{dns}) offers the possibility to directly compute the different terms, experimental uncertainty being replaced by the eventual influence of finite size of the
computational box and of convergence of statistics.

\cite{Kim_1989a}, in the context of pseudospectral \tsc{dns} of incompressible plane channel flow \citep{Kim_Moin_Moser_1987a,
                                                                                                         Moser_Kim_Mansour_1999a},
used a Green's function approach \citep{Ince_1926a_1D_Greenfunction_and_compatibility_relation,
                                        Courant_Hilbert_1953a_1D_Greenfunction,
                                        Bender_Orszag_1978a_1D_Greenfunction,
                                        Ockendon_Howison_Lacey_Movchan_2003a_Greenfuncion,
                                        Zauderer_2006a_1D_Greenfunction_and_compatibility_relation,
                                        MyintU_Debnath_2007a_1D_Greenfunction}
to solve, as a function of $y$ (normal-to-the-wall coordinate), the incompressible flow Poisson equation for $p'$ \eqref{Eq_WEPFTCF_s_I_001},
for each parallel-to-the-wall wavenumber $\kappa_x$ and $\kappa_z$.\footnote{\label{ff_WEPFTCF_s_001}
                                                                             Throughout the paper $x$ denotes the streamwise coordinate with corresponding velocity-component $u$,
                                                                             $y$ denotes the normal-to-the-wall coordinate with corresponding velocity-component $v$,
                                                                             and $z$ denotes the spanwise coordinate with corresponding velocity-component $w$.
                                                                             Finally, $y=0$ at the channel centerline,
                                                                             and the nondimensional distance from the wall is defined as
                                                                             $y^+:=(y-y_w)\bar u_\tau\nu^{-1}$, where $u_\tau$ is the friction-velocity and $\nu$ the kinematic viscosity.
                                                                            }
The separate solution for each source-term \eqref{Eq_WEPFTCF_s_I_001}, permits the separation of
slow $p'_{(\mathrm{s})}$ and rapid $p'_{(\mathrm{r})}$ parts.
In this way \citep{Kim_1989a,
                   Chang_Piomelli_Blake_1999a},
the slow and rapid contributions to $\phi_{ij}$ were computed \citep{Mansour_Kim_Moin_1988a}.
This procedure \citep{Kim_1989a} is now used in a standard way in incompressible plane channel flow \tsc{dns}, at least as far
as $\overline{p'^2_{(\mathrm{s})}}$, $\overline{p'^2_{(\mathrm{r})}}$, and $\overline{p'^2_{(\tau)}}$ are concerned,
and has also been extended to compressible flow studies \citep{Foysi_Sarkar_Friedrich_2004a}.
The term $\overline{p'^2_{(\tau)}}$,
usually called Stokes pressure~\citep{Mansour_Kim_Moin_1988a,
                                      Chang_Piomelli_Blake_1999a},
corresponds to the separately computed contribution of the wall-boundary condition \citep[(11.173), p. 439]{Pope_2000a}
$[\partial_y p']_w=[\mu\partial^2_{yy} v']_w$, $y$ being the normal-to-the-wall direction.
Notice that this boundary-condition is associated with the source-term $\partial^2_{x_ix_j}[\tau_{ij}']$ in the Poisson equation
for $p'$, which is equal to 0 at the incompressible flow limit \citep{Kim_Moin_Moser_1987a,
                                                                      Moser_Kim_Mansour_1999a,
                                                                      Hu_Sandham_2001a,
                                                                      Hoyas_Jimenez_2006a}.
This procedure was used by \cite{Chang_Piomelli_Blake_1999a} to study the detailed contributions from different locations in the flowfield to
wall-pressure wavenumber-frequency spectra.

\cite{Manceau_Wang_Laurence_2001a} have studied the wall-echo problem from the specific point-of-view of the so-called elliptic relaxation approach of \citet{Durbin_1993a},
which lumps together all wall-echo effects (both on pressure correlations and on dissipation-rate) into a single tensor.
These authors \citep{Manceau_Wang_Laurence_2001a}
used 2-point correlations (in the physical space) from the \tsc{dns} data of \citet{Moser_Kim_Mansour_1999a},
at wall-friction Reynolds-number $Re_{\tau_w}=590$.
The method used, which is related to classical work on wall-pressure spectra \citep{Panton_Linebarger_1974a} in that
it uses 2-point correlations in physical space, introduces wall-echo by an approximate method of images, one image-channel for the upper wall and another for the lower wall.
Furthermore, \citet{Manceau_Wang_Laurence_2001a} neglect Stokes pressure.
When analyzed under a formal Green's function framework, the approximation in the method of images used in \citet{Manceau_Wang_Laurence_2001a}
is equivalent to using an approximate Green's function, instead of the exact one. By comparison with the exact solution obtained in the present work
it is possible to quantify the approximation error of the method of images.

All known second-moment closures ({\sc smc}s) for wall-bounded flows treat pressure correlations appearing in Reynolds-stress transport equations by separating them into a quasi-homogeneous part and 
a second part accounting for inhomogeneity and/or wall-echo.
Concerning the quasi-homogeneous part of the redistribution term,
general tensorial representations are available~\citep{Ristorcelli_Lumley_Abid_1995a,
                                                       Craft_Launder_1996a,
                                                       Jakirlic_Hanjalic_2002a,
                                                       Gerolymos_Lo_Vallet_2012a}
which in order to satisfy the two-component limit ({\sc tcl})
realizability constraint~\citep[turbulence tends to {\sc tcl} as the wall is approached because of the strong damping of velocity fluctuations normal to the wall]{Shih_Lumley_1993a} should use
representation coefficients\footnote{\label{ff_WEPFTCF_s_002}
                                     The early closures~\citep{Rotta_1951a,
                                                               Launder_Reece_Rodi_1975a,
                                                               Speziale_Sarkar_Gatski_1991a}
which use constant coefficients do not satisfy the two-component realizability constraint~\citep{Shih_Lumley_1993a,
                                                                                                 Hanjalic_1994a,
                                                                                                 Schwarz_Bradshaw_1994a}}
that are function of the Reynolds-stress anisotropy tensor invariants~\citep{Lumley_1978a}.

Concerning inhomogeneity and wall-echo, these two mechanisms are not necessarily identical. Wall-echo, which is related to the surface integrals in~\eqref{Eq_WEPFTCF_s_I_002},
diminishes with distance-from-the-wall, whereas turbulence-inhomogeneity effects
may be active away from the wall, and even in the absence of walls. Indeed comparison of models including inhomogeneity-based correction~\citep{Craft_Launder_1996a,
                                                                                                                                                Gerolymos_Vallet_2001a,
                                                                                                                                                Suga_2004a},
and of models using wall-topology-based echo-terms,
suggests that inhomogeneity-based terms are active at the edge of the boundary-layer and improve the prediction of boundary-layer entrainment.
According to the way wall-proximity is handled, {\sc smc}s can be classified into three categories
\begin{enumerate}
\item Models which use wall-topology based (for instance distance from the wall or geometric normal to the wall) corrections of the quasi-homogeneous closures, tend to the homogeneous limit away from the wall ($y^+\gtrapprox 30$). 
Notice that elliptic blending models~\citep{Manceau_Hanjalic_2002a} use the same family of tensorial representations for the redistribution $\phi_{ij}$, the velocity/pressure-gradient correlation $\Pi_{ij}$ 
or the dissipation $\varepsilon_{ij}$ terms, and introduce the knowledge of wall-topology via the boundary conditions of the scalar Helmholtz equation used to define the scalar blending parameter.
Therefore, it can be argued that elliptic blending models belong to this category.
\item  Models free from any wall-topology parameters,
which use terms based on gradients of local turbulence quantities in the tensorial representations for  $\phi_{ij}$, $\Pi_{ij}$ or $\varepsilon_{ij}$, and
are therefore active both near and away from the wall. These models include wall-normal-free models based on a unit-vector in
the direction of inhomogeneity $\vec{e}_\tsc{I}$~\citep{Gerolymos_Vallet_2001a,
                                                        Gerolymos_Sauret_Vallet_2004a},
the proposal of \cite{Launder_Li_1994a} to Taylor-expand the mean-velocity gradient
in~\eqref{Eq_WEPFTCF_s_I_002a} leading to terms containing $\partial^2_{x_jx_k}\bar{u}_i$ in the tensorial representation for $\phi_{ij}$, and the work of \cite{Cormack_1975a} who suggested the use of extended 
tensorial representations including gradients of $\overline{u_i'u_j'}$ and of $\varepsilon_{ij}$. This last proposal \citep{Cormack_1975a} has not been applied to the development of a working model yet,
and recent work \citep{Gerolymos_Lo_Vallet_Younis_2012a} suggests that further terms can be profitably added to Cormack's basis.
\item The full (tensorial) elliptic relaxation approach of \cite{Durbin_1993a} where all wall-affected terms (both inhomogeneous and wall-echo) are lumped together into a correction of the homogeneous closures
which is obtained by the solution of six Helmholtz equations, one for each component of the Reynolds-stress tensor $\overline{u_i'u_j'}$. This approach is fundamentally different from elliptic blending in that it postulates no
particular tensorial representation for the near-wall corrections but, instead, determines these via appropriately chosen wall boundary conditions for the Helmholz equations.
Unfortunately, very few applications of this approach are available in the literature, where, for complex flows a simplified three-equation closure version is used instead \citep{Durbin_1995a}.
On the other hand, despite statements of the contrary,
this approach uses turbulent Reynolds number near-wall dampings.\footnote{\label{ff_WEPFTCF_s_003}
                                                                          Indeed, the Kolmogorov-scale clippings of the elliptic relaxation
lengthscale $L_\tsc{er}$ \citep[(9), p. 469]{Durbin_1993a} and
timescale $T_\tsc{er}$ \citep[(6), p. 468]{Durbin_1993a} can be rewritten as\\
$T_\tsc{er}:=\max\left[\dfrac{\rm k}{\varepsilon},C_{T_\tsc{er}}\sqrt{\dfrac{\nu}{\varepsilon}}\right]
            =\max\left[1,C_{T_\tsc{er}} Re_\tsc{t}^{-\frac{1}{2}}\right]\dfrac{\rm k}{\varepsilon}$\\
$L_\tsc{er}:=C_{L_\tsc{er}} \max\left[\dfrac{{\rm k}^\frac{3}{2}}{\varepsilon},C_{\eta_\tsc{er}}\left(\dfrac{\nu^3}{\varepsilon}\right)^\frac{1}{4}\right]
            =C_{L_\tsc{er}} \max\left[1,C_{\eta_\tsc{er}}Re_\tsc{t}^{-\frac{3}{4}}\right]\dfrac{{\rm k}^\frac{3}{2}}{\varepsilon}$\\
Hence the switches in $L_\tsc{er}$ and $T_\tsc{er}$ used in elliptic relaxation are in essence $Re_\tsc{t}$-dampings
(where $Re_\tsc{t}:=\mathrm{k}^2\varepsilon^{-1}\nu^{-1}$ is the turbulence Reynolds-number,
$\mathrm{k}$ is the turbulence kinetic energy, $\varepsilon$ is its dissipation-rate, and $\nu$ is the kinematic viscosity),
expectedly, since the ratios of large-to-Kolmogorov scales are invariably equal to some power of $Re_\tsc{t}$~\cite[p. 20]{Davidson_2004a}.
}
\end{enumerate}

None of the above three approaches is perfectly consistent with flow physics, which, ideally would require an inhomogeneous (including gradients of the Reynolds-stresses) tensorial representation coupled to a wall-echo correction.

In the present work we are interested in determining the wall-effects on correlations containing the fluctuating pressure.
We aim both at developing a methodology for creating databases of the contribution of each mechanism to $\phi_{ij}$, $\Pi_{ij}$ and $\overline{p'u'_i}$,
and at highlighting the differences between volume ($\mathfrak{V}$), wall ($w$), homogeneous ($\tsc{h}$) and inhomogeneous ($\tsc{i}$) contributions,
The first two contributions are extracted from the \tsc{dns} computations, while the latter two are used in modelling.
Homogeneous terms are not volume terms. They would correspond to hypothetical values of the pressure correlations of a series of 
independent homogeneous flows, each corresponding to the local values of anisotropy, lengthscale and mean-velocity-gradients.
These homogeneous terms are approximately equal to the volume terms away from the wall.
Whether such hypothetical homogeneous flows are obtainable for the anisotropy observed near the wall, in pure shear, is an open question.

In \S\ref{WEPFTCF_s_DNSCpS}, after a brief description of the present \tsc{dns} calculations (\S\ref{WEPFTCF_s_DNSCpS_ss_PCFCCM} and Appendix \ref{WEPFTCF_s_AppendixCELMNL}),
we revisit the Green's function (\S\ref{WEPFTCF_s_DNSCpS_ss_GFSPEq}) approach \citep{Kim_1989a} applied to \tsc{dns} computations of incompressible plane channel flow (\S\ref{WEPFTCF_s_DNSCpS_ss_PCFCCM}),
and we develop a procedure for separately evaluating (\S\ref{WEPFTCF_s_DNSCpS_ss_GFSPEq_sss_VWET})
the volume ($p'_{(\mathrm{r};\mathfrak{V})}$ and $p'_{(\mathrm{s};\mathfrak{V})}$) and the wall-echo ($p'_{(\mathrm{r};w)}$ and $p'_{(\mathrm{s};w)}$) terms in \eqref{Eq_WEPFTCF_s_I_002}.
By comparison with these exact results we also quantify the error of the approximate method of images of~\cite{Manceau_Wang_Laurence_2001a}, as a function of the wavenumber (\S\ref{WEPFTCF_s_DNSCpS_ss_GFSPEq_sss_AMI}),
showing that this method is a high-wavenumber approximation.
Analytical details on the Green's functions, with particular emphasis on the singular problem of $xz$-constant pressure fluctuations ($\kappa=0$), are given
in Appendix \ref{WEPFTCF_s_AppendixGF}.
In \S\ref{WEPFTCF_s_APC} we apply the algorithm (\S\ref{WEPFTCF_s_DNSCpS_ss_GFSPEq}) for computing the $p'$-splitting \eqref{Eq_WEPFTCF_s_I_002}
to low Reynolds-number plane channel flow and
study the contribution of different terms to correlations containing $p'$.

%-----------------------------------------------------------------------------------------------------------------------------------
%
%
%
%
%
%
%
%
%
\section{DNS computations and $p'$-splitting}\label{WEPFTCF_s_DNSCpS}
%
%
%
%
%
%
%
%
%
%-----------------------------------------------------------------------------------------------------------------------------------
The \tsc{dns} computations used to obtain the results analyzed in the present paper are discussed in \parrefnp{WEPFTCF_s_DNSCpS_ss_PCFCCM},
and the $p'$-splitting methodology is described in \parrefnp{WEPFTCF_s_DNSCpS_ss_GFSPEq}.

%-----------------------------------------------------------------------------------------------------------------------------------
%
%\subsection{Influence of $Re_{\tau_w}$ on pressure correlations in plane channel flow}\label{WEPFTCF_s_DNSCpS_ss_IRetauwPCPCF}
%
%-----------------------------------------------------------------------------------------------------------------------------------
%
%-----------------------------------------------------------------------------------------------------------------------------------
\begin{figure}
\begin{center}
\begin{picture}(400,360)
\put(-40,-180){\includegraphics[angle=0,width=460pt]{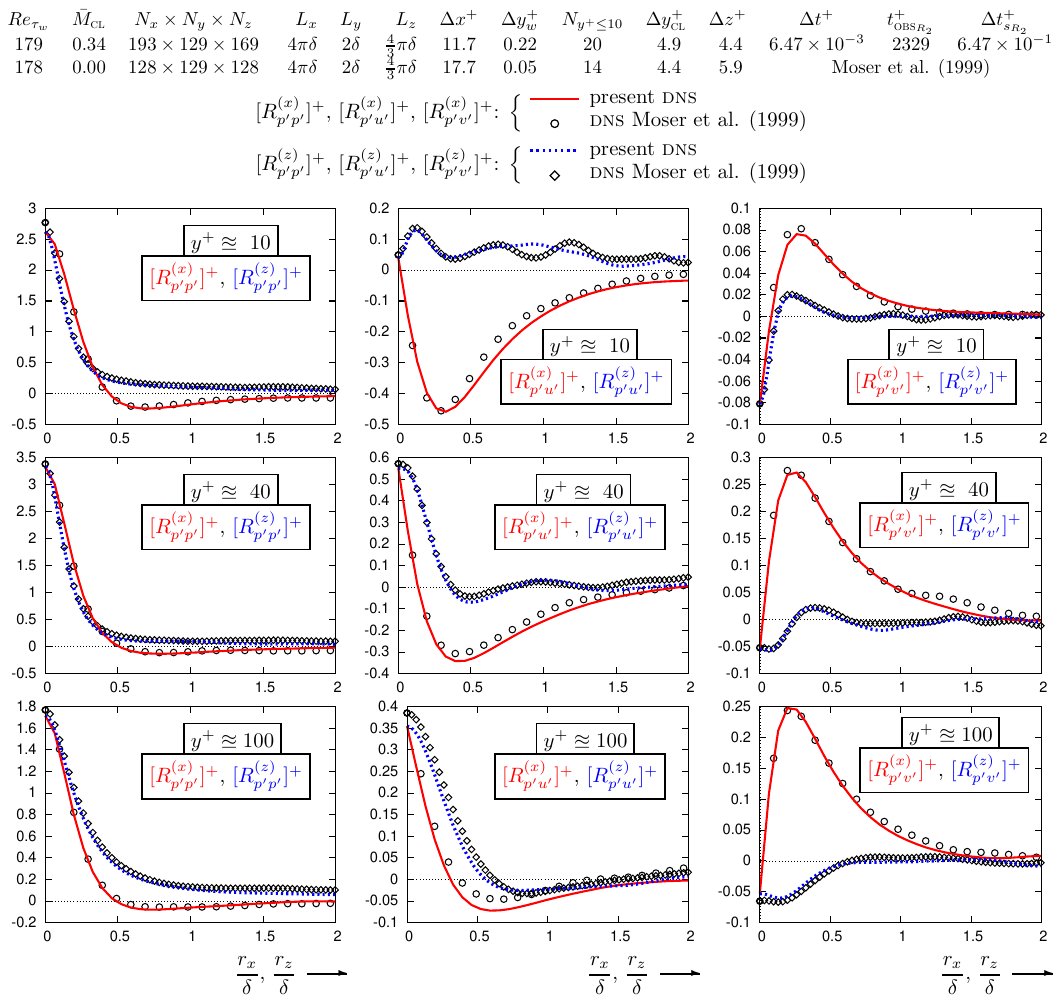}}
\end{picture}
\end{center}
\caption{Comparison of present \tsc{dns}-computed ($Re_{\tau_w}=179$; $\bar M_\tsc{cl}=0.34$; $193\times129\times169$ grid; \tabrefnp{Tab_WEPFTCF_s_DNSCpS_ss_PCFCCM_001})
2-point correlations \eqref{Eq_WEPFTCF_s_DNSCpS_ss_PCFCCM_001} containing the fluctuating pressure in wall-scaling
($[R^{(x)}_{p'p'}]^+$, $[R^{(x)}_{p'u'}]^+$, and $[R^{(x)}_{p'v'}]^+$ in the homogeneous streamwise ($x$) direction;
 $[R^{(z)}_{p'p'}]^+$, $[R^{(z)}_{p'u'}]^+$, and $[R^{(z)}_{p'v'}]^+$ in the homogeneous spanwise ($z$) direction),
as a function of distance in outer scaling ($\delta^{-1}r_x$ and $\delta^{-1}r_z$),
with reference results of incompressible pseudospectral \citep{Kim_Moin_Moser_1987a} \tsc{dns} computations
\citep[{$Re_{\tau_w}=178$, $M_\tsc{cl}=0$}]{Moser_Kim_Mansour_1999a}.}
\label{Fig_WEPFTCF_s_DNSCpS_ss_PCFCCM_001}
\end{figure}
%-----------------------------------------------------------------------------------------------------------------------------------
%
%-----------------------------------------------------------------------------------------------------------------------------------
\begin{figure}
\begin{center}
\begin{picture}(400,350)
\put(-40,-190){\includegraphics[angle=0,width=460pt]{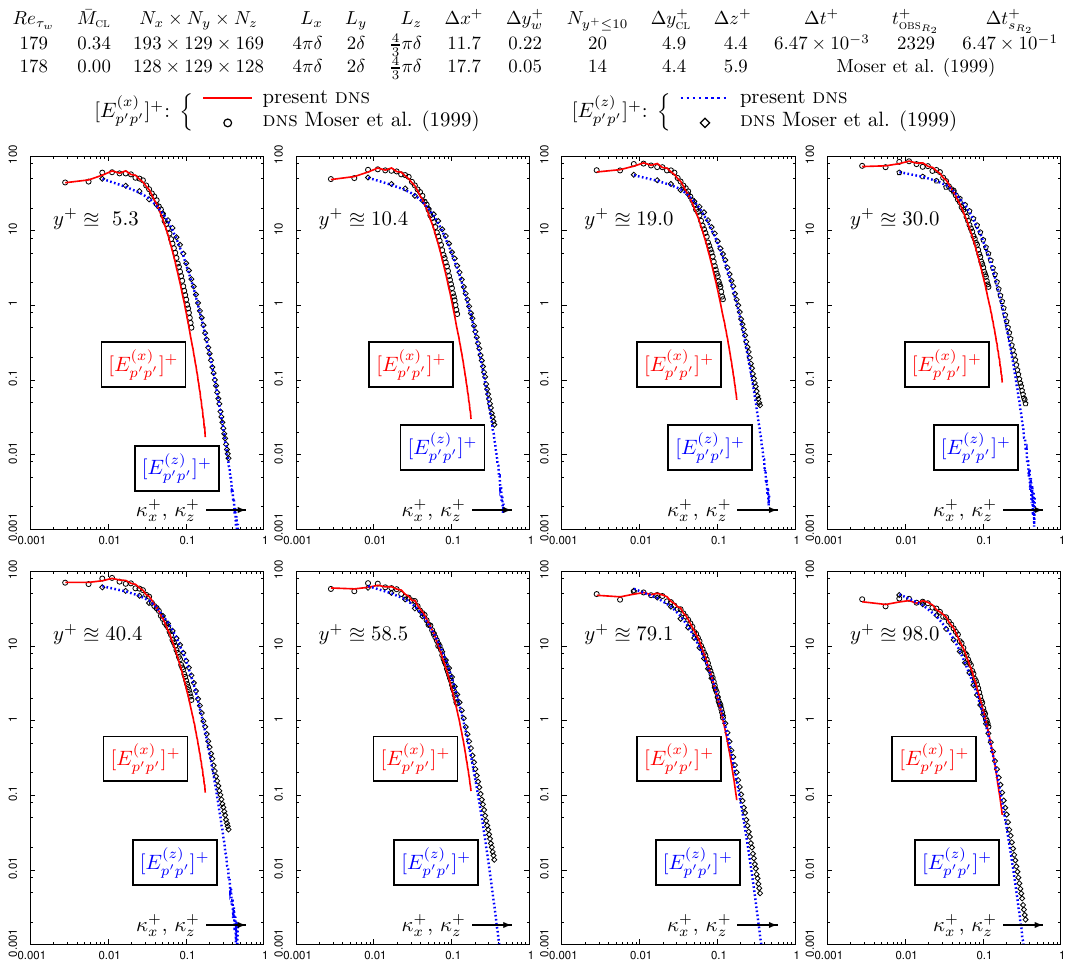}}
\end{picture}
\end{center}
\caption{Comparison of present {\sc dns}-computed ($Re_{\tau_w}=179$; $\bar M_\tsc{cl}=0.34$; $193\times129\times169$ grid; \tabrefnp{Tab_WEPFTCF_s_DNSCpS_ss_PCFCCM_001})
1-D pressure spectra in wall scaling ($[E^{(x)}_{p'p'}]^+$ and $[E^{(z)}_{p'p'}]^+$ in the homogeneous streamwise ($x$) and spanwise ($z$) directions),
as a function of wavenumber in wall units ($\kappa_x^+$ and $\kappa_z^+$),
with reference results of incompressible pseudospectral~\citep{Kim_Moin_Moser_1987a} \tsc{dns} computations
\citep[{$Re_{\tau_w}=178$, $M_\tsc{cl}=0$}]{Moser_Kim_Mansour_1999a}.}
\label{Fig_WEPFTCF_s_DNSCpS_ss_PCFCCM_002}
\end{figure}
%-----------------------------------------------------------------------------------------------------------------------------------
%
%-----------------------------------------------------------------------------------------------------------------------------------
\begin{table}
\begin{center}
%-----------------------------------------------------------------------
\scalebox{0.76}{
\begin{tabular}{cccrcrccccccccl}
$Re_{\tau_w}$&$\bar M_\tsc{cl}$&$N_x\times N_y\times N_z$&  $L_x$      &$L_y$    & $L_z$                  &$\Delta x^+$&$\Delta y_w^+$&$N_{y^+\leq10}$&$\Delta y_\tsc{cl}^+$&$\Delta z^+$&$\Delta t^+$       &$t_\tsc{obs}^+$&$\Delta t_s^+$     &$\Delta t_{s_{R_2}}^+$\\
  $179$      & $0.34$          &$193\times129\times169$  &$~4\pi\delta$&$2\delta$&$\tfrac{4}{3}\pi\delta$ &  $11.7$    &  $0.22$      &  $20$         &  $~4.9$             & $~4.4$     &$6.47\times10^{-3}$& $2329$        &$6.47\times10^{-3}$&$6.47\times10^{-1}$  \\
  $180$      & $0.34$          &$257\times129\times385$  &$~8\pi\delta$&$2\delta$&$4\pi\delta$            &  $17.6$    &  $0.23$      &  $20$         &  $~5.0$             & $~5.9$     &$6.56\times10^{-3}$& $~289$        &$6.57\times10^{-3}$&$6.57\times10^{-1}$  \\
  $180$      & $0.34$          &$129\times129\times129$  &$~4\pi\delta$&$2\delta$&$\tfrac{4}{3}\pi\delta$ &  $17.7$    &  $0.23$      &  $20$         &  $~5.0$             & $~5.9$     &$6.56\times10^{-3}$& $1539$        &$6.56\times10^{-3}$&$6.56\times10^{-1}$  \\
  $184$      & $0.34$          &$121\times161\times~81$  &$~4\pi\delta$&$2\delta$&$\tfrac{4}{3}\pi\delta$ &  $19.3$    &  $0.19$      &  $25$         &  $~4.4$             & $~9.6$     &$5.50\times10^{-3}$& $1014$        &$5.50\times10^{-3}$&$5.50\times10^{-1}$  \\
\end{tabular}
}
\caption{Parameters of the \tsc{dns} computations
         [$L_x$, $L_y$, $L_z$ ($N_x$, $N_y$, $N_z$) are the dimensions (number of grid-points) of the computational domain ($x=$ homogeneous streamwise, $y=$ normal-to-the-wall, $z=$ homogeneous spanwise direction);
          $\delta$ is the channel halfheight;
          $\Delta x^+$, $\Delta y_w^+$, $\Delta y_\tsc{cl}^+$, $\Delta z^+$ are the mesh-sizes in wall-units;
          $(\cdot)_w$ denotes wall and $(\cdot)_\tsc{cl}$ centerline values;
          $N_{y^+\leq10}$ is the number of grid points between the wall and $y^+=10$;
          $Re_{\tau_w}:=\bar u_\tau\delta\bar\nu_w^{-1}$;
          $\bar u_\tau$ is the friction velocity;
          $\delta$ is the channel halfheight;
          $\bar\nu_w=$ is the kinematic viscosity at the wall;
          $\bar M_\tsc{cl}$ is the centerline Mach-number;
          $\Delta t^+$ is the computational time-step in wall-units;
          $t_\tsc{obs}^+$ is the observation period in wall units over which statistics were computed;
          $\Delta t_s^+$ is the sampling time-step for the single-point statistics in wall-units;
          $\Delta t_{s_{R_2}}^+$ is the sampling time-step for the two-point statistics in wall-units].}
\label{Tab_WEPFTCF_s_DNSCpS_ss_PCFCCM_001}
%-----------------------------------------------------------------------
\end{center}
\end{table}
%-----------------------------------------------------------------------------------------------------------------------------------
%
%-----------------------------------------------------------------------------------------------------------------------------------
%
\subsection{Plane channel flow configuration and computational method}\label{WEPFTCF_s_DNSCpS_ss_PCFCCM}
%
%-----------------------------------------------------------------------------------------------------------------------------------

In the particular case of plane channel flow, the boundary surface consists of the upper and lower walls (${\mathfrak W}_l$ at $y=-\tfrac{1}{2}L_y=-\delta$
and ${\mathfrak W}_u$ at $y=+\tfrac{1}{2}L_y=+\delta$, respectively), and the periodic boundaries at $x=\pm\tfrac{1}{2}L_x$ and $z=\pm\tfrac{1}{2}L_z$.
The methodology described in the present paper is independent of the particular \tsc{dns} solver used. The \tsc{dns} database
was generated for $Re_{\tau_w}:=\bar u_\tau \delta \nu^{-1}=180$ (where $2\delta=L_y$ is the channel height, $u_\tau:=\sqrt{\bar\tau_w\rho^{-1}}$ is the friction velocity
and $\nu$ the kinematic viscosity) using the \tsc{dns} solver described and validated in \cite{Gerolymos_Senechal_Vallet_2010a}, which solves the flow-equations in physical space,
with $O(\Delta x^{17})$ discretization of the convective terms~\citep{Gerolymos_Senechal_Vallet_2009a}. The flow is modelled by the compressible Navier-Stokes equations with air as working medium,
but for the quasi-incompressible $\bar M_\tsc{cl}\approxeq0.34$ Mach-number considered in the present work mean density $\bar\rho$ variations do not exceed $1.5\%$,
and density fluctuations are negligibly small ($\rho'_{\rm rms}\lessapprox0.25\%\bar\rho$). For this reason
the turbulent correlations obtained can be considered as incompressible. This has been verified by systematic comparison with standard incompressible pseudospectral
\tsc{dns} data \citep{Kim_Moin_Moser_1987a,
                      Moser_Kim_Mansour_1999a,
                      delAlamo_Jimenez_2003a,
                      delAlamo_Jimenez_Zandonade_Moser_2004a,
                      Hoyas_Jimenez_2006a,
                      Hoyas_Jimenez_2008a},
both for single point statistics \citep[all \tsc{som}s\footnote{\label{ff_WEPFTCF_s_DNSCpS_ss_PCFCCM_001}
                                                                second-order moments \citep[$\overline{u_i' u_j'}$: Fig. 7, p. 797,
                                                                                            $\overline{p' u_i'}$: Fig. 8, p. 798,
                                                                                            $\phi_{ij}:=2\overline{p' S_{ij}'}$: Fig. 10, p. 800,
                                                                                            $\varepsilon^{(\mu)}_{ij}:=2\nu\overline{\partial_{x_\ell}u'_i\partial_{x_\ell}u'_j}$: Fig. 11, p. 801]{Gerolymos_Senechal_Vallet_2010a}
                                                               }
                                        and \tsc{tom}s\footnote{\label{ff_WEPFTCF_s_DNSCpS_ss_PCFCCM_002}
                                                                third-order moments \citep[$\overline{u_i' u_j' u_k'}$: Fig. 9, p. 799]{Gerolymos_Senechal_Vallet_2010a}
                                                               }
                                        appearing in the Reynolds-stress budgets]{Gerolymos_Senechal_Vallet_2010a}
and for spectra of velocity fluctuations in the homogeneous streamwise and spanwise directions \citep[Figs. 12--15, pp. 802--805]{Gerolymos_Senechal_Vallet_2010a}.
This very good agreement with pseudospectral incompressible \tsc{dns} \citep{Kim_Moin_Moser_1987a,
                                                                             Hoyas_Jimenez_2008a}
is also valid for 2-point correlations
\begin{equation}
R_{\mathrm{a}'\mathrm{b}'}(\vec{x},\vec{r}):=\overline{\mathrm{a}'(\vec{x},t)\mathrm{b}'(\vec{x}+\vec{r},t)}
                                                                                                                 \label{Eq_WEPFTCF_s_DNSCpS_ss_PCFCCM_001}
\end{equation}
containing the fluctuating pressure $p'$, {\em viz} $[R^{(x,z)}_{p'p'}]^+$, $[R^{(x,z)}_{p'u'}]^+$, and $[R^{(x,z)}_{p'v'}]^+$ \figref{Fig_WEPFTCF_s_DNSCpS_ss_PCFCCM_001}
and fluctuating pressure spectra\footnote{\label{ff_WEPFTCF_s_DNSCpS_ss_PCFCCM_003}
                                          Notice that the spectra given by~\cite{Moser_Kim_Mansour_1999a}
                                          are twice the \tsc{dft} of $R^+_{p'p'}$ (in wall units),
                                          and have to be appropriately rescaled~\citep{Briggs_Henson_1995a_2D_DFT} by $(2\pi)^{-1}L_x^+$ or $(2\pi)^{-1}L_z^+$ to get
                                          actual 1-D spectra $E^+_{u_iu_j}$~\citep[(6.206), p. 225]{Pope_2000a}, as those given by~\cite{delAlamo_Jimenez_Zandonade_Moser_2004a}.
                                         }
$[E^{(x)}_{p'p'}]^+$ and $[E^{(z)}_{p'p'}]^+$ \figref{Fig_WEPFTCF_s_DNSCpS_ss_PCFCCM_002}.
%-----------------------------------------------------------------------------------------------------------------------------------
\begin{figure}
\begin{center}
\begin{picture}(400,235)
\put(-40,-330){\includegraphics[angle=0,width=470pt]{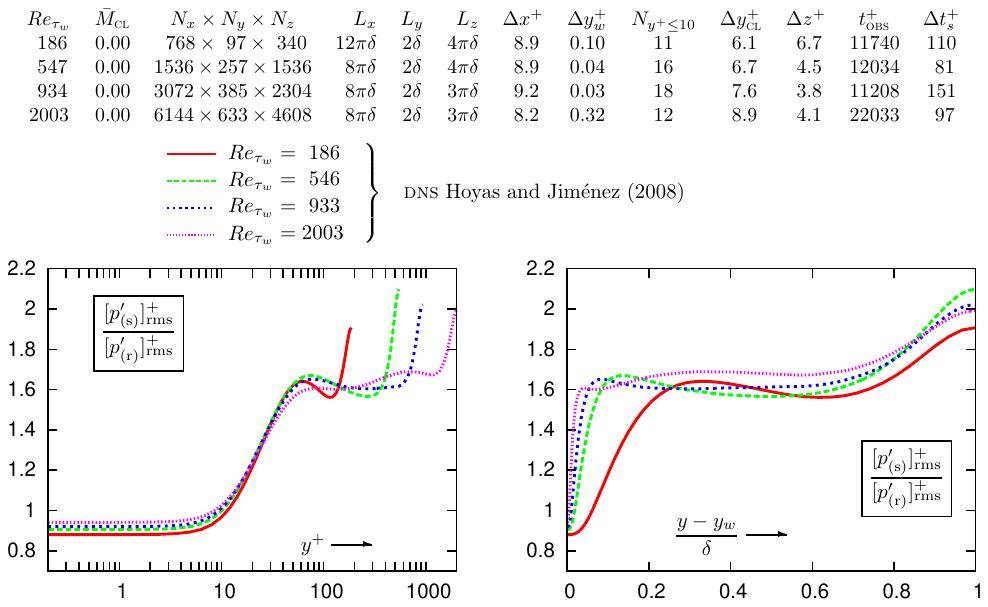}}
\end{picture}
\end{center}
\caption{Ratio of slow-to-rapid pressure fluctuations $[p'_{(r)}]^{-1}_\mathrm{rms}[p'_{(s)}]_\mathrm{rms}$ \eqref{Eq_WEPFTCF_s_I_001}, from the incompressible \tsc{dns} database of \citet{Hoyas_Jimenez_2006a,
                                                                                                                                                                                             Hoyas_Jimenez_2008a},
for different Reynolds numbers ($Re_{\tau_w}\in\{186,547,934,2003\}$),
plotted against the nondimensional distance-from-the-wall in inner ($y^+$) and outer ($\delta^{-1}(y-y_w)$) scaling.}
\label{Fig_WEPFTCF_s_DNSCpS_ss_PCFCCM_003}
\end{figure}
%-----------------------------------------------------------------------------------------------------------------------------------

In order to evaluate more precisely how the small compressibility in the \tsc{dns} computations affects the fluctuating pressure field,
an order-of-magnitude analysis of the complete compressible flow equation for $\nabla^2p'$ \eqref{Eq_WEPFTCF_s_AppendixCELMNL_ss_CFPEqp'_001d} was performed (Appendix \ref{WEPFTCF_s_AppendixCELMNL}),
indicating that the additional compressible terms in \eqref{Eq_WEPFTCF_s_AppendixCELMNL_ss_CFPEqp'_001d} scale with density fluctuations and gradients \eqref{Eq_WEPFTCF_s_AppendixCELMNL_ss_LMNFDPCF_004},
and may  be reasonably neglected for the present flow conditions \figref{Fig_WEPFTCF_s_AppendixCELMNL_ss_LMNFDPCF_001}.
The main effect of compressibility in the present computations, as far as $p'$-splitting is concerned, comes from the variations of mean density $\bar\rho$ ($\lessapprox1.5\%$),
which appears in the source-terms $Q_{(\rm s)}'+\textstyle Q_{(\rm r)}'$ \eqrefsab{Eq_WEPFTCF_s_I_001}{Eq_WEPFTCF_s_AppendixCELMNL_ss_CFPEqp'_001d},
in line with Morkovin's hypothesis \citep{So_Gatski_Sommer_1998a}.

The influence of the Reynolds-number needs consideration, especially as the data used in the present computations correspond to a rather low $Re_{\tau_w}\approxeq180$.
It is well known that wall-pressure fluctuations, in wall-units, $[p'_w]_\mathrm{rms}^+$, show a marked influence on Reynolds-number \citep{Goody_2004a,
                                                                                                                                            Hu_Morfey_Sandham_2006a,
                                                                                                                                            Tsuji_Fransson_Alfredsson_Johansson_2007a},
although alternative scalings exhibit a less pronounced dependence \citep[Fig. 14, p.21]{Tsuji_Fransson_Alfredsson_Johansson_2007a}.
However, the ratio of slow-to-rapid pressure fluctuations $[p'_{(r)}]^{-1}_\mathrm{rms}[p'_{(s)}]_\mathrm{rms}$ \figref{Fig_WEPFTCF_s_DNSCpS_ss_PCFCCM_003}
shows little dependence on $Re_{\tau_w}$ for $y^+\lessapprox100$, ranging from $\sim0.9$ at the wall to $\sim1.6$ at $y^+\approxeq50$,
and is approximately constant $\sim1.6$ in the region $\tfrac{2}{10}\delta\lessapprox y-y_w\lessapprox\tfrac{7}{10}\delta$ \figref{Fig_WEPFTCF_s_DNSCpS_ss_PCFCCM_003}.
At and near the centerline $y=\delta$, there is some scatter between different $Re_{\tau_w}$, but no clear trend is discernible \figref{Fig_WEPFTCF_s_DNSCpS_ss_PCFCCM_003}.
These results, based on the \tsc{dns} data of \citet{Hoyas_Jimenez_2008a}, suggest that, although $p'$ exhibits strong dependence on $Re_{\tau_w}$,
the relative importance of the slow and rapid mechanisms in the creation of $p'$ does not. Therefore, despite the low $Re_{\tau_w}\approxeq180$
of the present \tsc{dns} computations, it is believed that the obtained $p'$-splitting results provide useful information on the relative importance of different mechanisms
in the correlations containing $p'$. Nevertheless, \tsc{dns} data on volume ($p'_{(\mathrm{r};\mathfrak{V})}$ and $p'_{(\mathrm{s};\mathfrak{V})}$) and wall-echo
($p'_{(\mathrm{r};w)}$ and $p'_{(\mathrm{s};w)}$) terms at higher $Re_{\tau_w}$ are required to substantiate this.

%-----------------------------------------------------------------------------------------------------------------------------------
%
\subsection{Green's function solution of the Poisson equations}\label{WEPFTCF_s_DNSCpS_ss_GFSPEq}
%
%-----------------------------------------------------------------------------------------------------------------------------------

For fully developed ($x$-wise invariant in the mean) turbulent plane channel flow, the Poisson equation for $p'$ \eqref{Eq_WEPFTCF_s_I_001} can
be simplified to a system of independent \tsc{ode}s, one for each Fourier-component of $p'$, which can be efficiently solved by a Green's function approach \citep{Kim_1989a}.
In \parrefnp{WEPFTCF_s_DNSCpS_ss_GFSPEq_sss_ODEsFT}, we formalize the problem, in relation to previous work. Using mathematical tools and solutions detailed in Appendix \ref{WEPFTCF_s_AppendixGF},
we discuss, in \parrefnp{WEPFTCF_s_DNSCpS_ss_GFSPEq_sss_Kim}, the standard decomposition $p'=p'_{(\mathrm{r})}+p'_{(\mathrm{s})}+p'_{(\tau)}$ \eqref{Eq_WEPFTCF_s_DNSCpS_ss_GFSPEq_sss_ODEsFT_001},
with special emphasis on the singular case of the $xz$-averaged component $\overline{p'}^{xz}(y,t)$ \eqref{Eq_WEPFTCF_s_DNSCpS_ss_GFSPEq_sss_ODEsFT_005d}.
In \parrefnp{WEPFTCF_s_DNSCpS_ss_GFSPEq_sss_VWET}, we present the new methodology for the identification of volume and wall-echo terms
in the decomposition $p'=p'_{(\mathrm{r};\mathfrak{V})}+p'_{(\mathrm{r};w)}+p'_{(\mathrm{s};\mathfrak{V})}+p'_{(\mathrm{s};w)}+p'_{(\tau)}$ \eqref{Eq_WEPFTCF_s_DNSCpS_ss_GFSPEq_sss_VWET_001}.
In \parrefnp{WEPFTCF_s_DNSCpS_ss_GFSPEq_sss_AMI}, we show how the method-of-images approach applied to fully developed turbulent plane channel flow,
can be represented by an appropriate Green's function.
Finally, in \parrefnp{WEPFTCF_s_DNSCpS_ss_GFSPEq_sss_IWEbW}, we show that the approximation made in the method of images consists of neglecting the interaction of wall-echo between
the 2 walls, and demonstrate by comparison of the exact and approximate Green's functions that this is a high-wavenumber approximation.

%-----------------------------------------------------------------------
%
\subsubsection{ODEs for the Fourier transforms}\label{WEPFTCF_s_DNSCpS_ss_GFSPEq_sss_ODEsFT}
%
%-----------------------------------------------------------------------

The solution to \eqref{Eq_WEPFTCF_s_I_001}, with Neumann boundary-conditions at the walls \citep[p. 439]{Pope_2000a} and periodic boundary-conditions in the homogeneous directions,
can only be obtained up to an additive function of time $t$ \citep{Ince_1926a_1D_Greenfunction_and_compatibility_relation,
                                                                   Courant_Hilbert_1953a_1D_Greenfunction,
                                                                   MyintU_Debnath_2007a_1D_Greenfunction}.
We split the fluctuating pressure field as \citep{Kim_1989a}
\begin{alignat}{6}
p'(x,y,z,t)=p'_{(\mathrm{r})}(x,y,z,t)+p'_{(\mathrm{s})}(x,y,z,t)+p'_{(\tau)}(x,y,z,t)
                                                                                                                 \label{Eq_WEPFTCF_s_DNSCpS_ss_GFSPEq_sss_ODEsFT_001}
\end{alignat}
where the rapid $p'_{(\mathrm{r})}$, slow $p'_{(\mathrm{s})}$ and Stokes $p'_{(\tau)}$ pressure fluctuations
are solutions of~\citep{Kim_1989a,
                        Chang_Piomelli_Blake_1999a}\footnote{\label{ff_WEPFTCF_s_DNSCpS_ss_GFSPEq_sss_ODEsFT_001}
                                                             recall that, for strictly incompressible flow, the continuity equation for the fluctuating velocity field,
                                                             $\partial_{x_\ell}u'_\ell=0$, implies $\partial^2_{x_\ell x_k}(u_k'u_\ell')=\partial_{x_\ell}u_k'\;\partial_{x_k}u_\ell'$
                                                            }
\begin{subequations}
                                                                                                                 \label{Eq_WEPFTCF_s_DNSCpS_ss_GFSPEq_sss_ODEsFT_002}
\begin{alignat}{6}
\nabla^2\left[\begin{array}{c}p'_{(\mathrm{r})}\\
                              p'_{(\mathrm{s})}\\
                              p'_{(\tau)}\\\end{array}\right]=\left[\begin{array}{c}Q'_{(\mathrm{r})}\\
                                                                                    Q'_{(\mathrm{s})}\\
                                                                                    Q'_{(\tau)}      \\\end{array}\right]:=\left[\begin{array}{c}-2\rho\partial_{x_\ell}u_k'\;\partial_{x_k}\bar u_\ell            \\
                                                                                                                                                 -\rho(          \partial_{x_\ell}u_k'\;\partial_{x_k}u_\ell'
                                                                                                                                                      -\overline{\partial_{x_\ell}u_k'\;\partial_{x_k}u_\ell'})\\
                                                                                                                                                                                                            0\\\end{array}\right]
                                                                                                                 \label{Eq_WEPFTCF_s_DNSCpS_ss_GFSPEq_sss_ODEsFT_002a}
\end{alignat}
with boundary-conditions~\citep[pp. 390--392, 439--442]{Pope_2000a}
\begin{alignat}{6}
\dfrac{\partial  }
      {\partial y}\left[\begin{array}{c}p'_{(\mathrm{r})}\\
                                        p'_{(\mathrm{s})}\\
                                        p'_{(\tau)}\\\end{array}\right](x,y=\pm\tfrac{1}{2}L_y,z,t)=\left[\begin{array}{c}0              \\
                                                                                                                          0              \\
                                                                                                                          B'_{(\tau)_\pm}\\\end{array}\right]:=\left[\begin{array}{c}0                  \\
                                                                                                                                                                                     0                  \\
                                                                                                                                                                                    \mu\partial^2_{yy}v'\\\end{array}\right]
                                                                                                                 \label{Eq_WEPFTCF_s_DNSCpS_ss_GFSPEq_sss_ODEsFT_002b}
\end{alignat}
\end{subequations}

Since \eqref{Eq_WEPFTCF_s_DNSCpS_ss_GFSPEq_sss_ODEsFT_002} is linear in $p'_{(m)}$ ($m\in\{\mathrm{r},\mathrm{s},\tau\}$) the 3 problems can be solved independently
to compute the 3 fields in \eqref{Eq_WEPFTCF_s_DNSCpS_ss_GFSPEq_sss_ODEsFT_001}. The last field in the superposition \eqref{Eq_WEPFTCF_s_DNSCpS_ss_GFSPEq_sss_ODEsFT_001}
is introduced to satisfy the wall boundary-conditions associated with the normal-to-the-wall ($y$) momentum equation
\begin{alignat}{6}
\dfrac{\partial p'}
      {\partial y }\Big\rvert_{y=\pm\frac{1}{2}L_y}=\dfrac{\partial \tau_{y\ell}'}
                                                          {\partial      x_\ell  }\Big\rvert_{y=\pm\frac{1}{2}L_y}
                                                   =\mu\nabla^2 v'\Big\rvert_{y=\pm\frac{1}{2}L_y}
                                                   =\mu\dfrac{\partial^2 v'}
                                                             {\partial  y^2}\Big\rvert_{y=\pm\frac{1}{2}L_y}
                                                                                                                 \label{Eq_WEPFTCF_s_DNSCpS_ss_GFSPEq_sss_ODEsFT_003}
\end{alignat}
because of the no-slip wall boundary-condition $u_i'(x,y=\pm\tfrac{1}{2}L_y,z,t)=0\;\forall x,z, t$. By \eqref{Eq_WEPFTCF_s_DNSCpS_ss_GFSPEq_sss_ODEsFT_003} the field $p'_{(\tau)}$ is obviously
related to the fluctuating wall-shear-stress, and is usually called Stokes pressure \citep{Mansour_Kim_Moin_1988a,
                                                                                           Kim_1989a,
                                                                                           Chang_Piomelli_Blake_1999a}
although \citet[p. 439]{Pope_2000a} suggests the alternative term harmonic pressure,
because in incompressible flow $\nabla^2p'_\tau=0$.\footnote{\label{ff_WEPFTCF_s_DNSCpS_ss_GFSPEq_sss_ODEsFT_002}
                                                             This is however no longer true in the compressible flow case \citep{Foysi_Sarkar_Friedrich_2004a}.
                                                            }
                                                                                           
The directions $x$ and $z$ being homogeneous (periodic in the computational model), following \citet{Kim_1989a}, we replace the
$xz$-Fourier-transforms\footnote{\label{ff_WEPFTCF_s_DNSCpS_ss_GFSPEq_sss_ODEsFT_003}
                                 The Fourier-transforms $\hat p_{(m)}', \hat Q_{(m)}', \hat B_{(m)_\pm}'\in\mathbb{C}$ are computed
                                 using standard {\sc dft} (discrete Fourier transform) techniques~\citep{Briggs_Henson_1995a_2D_DFT}
                                 in the periodic directions $x$ and $z$, with maximum computable wavenumbers $\kappa_{x_{\rm max}}=\pi(\Delta x)^{-1}$ and $\kappa_{z_{\rm max}}=\pi(\Delta z)^{-1}$.
                                }
of $p'_{(m)}$, $Q'_{(m)}$ ($m\in\{\mathrm{r},\mathrm{s},\tau\}$) and $B_{(\tau)_\pm}'$
\begin{subequations}
                                                                                                                 \label{Eq_WEPFTCF_s_DNSCpS_ss_GFSPEq_sss_ODEsFT_004}
\begin{alignat}{6}
p_{(m)}'(x,y,z,t)     &=\int_{-\infty}^{+\infty}
                        \int_{-\infty}^{+\infty}\hat p_{(m)}'    (\kappa_x,y,\kappa_z,t)\;e^{i\kappa_x x+i\kappa_z z}\;d\kappa_x\;d\kappa_z
                                                                                                                 \label{Eq_WEPFTCF_s_DNSCpS_ss_GFSPEq_sss_ODEsFT_004a}\\
Q_{(m)}'(x,y,z,t)     &=\int_{-\infty}^{+\infty}
                        \int_{-\infty}^{+\infty}\hat Q_{(m)}'    (\kappa_x,y,\kappa_z,t)\;e^{i\kappa_x x+i\kappa_z z}\;d\kappa_x\;d\kappa_z
                                                                                                                 \label{Eq_WEPFTCF_s_DNSCpS_ss_GFSPEq_sss_ODEsFT_004b}\\
B_{(\tau)_\pm}'(x,z,t)&=\int_{-\infty}^{+\infty}
                        \int_{-\infty}^{+\infty}\hat B_{(\tau)_\pm}'(\kappa_x,\kappa_z,t)\;e^{i\kappa_x x+i\kappa_z z}\;d\kappa_x\;d\kappa_z
                                                                                                                 \label{Eq_WEPFTCF_s_DNSCpS_ss_GFSPEq_sss_ODEsFT_004c}
\end{alignat}
\end{subequations}
in the \tsc{pde}\footnote{\label{ff_WEPFTCF_s_DNSCpS_ss_GFSPEq_sss_ODEsFT_004}
                          \tsc{pde}: partial differential equation; \tsc{ode}: ordinary differential equation
                         }
\eqref{Eq_WEPFTCF_s_DNSCpS_ss_GFSPEq_sss_ODEsFT_002}
to obtain the \tsc{ode}\ref{ff_WEPFTCF_s_DNSCpS_ss_GFSPEq_sss_ODEsFT_004}
\begin{subequations}
                                                                                                                 \label{Eq_WEPFTCF_s_DNSCpS_ss_GFSPEq_sss_ODEsFT_005}
\begin{alignat}{6}
&\left[\dfrac{\partial^2}{\partial y^2}-\kappa^2\right]\left[\begin{array}{c}\hat p'_{(\mathrm{r})}\\
                                                                             \hat p'_{(\mathrm{s})}\\
                                                                             \hat p'_{(\tau)}\\\end{array}\right](\kappa_x,y,\kappa_z,t)=\left[\begin{array}{c}\hat Q'_{(\mathrm{r})}\\
                                                                                                                                                               \hat Q'_{(\mathrm{s})}\\
                                                                                                                                                               \hat 0                \\\end{array}\right](\kappa_x,y,\kappa_z,t)
                                                                                                                 \label{Eq_WEPFTCF_s_DNSCpS_ss_GFSPEq_sss_ODEsFT_005a}\\
&\dfrac{\partial\hat p_{(m)}'}{\partial y}(\kappa_x,y=\pm\tfrac{1}{2}L_y,\kappa_z,t)=\left[\begin{array}{c}0                                        \\
                                                                                                           0                                        \\
                                                                                                           \hat B'_{(\tau)_\pm}(\kappa_x,\kappa_z,t)\\\end{array}\right]
                                                                                                                 \label{Eq_WEPFTCF_s_DNSCpS_ss_GFSPEq_sss_ODEsFT_005b}\\
&\kappa:=\sqrt{\kappa_x^2+\kappa_z^2}\in\mathbb{R}_{\geq0}
                                                                                                                 \label{Eq_WEPFTCF_s_DNSCpS_ss_GFSPEq_sss_ODEsFT_005c}
\end{alignat}

The classical solution of \eqref{Eq_WEPFTCF_s_DNSCpS_ss_GFSPEq_sss_ODEsFT_005} by \cite{Kim_1989a}, using a Green's function approach, provides detailed information
on the structure of the rapid and slow fields, and was extended by \cite{Chang_Piomelli_Blake_1999a} to study the detailed influence of the sources, both in type (slow and rapid)
and $y$-location in the channel.
In the present work we extend this analysis by identifying and studying the wall-echo influence.
Notice that, by \eqref{Eq_WEPFTCF_s_DNSCpS_ss_GFSPEq_sss_ODEsFT_004},
\begin{alignat}{6}
\hat p'(\kappa_x=0,y,\kappa_z=0,t)\stackrel{\eqref{Eq_WEPFTCF_s_DNSCpS_ss_GFSPEq_sss_ODEsFT_004a}}{=}\overline{p'(x,y,z,t)}^{xz}\in\mathbb{R}
                                                                                                                 \label{Eq_WEPFTCF_s_DNSCpS_ss_GFSPEq_sss_ODEsFT_005d}
\end{alignat}
the case $\kappa=0$ corresponds to the time-fluctuation of the $xz$-averaged pressure at each $y$-location,
while
\begin{alignat}{6}
p'_\tsc{b}(t):=\overline{p'(x,y,z,t)}^{xyz}\stackrel{\eqrefsab{Eq_WEPFTCF_s_DNSCpS_ss_GFSPEq_sss_ODEsFT_004a}
                                                              {Eq_WEPFTCF_s_DNSCpS_ss_GFSPEq_sss_ODEsFT_005d}}{=}\dfrac{1}{L_y}\int_{-\tfrac{1}{2}L_y}^{+\tfrac{1}{2}L_y}\hat p'(\kappa_x=0,y,\kappa_z=0,t)\;dy
                                                                                                                 \label{Eq_WEPFTCF_s_DNSCpS_ss_GFSPEq_sss_ODEsFT_005e}
\end{alignat}
is the time-fluctuation of the bulk (volume-averaged) pressure.
\end{subequations}

%-----------------------------------------------------------------------
%
\subsubsection{Exact solution of \eqref{Eq_WEPFTCF_s_DNSCpS_ss_GFSPEq_sss_ODEsFT_005}}\label{WEPFTCF_s_DNSCpS_ss_GFSPEq_sss_Kim}
%
%-----------------------------------------------------------------------

\begin{figure}
\begin{center}
\begin{picture}(400,150)
\put(-20,-310){\includegraphics[angle=0,width=480pt]{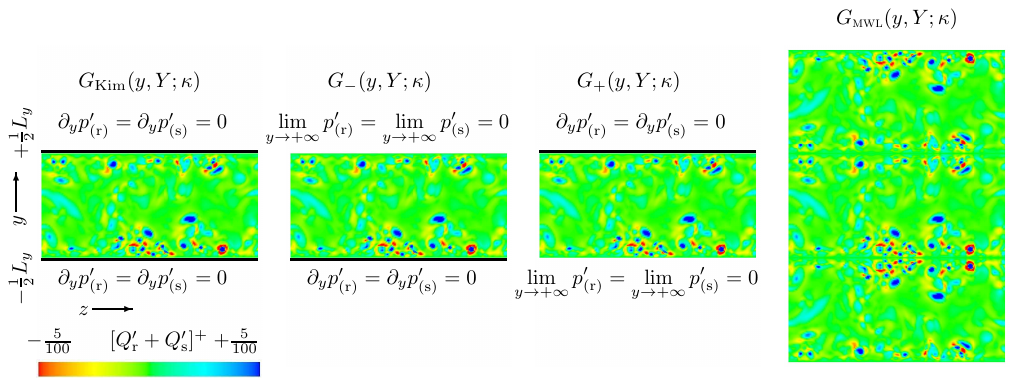}}
\end{picture}
\end{center}
\caption{Instantaneous pressure fluctuation sources (sum $[Q_\mathrm{(r)}'+Q_\mathrm{(s)}']^+$ of rapid and slow sources in wall units) at $x=0$
(\tsc{dns}; $Re_{\tau_w}=179$; $\bar M_\tsc{cl}=0.34$; $193\times129\times169$ grid; \tabrefnp{Tab_WEPFTCF_s_DNSCpS_ss_PCFCCM_001}), and boundary conditions
for the exact problem \eqref{Eq_WEPFTCF_s_DNSCpS_ss_GFSPEq_sss_ODEsFT_005} whose solution \citep{Kim_1989a} is given by $G_\mathrm{Kim}(y,Y;\kappa)$ \eqref{Eq_WEPFTCF_s_AppendixGF_ss_kappaneq0_sss_Kim_002b},
and for the two virtual halfspace problems with only one wall present defining $G_\pm(y,Y;\kappa)$ \eqref{Eq_WEPFTCF_s_AppendixGF_ss_kappaneq0_sss_ULWGF_004},
which define the local wall-corrections appearing in the high-wavenumber approximation of the method of images \citep{Manceau_Wang_Laurence_2001a},
corresponding to the approximate Green's function $G_\tsc{mwl}(y,Y;\kappa)$ \eqref{Eq_WEPFTCF_s_AppendixGF_ss_AEMI_005c}.}
\label{Fig_WEPFTCF_s_DNSCpS_ss_GFSPEq_sss_ODEsFT_001}
\end{figure}

The exact solution to \eqref{Eq_WEPFTCF_s_DNSCpS_ss_GFSPEq_sss_ODEsFT_005a}
with boundary conditions \eqref{Eq_WEPFTCF_s_DNSCpS_ss_GFSPEq_sss_ODEsFT_005b},
for the $xz$-Fourier-components \eqref{Eq_WEPFTCF_s_DNSCpS_ss_GFSPEq_sss_ODEsFT_004a} of the slow $p'_{(\mathrm{s})}$ and rapid $p'_{(\mathrm{r})}$ fields,
is given in \citet{Kim_1989a}, and has been widely used~\citep{Mansour_Kim_Moin_1988a,
                                                               Chang_Piomelli_Blake_1999a,
                                                               Foysi_Sarkar_Friedrich_2004a}.
Following the analysis in \parrefnp{WEPFTCF_s_AppendixGF_ss_kappaneq0_sss_Kim} for $\kappa\neq0$ \eqref{Eq_WEPFTCF_s_AppendixGF_ss_kappaneq0_sss_Kim_002b}
and in \parrefnp{WEPFTCF_s_AppendixGF_ss_kappaeq0_sss_Kim} for $\kappa=0$ \eqref{Eq_WEPFTCF_s_AppendixGF_ss_kappaeq0_sss_Kim_004},
it reads
\begin{subequations}
                                                                                                                 \label{Eq_WEPFTCF_s_DNSCpS_ss_GFSPEq_sss_Kim_001}
\begin{alignat}{6}
\left[\begin{array}{c}\hat p'_{(\mathrm{r})}\\
                      \hat p'_{(\mathrm{s})}\\\end{array}\right](\kappa_x,y,\kappa_z,t)=
\int_{-\tfrac{1}{2}L_y}^{+\tfrac{1}{2}L_y}G_\mathrm{Kim}(y,Y;\kappa)\left[\begin{array}{c}\hat Q'_{(\mathrm{r})}\\
                                                                                          \hat Q'_{(\mathrm{s})}\\\end{array}\right](\kappa_x,Y,\kappa_z,t)\;dY
                                                                                                                 \label{Eq_WEPFTCF_s_DNSCpS_ss_GFSPEq_sss_Kim_001a}
\end{alignat}
where the Green's function
\begin{alignat}{6}
G_{\rm Kim}(y,Y;\kappa)\stackrel{\eqrefsab{Eq_WEPFTCF_s_AppendixGF_ss_kappaneq0_sss_Kim_002b}
                                          {Eq_WEPFTCF_s_AppendixGF_ss_kappaeq0_sss_Kim_004}  }{=}\left\{
                                                                                                 \begin{array}{lc}-\dfrac{\cosh[\kappa(L_y-\abs{y-Y})]+\cosh[\kappa(y+Y)]}
                                                                                                                         {2\kappa\sinh{\kappa L_y}                       }&;\;\kappa\neq0\\
                                                                                                                   \tfrac{1}{2}|y-Y|                                      &;\;\kappa=   0\\\end{array}\right.
                                                                                                                 \label{Eq_WEPFTCF_s_DNSCpS_ss_GFSPEq_sss_Kim_001b}
\end{alignat}
\end{subequations}
and satisfies the correct homogeneous Neumann boundary-conditions on both walls \figref{Fig_WEPFTCF_s_DNSCpS_ss_GFSPEq_sss_ODEsFT_001}.
The exact solution to \eqref{Eq_WEPFTCF_s_DNSCpS_ss_GFSPEq_sss_ODEsFT_005a}, for the $xz$-Fourier-components \eqref{Eq_WEPFTCF_s_DNSCpS_ss_GFSPEq_sss_ODEsFT_004a} of the Stokes (harmonic) pressure,
with boundary-conditions \eqref{Eq_WEPFTCF_s_DNSCpS_ss_GFSPEq_sss_ODEsFT_005b} was given in \citet{Chang_Piomelli_Blake_1999a},
and following the analysis in \parrefnp{WEPFTCF_s_AppendixGF_ss_kappaneq0_sss_ABCs} for $\kappa\neq0$ \eqref{Eq_WEPFTCF_s_AppendixGF_ss_kappaneq0_sss_ABCs_002}
and in \parrefnp{WEPFTCF_s_AppendixGF_ss_kappaeq0_sss_ABCs} for $\kappa=0$ \eqref{Eq_WEPFTCF_s_AppendixGF_ss_kappaeq0_sss_ABCs_003},
reads
\begin{alignat}{6}
&\hat p'_{(\tau)}(\kappa_x,y,\kappa_z,t) \stackrel{
                                         \eqrefsab{Eq_WEPFTCF_s_AppendixGF_ss_kappaneq0_sss_ABCs_002}
                                                  {Eq_WEPFTCF_s_AppendixGF_ss_kappaeq0_sss_ABCs_003}}{=}
                                                                                                                 \label{Eq_WEPFTCF_s_DNSCpS_ss_GFSPEq_sss_Kim_002}\\
                                                                              &\left\{\begin{array}{lc}\dfrac{\hat B_{(\tau)_+}'(\kappa_x,\kappa_z,t)\cosh[\kappa(\tfrac{1}{2}L_y+y)]
                                                                                                             -\hat B_{(\tau)_-}'(\kappa_x,\kappa_z,t)\cosh[\kappa(\tfrac{1}{2}L_y-y)]}
                                                                                                             {\kappa\sinh{\kappa L_y}                                                }&;\;\kappa\neq0\\
                                                                                                       \tfrac{1}{2}\hat B_{(\tau)_-}'(\kappa_x,\kappa_z,t)\;y 
                                                                                                     + \tfrac{1}{2}\hat B_{(\tau)_+}'(\kappa_x,\kappa_z,t) y +p'_{\tsc{bc}_0}(t)       &;\;\kappa=   0\\\end{array}\right.
                                                                                                                 \notag
\end{alignat}
where the choice of the additive constant $p'_{\tsc{bc}_0}\in\mathbb{R}$, up to which the harmonic pressure can be defined \parref{WEPFTCF_s_AppendixGF_ss_kappaeq0_sss_ABCs},
was chosen to satisfy the constraint $p'_\tsc{b}(t):=\overline{p'}^{xyz}(t)=0$ ({\em ie} constant bulk pressure).

As discussed in \parrefnp{WEPFTCF_s_AppendixGF_ss_kappaeq0}, the case $\kappa=0$ is singular, and a solution exists iff the compatibility conditions \eqrefsab{Eq_WEPFTCF_s_AppendixGF_ss_kappaeq0_sss_ABCs_002}
                                                                                                                                                              {Eq_WEPFTCF_s_AppendixGF_ss_kappaeq0_sss_Kim_001}
\begin{subequations}
                                                                                                                 \label{Eq_WEPFTCF_s_DNSCpS_ss_GFSPEq_sss_Kim_003}
\begin{alignat}{6}
L_y
\left[\begin{array}{c}\overline{Q'_{(\mathrm{r})}}^{xyz}\\
                      \overline{Q'_{(\mathrm{s})}}^{xyz}\\\end{array}\right](t)=
\int_{-\tfrac{1}{2}L_y}^{+\tfrac{1}{2}L_y}\left[\begin{array}{c}\hat Q'_{(\mathrm{r})}\\
                                                                \hat Q'_{(\mathrm{s})}\\\end{array}\right](\kappa_x=0,Y,\kappa_z=0,t)\;dY=\left[\begin{array}{c}0\\
                                                                                                                                                                0\\\end{array}\right]
                                                                                                                 \label{Eq_WEPFTCF_s_DNSCpS_ss_GFSPEq_sss_Kim_003a}
\end{alignat}
\begin{alignat}{6}
     \hat B_{(\tau)_+}'(\kappa_x=0,\kappa_z=0,t)=     \hat B_{(\tau)_-}'(\kappa_x=0,\kappa_z=0,t)\iff
                                                                                                                 \label{Eq_WEPFTCF_s_DNSCpS_ss_GFSPEq_sss_Kim_003b}\\
\overline{B_{(\tau)_+}'}^{xz}(t)                =\overline{B_{(\tau)_-}'}^{xz}(t)
                                                                                                                 \notag
\end{alignat}
\end{subequations}
hold,
where $\overline{(\cdot)}^{xyz}$ denotes the volume average\footnote{\label{ff_WEPFTCF_s_DNSCpS_ss_GFSPEq_sss_Kim_001}
                                                                    $\displaystyle\overline{(\cdot)}^{xyz}:=\dfrac{1            }
                                                                                                                  {L_x\;L_y\;L_z}\int_{-\tfrac{1}{2}L_x}^{+\tfrac{1}{2}L_x}
                                                                                                                                 \int_{-\tfrac{1}{2}L_y}^{+\tfrac{1}{2}L_y}
                                                                                                                                 \int_{-\tfrac{1}{2}L_z}^{+\tfrac{1}{2}L_z}(\cdot)dx\;dy\;dz$\\
                                                                    $\displaystyle\overline{(\cdot)}^{xz}:=\dfrac{1       }
                                                                                                                 {L_x\;L_z}\int_{-\tfrac{1}{2}L_x}^{+\tfrac{1}{2}L_x}
                                                                                                                           \int_{-\tfrac{1}{2}L_z}^{+\tfrac{1}{2}L_z}(\cdot)dx\;dz$
                                                                    }
(bulk average) and $\overline{(\cdot)}^{xz}$ is the surface average\ref{ff_WEPFTCF_s_DNSCpS_ss_GFSPEq_sss_Kim_001}
in the homogeneous directions $x$ and $z$. If the computational box is large enough in the homogeneous directions for the ergodic hypothesis \citep[p. 243--256]{Monin_Yaglom_1971a} to hold,
\eqref{Eq_WEPFTCF_s_DNSCpS_ss_GFSPEq_sss_Kim_003} are satisfied, at least approximately.
In practice, the compatibility condition \eqref{Eq_WEPFTCF_s_DNSCpS_ss_GFSPEq_sss_Kim_003a} was enforced numerically when computing the Fourier-transforms \eqref{Eq_WEPFTCF_s_DNSCpS_ss_GFSPEq_sss_ODEsFT_004b}.
The form \eqref{Eq_WEPFTCF_s_AppendixGF_ss_kappaeq0_sss_ABCs_003} of $q_\tsc{bc}(y;\kappa=0,B_\pm)$ automatically enforces
the compatibility condition, by taking the average gradient $\tfrac{1}{2}(B_-+B_+)$.
In practice $\hat B_{(\tau)_-}'(\kappa_x=0,\kappa_z=0,t)\approxeq\hat B_{(\tau)_+}'(\kappa_x=0,\kappa_z=0,t)\approxeq0$ in \eqref{Eq_WEPFTCF_s_DNSCpS_ss_GFSPEq_sss_Kim_003b},
especially with increasing box size, so that taking the average is a good choice. Of course, \eqref{Eq_WEPFTCF_s_I_001}
with Neumann boundary conditions at the walls~\citep[p. 439]{Pope_2000a}, can only be solved up to an additive constant, which is fixed by the choice of $p'_{\tsc{bc}_0}$ in the
Stokes field $p'_{(\tau)}$ \eqref{Eq_WEPFTCF_s_DNSCpS_ss_GFSPEq_sss_Kim_002} ({\em ie} constant bulk pressure).
%-----------------------------------------------------------------------------------------------------------------------------------
%-----------------------------------------------------------------------------------------------------------------------------------

%-----------------------------------------------------------------------
%
\subsubsection{Volume and wall-echo terms}\label{WEPFTCF_s_DNSCpS_ss_GFSPEq_sss_VWET}
%
%-----------------------------------------------------------------------

The main purpose of the present algorithm is to distinguish in \eqref{Eq_WEPFTCF_s_I_002} the contributions of the volume integral (volume terms denoted by $(\cdot)_\mathfrak{V}$)
from the contributions of the surface integral (wall-echo terms denoted by $(\cdot)_w$), {\em ie} to extend \eqref{Eq_WEPFTCF_s_DNSCpS_ss_GFSPEq_sss_ODEsFT_001} as
\begin{alignat}{6}
p'(x,y,z,t)=&\underbrace{p'_{(\mathrm{r};\mathfrak{V})}(x,y,z,t)+p'_{(\mathrm{r};w)}(x,y,z,t)}_{\displaystyle p'_{(\mathrm{r})}(x,y,z,t)}
                                                                                                                 \notag\\
           +&\underbrace{p'_{(\mathrm{s};\mathfrak{V})}(x,y,z,t)+p'_{(\mathrm{s};w)}(x,y,z,t)}_{\displaystyle p'_{(\mathrm{s})}(x,y,z,t)}+p'_{(\tau)}(x,y,z,t)
                                                                                                                 \label{Eq_WEPFTCF_s_DNSCpS_ss_GFSPEq_sss_VWET_001}
\end{alignat}
where
$p'_{(\mathrm{r};\mathfrak{V})}$ corresponds to the contribution of the volume  integral in \eqref{Eq_WEPFTCF_s_I_002a},
$p'_{(\mathrm{r};w)}$            corresponds to the contribution of the surface integral in \eqref{Eq_WEPFTCF_s_I_002a},
$p'_{(\mathrm{s};\mathfrak{V})}$ corresponds to the contribution of the volume  integral in \eqref{Eq_WEPFTCF_s_I_002b}, and
$p'_{(\mathrm{s};w)}$            corresponds to the contribution of the surface integral in \eqref{Eq_WEPFTCF_s_I_002b}.

Working directly with the surface integrals in \eqref{Eq_WEPFTCF_s_I_002} would have been a complex task, because they are implicit, {\em ie} they contain the
value of the corresponding field at the wall. On the other hand, it is straightforward to evaluate the volume integrals in \eqref{Eq_WEPFTCF_s_I_002},
which contain only the sources and the 3-D freespace Green's function, which satisfies $\lim_{\abs{\vec{\mathfrak{x}}-\vec{x}}\to\infty}\abs{\vec{\mathfrak{x}}-\vec{x}}^{-1}=0$.
Therefore the $xz$-Fourier-components of the volume terms should satisfy \eqref{Eq_WEPFTCF_s_DNSCpS_ss_GFSPEq_sss_ODEsFT_005a},
corresponding to the same distribution of sources as the complete physical problem, for $y\in[-\tfrac{1}{2}L_y,+\tfrac{1}{2}L_y]$
\begin{subequations}
                                                                                                                 \label{Eq_WEPFTCF_s_DNSCpS_ss_GFSPEq_sss_VWET_002}
\begin{alignat}{6}
&\left[\dfrac{\partial^2}{\partial y^2}-\kappa^2\right]\left[\begin{array}{c}\hat p'_{(\mathrm{r};\mathfrak{V})}\\
                                                                             \hat p'_{(\mathrm{s};\mathfrak{V})}\\
                                                                                               \end{array}\right](\kappa_x,y,\kappa_z,t)=\left[\begin{array}{c}\hat Q'_{(\mathrm{r})}\\
                                                                                                                                                               \hat Q'_{(\mathrm{s})}\\
                                                                                                                                                                                     \\\end{array}\right](\kappa_x,y,\kappa_z,t)
                                                                                                                 \label{Eq_WEPFTCF_s_DNSCpS_ss_GFSPEq_sss_VWET_002a}
\end{alignat}
but not the homogeneous Neumann wall-boundary-conditions \eqref{Eq_WEPFTCF_s_DNSCpS_ss_GFSPEq_sss_ODEsFT_005b}, which are directly related to the surface integrals in \eqref{Eq_WEPFTCF_s_I_002}.
They are required instead to decay as $\abs{y}\to\infty$,
\begin{alignat}{6}
\lim_{\abs{y}\to\infty}\left[\begin{array}{c}\hat p'_{(\mathrm{r};\mathfrak{V})}\\
                                                                             \hat p'_{(\mathrm{s};\mathfrak{V})}\\
                                                                                               \end{array}\right](\kappa_x,y,\kappa_z,t)=0
                                                                                                                 \label{Eq_WEPFTCF_s_DNSCpS_ss_GFSPEq_sss_VWET_002b}
\end{alignat}
{\em ie} they correspond to the solution of the hypothetical problem where $p'$ is generated by the same distribution of sources in $y\in[-\tfrac{1}{2}L_y,+\tfrac{1}{2}L_y]$ with the walls absent.
\end{subequations}
Following the analysis in \parrefnp{WEPFTCF_s_AppendixGF_ss_kappaneq0_sss_FSGF} for $\kappa\neq0$ \eqref{Eq_WEPFTCF_s_AppendixGF_ss_kappaneq0_sss_FSGF_002}
and in \parrefnp{WEPFTCF_s_AppendixGF_ss_kappaeq0_sss_FSGF} for $\kappa=0$ \eqref{Eq_WEPFTCF_s_AppendixGF_ss_kappaeq0_sss_FSGF_002},
the solution of \eqref{Eq_WEPFTCF_s_DNSCpS_ss_GFSPEq_sss_VWET_002} reads
\begin{subequations}
                                                                                                                 \label{Eq_WEPFTCF_s_DNSCpS_ss_GFSPEq_sss_VWET_003}
\begin{alignat}{6}
\left[\begin{array}{c}\hat p'_{(\mathrm{r};\mathfrak{V})}\\
                      \hat p'_{(\mathrm{s};\mathfrak{V})}\\\end{array}\right](\kappa_x,y,\kappa_z,t)=
\int_{-\tfrac{1}{2}L_y}^{+\tfrac{1}{2}L_y}G_\mathfrak{V}(y,Y;\kappa)\left[\begin{array}{c}\hat Q'_{(\mathrm{r})}\\
                                                                                          \hat Q'_{(\mathrm{s})}\\\end{array}\right](\kappa_x,Y,\kappa_z,t)\;dY
                                                                                                                 \label{Eq_WEPFTCF_s_DNSCpS_ss_GFSPEq_sss_VWET_003a}
\end{alignat}
where the freespace Green's function
\begin{alignat}{6}
G_\mathfrak{V}(y,Y;\kappa)\stackrel{\eqrefsab{Eq_WEPFTCF_s_AppendixGF_ss_kappaneq0_sss_FSGF_002}
                                             {Eq_WEPFTCF_s_AppendixGF_ss_kappaeq0_sss_FSGF_002} }{=}\left\{
                                                                                                 \begin{array}{lc}-\dfrac{\mathrm{e}^{-\kappa|y-Y|}}{2\kappa}             &;\;\kappa\neq0\\
                                                                                                                                                                          &              \\
                                                                                                                   \tfrac{1}{2}|y-Y|                                      &;\;\kappa=   0\\\end{array}\right.
                                                                                                                 \label{Eq_WEPFTCF_s_DNSCpS_ss_GFSPEq_sss_Kim_001b}
\end{alignat}
ensures boundedness at infinity. The same compatibility relation for the distribution of sources, as for the complete problem \eqref{Eq_WEPFTCF_s_DNSCpS_ss_GFSPEq_sss_Kim_003a},
is required for the case $\kappa=0$ \parref{WEPFTCF_s_AppendixGF_ss_kappaeq0_sss_FSGF}.
\end{subequations}
 
Because of the linearity of \eqrefsab{Eq_WEPFTCF_s_I_001}
                                                        {Eq_WEPFTCF_s_I_002},
the splitting \eqref{Eq_WEPFTCF_s_DNSCpS_ss_GFSPEq_sss_VWET_001}
readily implies
\begin{alignat}{6}
\left[\begin{array}{c}\hat p'_{(\mathrm{r};w)}\\
                      \hat p'_{(\mathrm{s};w)}\\\end{array}\right](\kappa_x,y,\kappa_z,t)=\left[\begin{array}{c}\hat p'_{(\mathrm{r}             )}\\
                                                                                                                \hat p'_{(\mathrm{s}             )}\\\end{array}\right](\kappa_x,y,\kappa_z,t)
                                                                                         -\left[\begin{array}{c}\hat p'_{(\mathrm{r};\mathfrak{V})}\\
                                                                                                                \hat p'_{(\mathrm{s};\mathfrak{V})}\\\end{array}\right](\kappa_x,y,\kappa_z,t)
                                                                                                                 \label{Eq_WEPFTCF_s_DNSCpS_ss_GFSPEq_sss_VWET_004}
\end{alignat}
Combining \eqrefsabc{Eq_WEPFTCF_s_DNSCpS_ss_GFSPEq_sss_Kim_001}
                    {Eq_WEPFTCF_s_DNSCpS_ss_GFSPEq_sss_VWET_003}
                    {Eq_WEPFTCF_s_DNSCpS_ss_GFSPEq_sss_VWET_004}
we may define the wall-echo Green's function
\begin{alignat}{6}
G_w(y,Y;\kappa):=&G_\mathrm{Kim}(y,Y;\kappa)-G_\mathfrak{V}(y,Y;\kappa)
                                                                                                                 \label{Eq_WEPFTCF_s_DNSCpS_ss_GFSPEq_sss_VWET_005}
\end{alignat}
corresponding to the surface integrals (wall-echo) in \eqref{Eq_WEPFTCF_s_I_002}, which satisfies
\begin{alignat}{6}
\left[\begin{array}{c}\hat p'_{(\mathrm{r};w)}\\
                      \hat p'_{(\mathrm{s};w)}\\\end{array}\right](\kappa_x,y,\kappa_z,t)=
\int_{-\tfrac{1}{2}L_y}^{+\tfrac{1}{2}L_y}G_w(y,Y;\kappa)\left[\begin{array}{c}\hat Q'_{(\mathrm{r})}\\
                                                                                \hat Q'_{(\mathrm{s})}\\\end{array}\right](\kappa_x,Y,\kappa_z,t)\;dY
                                                                                                                 \label{Eq_WEPFTCF_s_DNSCpS_ss_GFSPEq_sss_VWET_006}
\end{alignat}
Notice that since, for $\kappa=0$,  $G_\mathrm{Kim}(y,Y;\kappa=0)=G_\mathfrak{V}(y,Y;\kappa=0)$ \eqref{Eq_WEPFTCF_s_AppendixGF_ss_kappaeq0_sss_ULWGF_001}, we have $G_w(y,Y;\kappa=0)=0$,
{\em ie} wall-echo applies only on fields varying in at least one of the homogeneous directions $x$ or $z$ ($\kappa\neq 0$).

%-----------------------------------------------------------------------
%
\subsubsection{The approximate method of images}\label{WEPFTCF_s_DNSCpS_ss_GFSPEq_sss_AMI}
%
%-----------------------------------------------------------------------

With the above developments we can directly evaluate the wall-echo terms for the slow and rapid parts in each of the correlations containing $p'$ (\S\ref{WEPFTCF_s_APC}).
By comparing with the exact solution obtained in the present work, it is possible to evaluate
the approximation error of the method of images of \cite{Manceau_Wang_Laurence_2001a}.

The approximate method of images \citep{Manceau_Wang_Laurence_2001a} simply adds two mirror images of the channel,
one above the upper wall and another below the lower wall \figref{Fig_WEPFTCF_s_DNSCpS_ss_GFSPEq_sss_ODEsFT_001},
to account for the presence of the walls, postulating that an approximation to the solution is obtained by using the freespace Green's function $G_\mathfrak{V}(y,Y;\kappa)$
in the extended domain $y\in[-\tfrac{3}{2}L_y,+\tfrac{3}{2}L_y]$. Taking into account the mirror symmetry of the ghost channels with respect to the corresponding wall \figref{Fig_WEPFTCF_s_DNSCpS_ss_GFSPEq_sss_ODEsFT_001},
this is equivalent (\S\ref{WEPFTCF_s_AppendixGF_ss_AEMI}) to using an appropriate Green's function $G_\tsc{mwl}(y,Y;\kappa)$ \eqref{Eq_WEPFTCF_s_AppendixGF_ss_AEMI_005c}
on the actual sources between the channel walls, {\em ie} for $y\in[-\tfrac{1}{2}L_y,+\tfrac{1}{2}L_y]$
\begin{alignat}{6}
\left[\begin{array}{c}\hat p'_{(\mathrm{r};\tsc{mwl})}]\\
                      \hat p'_{(\mathrm{s};\tsc{mwl})}]\\\end{array}\right](\kappa_x,y,\kappa_z,t)=
\int_{-\tfrac{1}{2}L_y}^{+\tfrac{1}{2}L_y}G_\tsc{mwl}(y,Y;\kappa)\left[\begin{array}{c}\hat Q'_{(\mathrm{r})}\\
                                                                                       \hat Q'_{(\mathrm{s})}\\\end{array}\right](\kappa_x,Y,\kappa_z,t)\;dY
                                                                                                                 \label{Eq_WEPFTCF_s_DNSCpS_ss_GFSPEq_sss_AMI_001}
\end{alignat}
By \eqref{Eq_WEPFTCF_s_AppendixGF_ss_AEMI_005c}
\begin{alignat}{6}
G_\tsc{mwl}(y,Y;\kappa)=G_\mathfrak{V}(y,Y;\kappa)+G_{w_-}(y,Y;\kappa)+G_{w_+}(y,Y;\kappa)
                                                                                                                 \label{Eq_WEPFTCF_s_DNSCpS_ss_GFSPEq_sss_AMI_002}
\end{alignat}
where $G_{w_\pm}(y,Y;\kappa)$ \eqref{Eq_WEPFTCF_s_AppendixGF_ss_kappaneq0_sss_ULWGF_005}
 are the wall-echo Green's functions corresponding to the virtual halfspace problems with only one of the walls present \figref{Fig_WEPFTCF_s_DNSCpS_ss_GFSPEq_sss_ODEsFT_001}.
Notice that the halfspace problems are solved exactly
(\S\ref{WEPFTCF_s_AppendixGF_ss_kappaneq0_sss_ULWGF}, \S\ref{WEPFTCF_s_AppendixGF_ss_kappaeq0_sss_ULWGF}).
Inherently, this is tantamount to assuming that there is no interaction between the echo effects of the upper and lower walls,
since in the case of an isolated wall (halfspace problem; \figrefnp{Fig_WEPFTCF_s_DNSCpS_ss_GFSPEq_sss_ODEsFT_001}) the method of images yields
the exact solution \citep[pp. 439--442]{Pope_2000a}.
%-----------------------------------------------------------------------------------------------------------------------------------
\begin{figure}
\begin{center}
\begin{picture}(400,380)
\put(-40,-110){\includegraphics[angle=0,width=470pt]{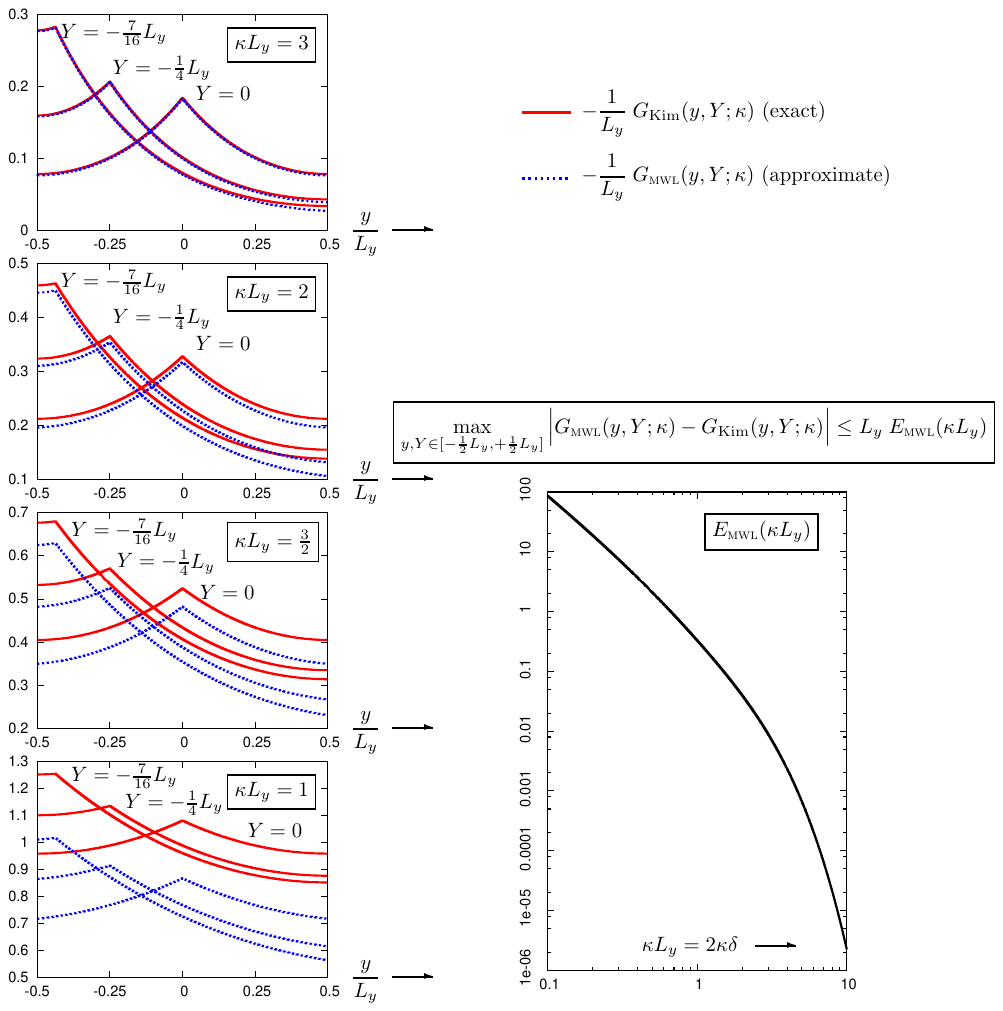}}
\end{picture}
\end{center}
\caption{Comparison of the exact Green's function \eqref{Eq_WEPFTCF_s_AppendixGF_ss_kappaneq0_sss_Kim_002b}, $G_\mathrm{Kim}(y,Y;\kappa)$ \citep{Kim_1989a},
with the method-of-images approximation \eqref{Eq_WEPFTCF_s_AppendixGF_ss_AEMI_005}, $G_\tsc{mwl}(y,Y;\kappa)$ \citep{Manceau_Wang_Laurence_2001a},
for $Y\in\{-\tfrac{7}{16}L_y,-\tfrac{1}{4}L_y,0\}$ and $\kappa L_y\in\{1,\tfrac{3}{2},2,3\}$, plotted as a function of $yL_y^{-1}$,
and log-plot of the upper bound estimate of the error made by the method-of-images approximation \eqref{Eq_WEPFTCF_s_AppendixGF_ss_AEMI_008}
as a function of the nondimensional wavenumber $\kappa L_y=2\kappa\delta$ ($L_y=2\delta$ is the channel height).}
\label{Fig_WEPFTCF_s_DNSCpS_ss_GFSPEq_sss_AMI_001}
\end{figure}
%-----------------------------------------------------------------------------------------------------------------------------------

The approximation error of the method of images comes from the fact that the direct influence of the upper wall,
approximated by the halfspace problem $G_{w_+}(y,Y;\kappa)$ \figref{Fig_WEPFTCF_s_DNSCpS_ss_GFSPEq_sss_ODEsFT_001},
induces a nonzero gradient $\partial_y [\hat p'_{(\mathrm{r};+)}, \hat p'_{(\mathrm{s};+)}]^\tsc{t}(\kappa_x,y=-\tfrac{1}{2}L_y,\kappa_z,t)\neq0$ at the lower wall,
and that there is no feedback from the lower wall to correct this (and {\em vice versa}). The nondimensional approximation error can be evaluated (\S\ref{WEPFTCF_s_AppendixGF_ss_AEMI})
by comparison with the exact solution \eqref{Eq_WEPFTCF_s_AppendixGF_ss_kappaneq0_sss_Kim_002b} of \cite{Kim_1989a},
and is a function of the nondimensional wavenumber $\kappa L_y=2\kappa\delta$ \eqref{Eq_WEPFTCF_s_AppendixGF_ss_AEMI_008}.
For large nondimensional wavenumbers $\kappa L_y$, {\em ie} structures with small streamwise and spanwise extent compared to the channel height,
this error is small, and rapidly decreases with increasing wavenumber \figref{Fig_WEPFTCF_s_DNSCpS_ss_GFSPEq_sss_AMI_001}.

Let $\ell_\kappa:=2\pi\kappa^{-1}$ be the representative size (wavelength) in the $xz$ plane of structures corresponding to 
$\kappa:=\sqrt{\kappa_x^2+\kappa_z^2}$ \eqref{Eq_WEPFTCF_s_DNSCpS_ss_GFSPEq_sss_ODEsFT_005c}.
For $\kappa L_y\gtrapprox3\iff\ell_\kappa\lessapprox4\delta$,
the approximation of $G_\mathrm{Kim}(y,Y;\kappa)$ \eqref{Eq_WEPFTCF_s_AppendixGF_ss_kappaneq0_sss_Kim_002b} by $G_\tsc{mwl}(y,Y;\kappa)$ \eqref{Eq_WEPFTCF_s_AppendixGF_ss_AEMI_005c}
is satisfactory \figref{Fig_WEPFTCF_s_DNSCpS_ss_GFSPEq_sss_AMI_001}, the nondimensional error being $\lessapprox1\%$. For the present case with $Re_{\tau_w}\approxeq180$ this corresponds roughly to $\kappa^+\gtrapprox\tfrac{1}{100}$,
which includes the wavenumbers where most of the energy of the $p'$-spectra is contained \figref{Fig_WEPFTCF_s_DNSCpS_ss_PCFCCM_002},
implying that the method of images is a satisfactory engineering approximation. However, for
larger structures, {\em eg} $\kappa L_y\approxeq\tfrac{3}{2}\iff\ell_\kappa\approxeq8\delta$,
such as the superstructures observed in higher-$Re_{\tau_w}$ wall turbulence \citep{Hutchins_Marusic_2007a,
                                                                                    Balakumar_Adrian_2007a},
the approximation error is roughly $10\%$, and then grows exponentially with increasing size.
%-----------------------------------------------------------------------------------------------------------------------------------
\begin{figure}
\begin{center}
\begin{picture}(400,200)
\put(-50,-300){\includegraphics[angle=0,width=480pt]{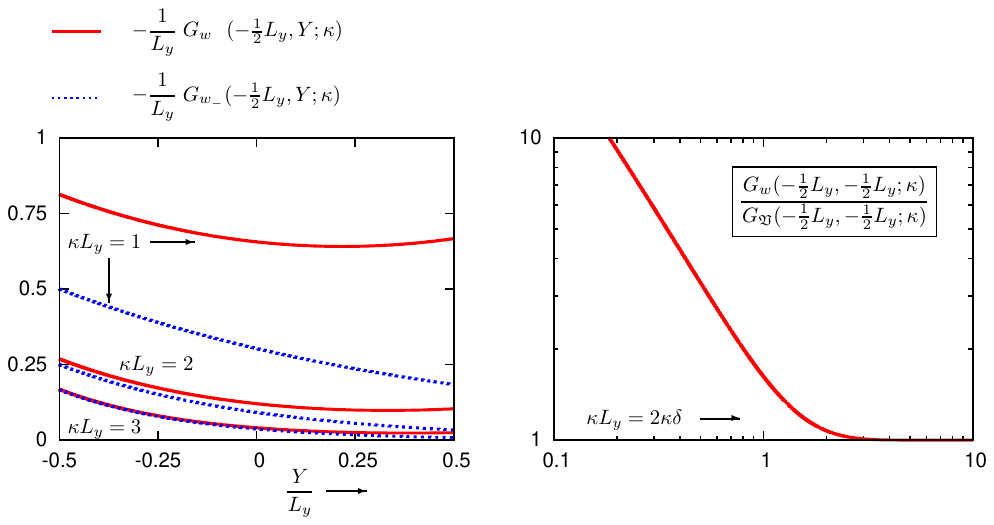}}
\end{picture}
\end{center}
\caption{Comparison of the exact wall-echo Green's function $G_w(y=-\tfrac{1}{2}L_y,Y;\kappa)$ \eqref{Eq_WEPFTCF_s_DNSCpS_ss_GFSPEq_sss_AMI_002},
for the computation of the solution at the lower wall $y=-\tfrac{1}{2}L_y$,
with the wall-echo Green's function $G_{w_-}(y=-\tfrac{1}{2}L_y,Y;\kappa)$ \eqref{Eq_WEPFTCF_s_AppendixGF_ss_AEMI_005},
of the corresponding halfspace problem with only the lower wall present \figref{Fig_WEPFTCF_s_DNSCpS_ss_GFSPEq_sss_ODEsFT_001},
for $\kappa L_y\in\{1,2,3\}$, plotted as a function of $YL_y^{-1}$,
and log-plot of the ratio at the wall ($y=Y=-\tfrac{1}{2}L_y$)
of the exact wall-echo Green's function $G_w(y=-\tfrac{1}{2}L_y,Y=-\tfrac{1}{2}L_y;\kappa)$ \eqref{Eq_WEPFTCF_s_DNSCpS_ss_GFSPEq_sss_AMI_002}
on the corresponding freespace Green's function $G_\mathfrak{V}(y=-\tfrac{1}{2}L_y,Y=-\tfrac{1}{2}L_y;\kappa)$ \eqref{Eq_WEPFTCF_s_AppendixGF_ss_kappaneq0_sss_FSGF_002}
plotted against the nondimensional wavenumber $\kappa L_y=2\kappa\delta$ ($L_y=2\delta$ is the channel height).}
\label{Fig_WEPFTCF_s_DNSCpS_ss_GFSPEq_sss_IWEbW_001}
\end{figure}
%-----------------------------------------------------------------------------------------------------------------------------------

%-----------------------------------------------------------------------------------------------------------------------------------
%
\subsubsection{Interaction of wall-echo between walls}\label{WEPFTCF_s_DNSCpS_ss_GFSPEq_sss_IWEbW}
%
%-----------------------------------------------------------------------------------------------------------------------------------

Since the freespace Green's function $G_\mathfrak{V}(y,Y;\kappa)$ \eqrefsab{Eq_WEPFTCF_s_AppendixGF_ss_kappaneq0_sss_FSGF_002}{Eq_WEPFTCF_s_AppendixGF_ss_kappaeq0_sss_ULWGF_001}
is the same for all problems \figref{Fig_WEPFTCF_s_DNSCpS_ss_GFSPEq_sss_ODEsFT_001}, the differences \figref{Fig_WEPFTCF_s_DNSCpS_ss_GFSPEq_sss_AMI_001} between
the exact solution $G_\mathrm{Kim}(y,Y;\kappa)$ \eqref{Eq_WEPFTCF_s_AppendixGF_ss_kappaneq0_sss_Kim_002b} and
the method-of-images approximation $G_\tsc{mwl}(y,Y;\kappa)$ \eqref{Eq_WEPFTCF_s_AppendixGF_ss_AEMI_005},
corresponds to increasingly strong interaction between the two walls with decreasing nondimensional wavenumber $\kappa L_y=2\kappa\delta$,
leading to amplification of the echo effect.
To further explain this phenomenon notice the identity
\begin{alignat}{6}
G_\mathfrak{V}(y=-\tfrac{1}{2}L_y,Y;\kappa\neq0)\stackrel{\eqref{Eq_WEPFTCF_s_AppendixGF_ss_kappaneq0_sss_FSGF_002}}{=}&
                                                -\dfrac{\mathrm{e}^{\kappa(Y+\tfrac{1}{2}L_y)}}{2\kappa}
                                                                                                                              \notag\\
                                                \stackrel{\eqref{Eq_WEPFTCF_s_AppendixGF_ss_kappaneq0_sss_ULWGF_004a}}{=}&
                                                G_-(y=-\tfrac{1}{2}L_y,Y;\kappa\neq0)-G_\mathfrak{V}(y=-\tfrac{1}{2}L_y,Y;\kappa\neq0)
                                                                                                                              \notag\\
                                                \stackrel{\eqref{Eq_WEPFTCF_s_AppendixGF_ss_AEMI_005c}}{=:}&G_{w_-}(y=-\tfrac{1}{2}L_y,Y;\kappa\neq0)\quad\forall Y\in(-\tfrac{1}{2}L_y,+\tfrac{1}{2}L_y)
                                                                                                                              \label{Eq_Eq_WEPFTCF_s_DNSCpS_ss_GFSPEq_sss_IWEbW_001}
\end{alignat}
which implies by \eqref{Eq_WEPFTCF_s_AppendixGF_ss_GFSMHEq_003} that for the halfspace problems (\S\ref{WEPFTCF_s_AppendixGF_ss_kappaneq0_sss_ULWGF}) the wall-echo, at the wall ($y=-\tfrac{1}{2}L_y$), exactly equals the volume term at the wall.
\begin{subequations}
                                                                                                                              \label{Eq_Eq_WEPFTCF_s_DNSCpS_ss_GFSPEq_sss_IWEbW_002}
\begin{alignat}{6}
p'_{(\mathrm{s};w_-)}(x,y=-\tfrac{1}{2}L_y,z,t)\stackrel{\eqrefsab{Eq_Eq_WEPFTCF_s_DNSCpS_ss_GFSPEq_sss_IWEbW_001}{Eq_WEPFTCF_s_AppendixGF_ss_GFSMHEq_003}}{=}&p'_{(\mathrm{s};\mathfrak{V})}(x,y=-\tfrac{1}{2}L_y,z,t)
                                                                                                                              \label{Eq_Eq_WEPFTCF_s_DNSCpS_ss_GFSPEq_sss_IWEbW_002a}\\
p'_{(\mathrm{r};w_-)}(x,y=-\tfrac{1}{2}L_y,z,t)\stackrel{\eqrefsab{Eq_Eq_WEPFTCF_s_DNSCpS_ss_GFSPEq_sss_IWEbW_001}{Eq_WEPFTCF_s_AppendixGF_ss_GFSMHEq_003}}{=}&p'_{(\mathrm{r};\mathfrak{V})}(x,y=-\tfrac{1}{2}L_y,z,t)
                                                                                                                              \label{Eq_Eq_WEPFTCF_s_DNSCpS_ss_GFSPEq_sss_IWEbW_002b}
\end{alignat}
\end{subequations}
This result \eqref{Eq_Eq_WEPFTCF_s_DNSCpS_ss_GFSPEq_sss_IWEbW_002} following from the equality of the corresponding Green's functions \eqref{Eq_Eq_WEPFTCF_s_DNSCpS_ss_GFSPEq_sss_IWEbW_001}
is valid independently of the particular sources $\hat{Q}(y;\kappa)$ \eqref{Eq_WEPFTCF_s_DNSCpS_ss_GFSPEq_sss_ODEsFT_005a}.

At high nondimensional wavenumbers $\kappa L_y$ the exact wall-echo Green's function $G_w(y=-\tfrac{1}{2}L_y,Y;\kappa)$ \eqref{Eq_WEPFTCF_s_DNSCpS_ss_GFSPEq_sss_AMI_002},
for the computation of $p_w'$ at the lower wall ($y=-\tfrac{1}{2}L_y$),
is approximately equal ($\kappa L_y=3$; \figrefnp{Fig_WEPFTCF_s_DNSCpS_ss_GFSPEq_sss_IWEbW_001})
to the wall-echo Green's function $G_{w_-}(y=-\tfrac{1}{2}L_y,Y;\kappa)$ \eqref{Eq_WEPFTCF_s_AppendixGF_ss_AEMI_005} of the halfspace problem with
only the lower wall present \figref{Fig_WEPFTCF_s_DNSCpS_ss_GFSPEq_sss_ODEsFT_001},
except at the upper part of the channel $Y\in[\tfrac{1}{4}L_y,\tfrac{1}{2}L_y]$, where the influence of the upper wall on the Green's function used
for the computation of $p_w'$ at the lower wall ($y=-\tfrac{1}{2}L_y$) is felt.
As the nondimensional wavenumber $\kappa L_y$ further decreases ({\em ie} the corresponding structure size $\ell_\kappa\delta^{-1}:=2\pi\kappa^{-1}\delta^{-1}$ increases)
the influence of the upper wall is felt further down \figref{Fig_WEPFTCF_s_DNSCpS_ss_GFSPEq_sss_ODEsFT_001}. Of course there is an analogous effect concerning the influence of the lower wall on the upper wall.
As the nondimensional wavenumber $\kappa L_y$ decreases this mutual interaction between the two walls amplifies the echo effect,
the amplification growing exponentially with structure size \figref{Fig_WEPFTCF_s_DNSCpS_ss_GFSPEq_sss_ODEsFT_001}.
%-----------------------------------------------------------------------------------------------------------------------------------
\begin{figure}
\begin{center}
\begin{picture}(500,160)
\put(-40,-430){\includegraphics[angle=0,width=470pt]{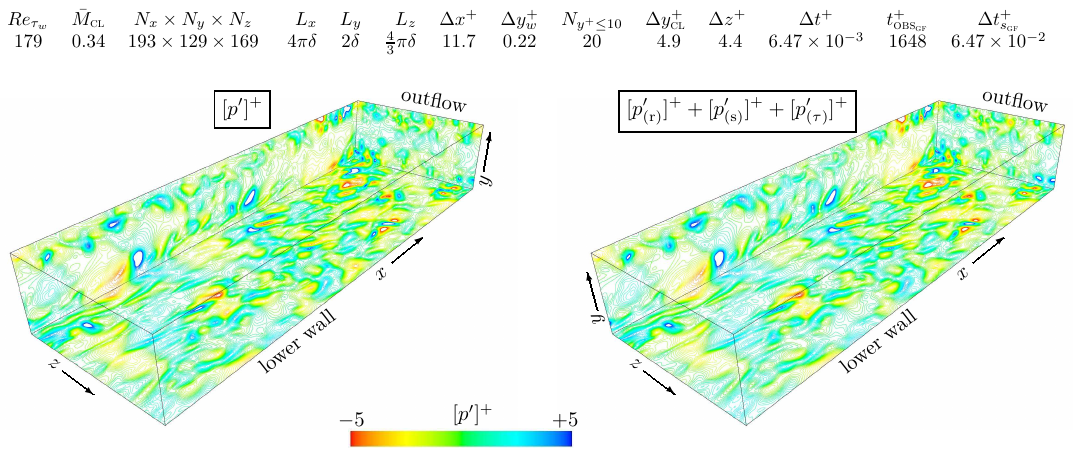}}
\end{picture}
\end{center}
\caption{Instantaneous fluctuating pressure in wall units $[p']^+$ obtained directly ($Re_{\tau_w}=179$; $\bar M_\tsc{cl}=0.34$; $193\times129\times169$ grid; \tabrefnp{Tab_WEPFTCF_s_DNSCpS_ss_PCFCCM_001})
by the compressible \tsc{dns} solver \citep{Gerolymos_Senechal_Vallet_2010a}
compared to the superposition of the three fields, rapid $[p_{(\mathrm{r})}']^+$ and slow $[p_{(\mathrm{s})}']^+$ obtained by the Green's function solution \eqref{Eq_WEPFTCF_s_DNSCpS_ss_GFSPEq_sss_Kim_001},
and Stokes field $[p_{(\tau)}']^+$ \eqref{Eq_WEPFTCF_s_DNSCpS_ss_GFSPEq_sss_Kim_002}, in wall units
(50 contours in the range $[-5,+5]$ on the lower wall, the outflow $x$-periodic interface, and the $z$-periodicity interface).}
\label{Fig_WEPFTCF_s_APC_ss_FPF_001}
\end{figure}
%-----------------------------------------------------------------------------------------------------------------------------------
%
%-----------------------------------------------------------------------------------------------------------------------------------
\begin{figure}
\begin{center}
\begin{picture}(500,240)
\put(-40,-315){\includegraphics[angle=0,width=470pt]{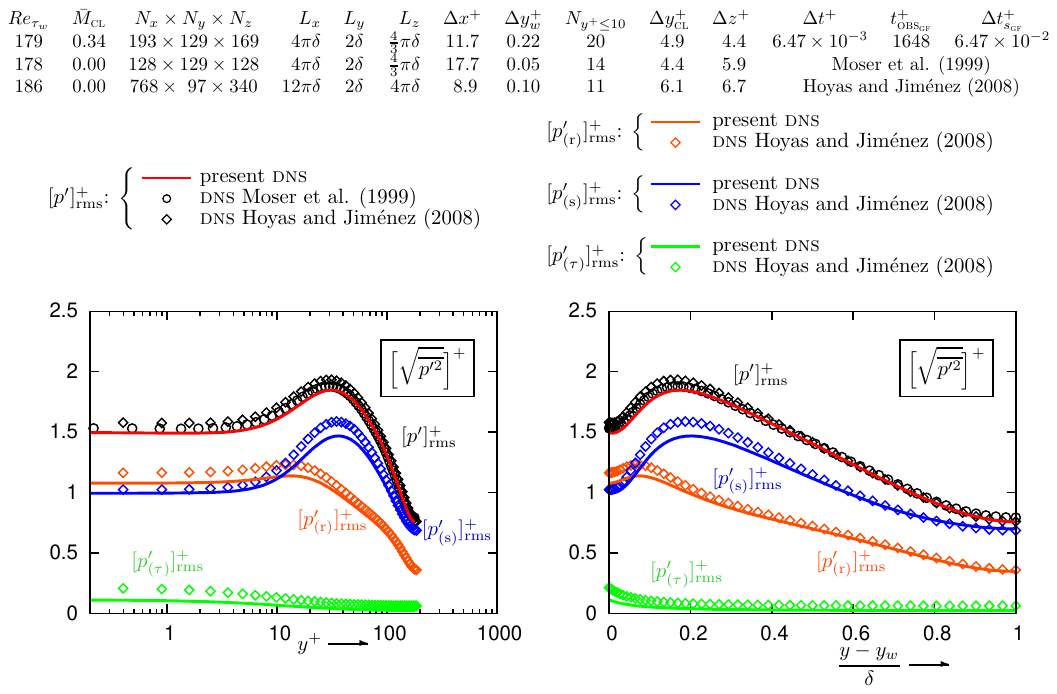}}
\end{picture}
\end{center}
\caption{Comparison of ${\rm rms}$-values $p'_\mathrm{rms}:=\sqrt{\overline{p'^2}}$ of the fluctuating pressure $p'$
and of the fields $p_{(\mathrm{r})}'+p_{(\mathrm{s})}'+p_{(\tau)}'=p'$ \eqref{Eq_WEPFTCF_s_DNSCpS_ss_GFSPEq_sss_ODEsFT_001},
from the present {\sc dns} computations ($Re_{\tau_w}=179$; $\bar M_\tsc{cl}=0.34$; $193\times129\times169$ grid; \tabrefnp{Tab_WEPFTCF_s_DNSCpS_ss_PCFCCM_001})
with reference results of incompressible pseudospectral~\citep{Kim_Moin_Moser_1987a} \tsc{dns} computations
of \citet[$Re_{\tau_w}=186$, $M_\tsc{cl}=0$]{Hoyas_Jimenez_2008a} and of \citet[$Re_{\tau_w}=178$, $M_\tsc{cl}=0$]{Moser_Kim_Mansour_1999a},
in wall units, plotted against the nondimensional distance-from-the-wall in inner ($y^+$) and outer ($\delta^{-1}(y-y_w)$) scaling.}
\label{Fig_WEPFTCF_s_APC_ss_FPF_002}
\end{figure}
%-----------------------------------------------------------------------------------------------------------------------------------
%
%-----------------------------------------------------------------------------------------------------------------------------------
\begin{figure}
\begin{center}
\begin{picture}(500,235)
\put(-40,-325){\includegraphics[angle=0,width=470pt]{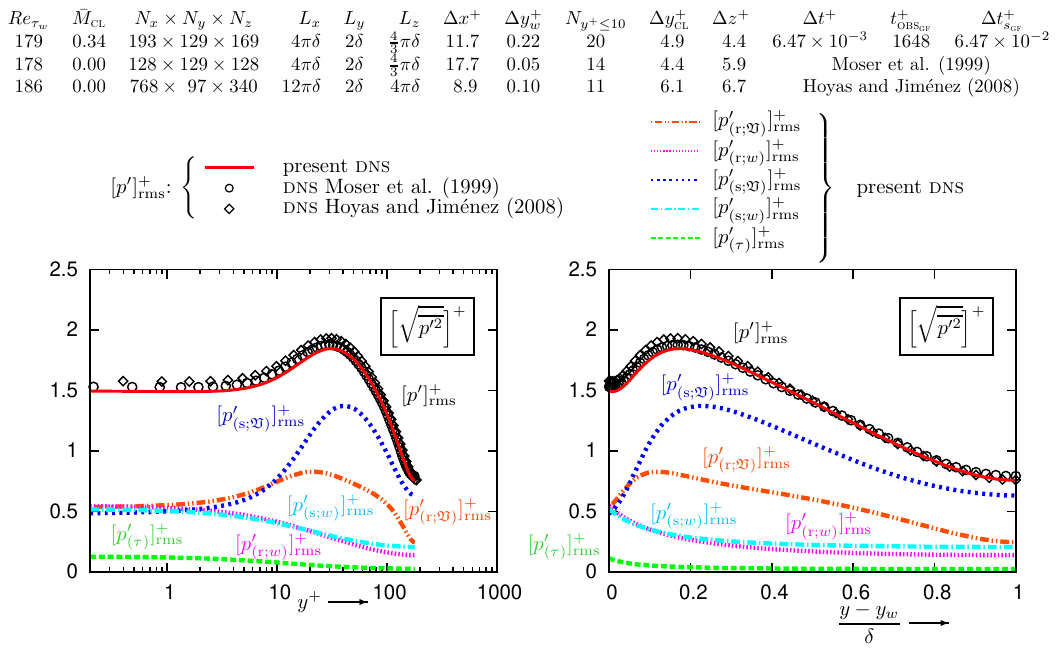}}
\end{picture}
\end{center}
\caption{Rms-values of the 5 fluctuating pressure fields in the decomposition $p'=p_{(\mathrm{r};\mathfrak{V})}'+p_{(\mathrm{r};w)}'+p_{(\mathrm{s};\mathfrak{V})}'+p_{(\mathrm{s};w)}'+p_{(\tau)}'$
\eqref{Eq_WEPFTCF_s_DNSCpS_ss_GFSPEq_sss_VWET_001}, from the present \tsc{dns} computations ($Re_{\tau_w}=179$; $\bar M_\tsc{cl}=0.34$; $193\times129\times169$ grid; \tabrefnp{Tab_WEPFTCF_s_DNSCpS_ss_PCFCCM_001}),
plotted against the nondimensional distance-from-the-wall in inner ($y^+$) and outer ($\delta^{-1}(y-y_w)$) scaling,
and $p'_\mathrm{rms}$ from various \tsc{dns} databases \citep{Moser_Kim_Mansour_1999a,
                                                              Hoyas_Jimenez_2008a,
                                                              Gerolymos_Senechal_Vallet_2010a}.}
\label{Fig_WEPFTCF_s_APC_ss_FPF_003}
\end{figure}
%-----------------------------------------------------------------------------------------------------------------------------------
%
%-----------------------------------------------------------------------------------------------------------------------------------
\begin{figure}
\begin{center}
\begin{picture}(500,540)
\put(-40,-60){\includegraphics[angle=0,width=460pt]{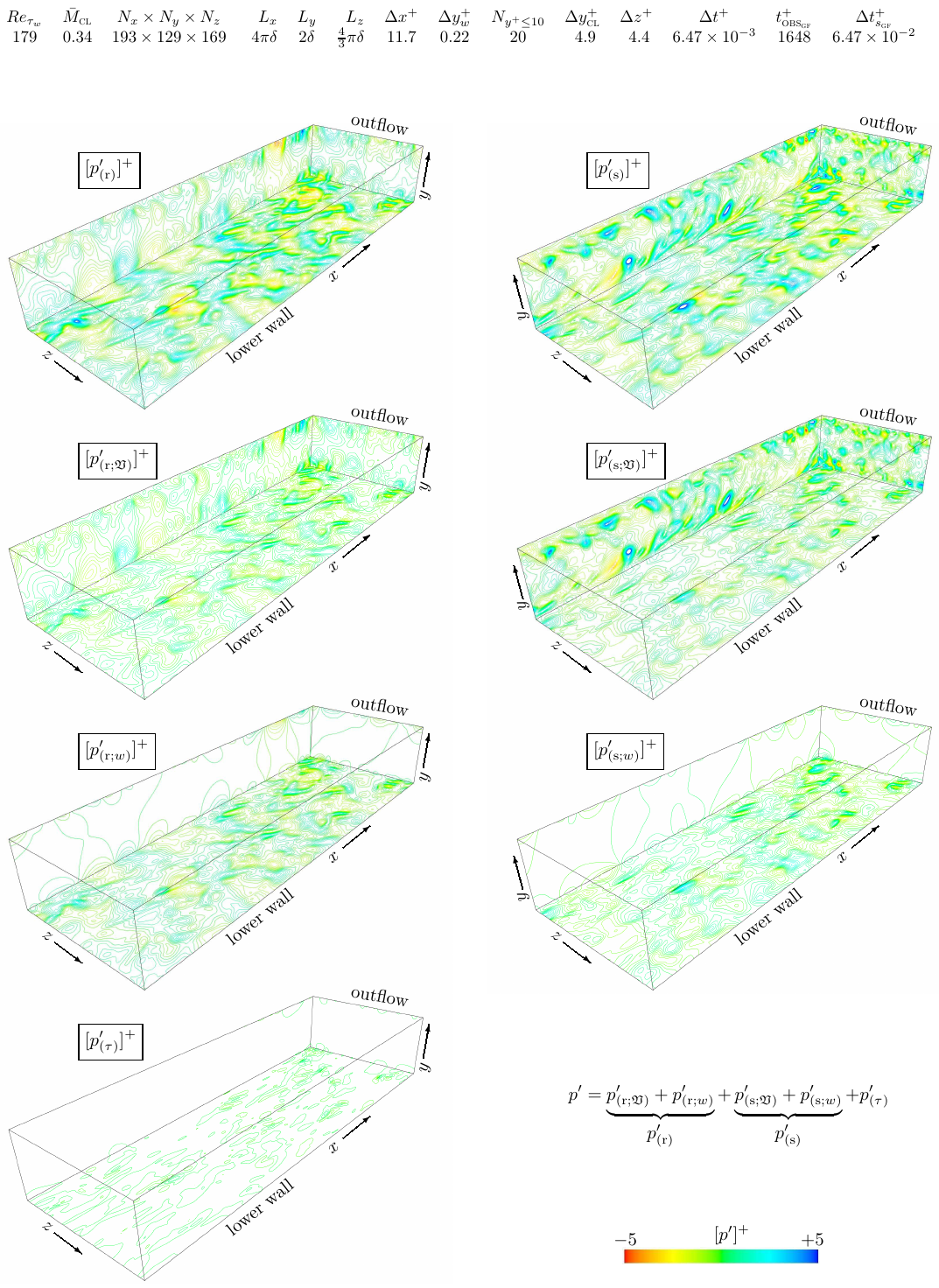}}
\end{picture}
\end{center}
\caption{Instantaneous fluctuating pressure fields \eqref{Eq_WEPFTCF_s_DNSCpS_ss_GFSPEq_sss_VWET_001}
from \tsc{dns} computations ($Re_{\tau_w}=179$; $\bar M_\tsc{cl}=0.34$; $193\times129\times169$ grid; \tabrefnp{Tab_WEPFTCF_s_DNSCpS_ss_PCFCCM_001}),
Stokes $[p_{(\tau)}']^+$ \eqref{Eq_WEPFTCF_s_DNSCpS_ss_GFSPEq_sss_Kim_002}, and
rapid $[p_{(\mathrm{r})}']^+$ and slow $[p_{(\mathrm{s})}']^+$ \eqref{Eq_WEPFTCF_s_DNSCpS_ss_GFSPEq_sss_Kim_001}, and their decomposition into weakly inhomogeneous volume fields,
$[p_{(\mathrm{r};\mathfrak{V})}']^+$ and $[p_{(\mathrm{s};\mathfrak{V})}']^+$ \eqref{Eq_WEPFTCF_s_DNSCpS_ss_GFSPEq_sss_VWET_003},
and strongly inhomogeneous wall-echo fields,
$[p_{(\mathrm{r};w)}']^+$ and $[p_{(\mathrm{s};w)}']^+$ \eqref{Eq_WEPFTCF_s_DNSCpS_ss_GFSPEq_sss_VWET_006}, in wall units
(50 contours in the range $[-5,+5]$ on the lower wall, the outflow $x$-periodic interface, and the $z$-periodicity interface).}
\label{Fig_WEPFTCF_s_APC_ss_FPF_004}
\end{figure}
%-----------------------------------------------------------------------------------------------------------------------------------
 
%-----------------------------------------------------------------------------------------------------------------------------------
%
%
%
%
%
%
%
%
%
\section{Analysis of pressure correlations}\label{WEPFTCF_s_APC}
%
%
%
%
%
%
%
%
%
%-----------------------------------------------------------------------------------------------------------------------------------

The algorithm (\S\ref{WEPFTCF_s_DNSCpS_ss_GFSPEq_sss_VWET}) for separating the rapid and slow contributions to $p'$ (\S\ref{WEPFTCF_s_DNSCpS_ss_GFSPEq_sss_Kim})
into weakly inhomogeneous volume ($p_{(\mathrm{r};\mathfrak{V})}'$ and $p_{(\mathrm{s};\mathfrak{V})}'$) and
strongly inhomogeneous wall-echo terms ($p_{(\mathrm{r};w)}'$ and $p_{(\mathrm{s};w)}'$), was applied to a low-Reynolds-number ($Re_{\tau_w}\approxeq180$)
well-resolved \tsc{dns} (grid $193\times129\times169$; \tabrefnp{Tab_WEPFTCF_s_DNSCpS_ss_PCFCCM_001}).
The simulation was started at $t=t_0$ by interpolation of a well-converged simulation on a coarser grid (grid $129\times129\times129$; \tabrefnp{Tab_WEPFTCF_s_DNSCpS_ss_PCFCCM_001}),
and continued for $t^+_{\tsc{gf}_0}-t^+_0=681$. During this interval $t^+\in[t^+_0,t^+_{\tsc{gf}_0}]$ statistics for $\bar u_i$ and $\overline{u_i'u_j'}$,
necessary for the computation of the source-terms in \eqref{Eq_WEPFTCF_s_DNSCpS_ss_GFSPEq_sss_ODEsFT_002a}, were computed with sampling at every iteration
($\Delta t_s^+=\Delta t^+\approxeq6.47\times10^{-3}\iff f_s^+\approxeq154$). Then the Green's function algorithm for $p'$-splitting \eqref{Eq_WEPFTCF_s_DNSCpS_ss_GFSPEq_sss_VWET_001}
was applied, and statistics of pressure correlations were computed for an observation time $t_{\tsc{obs}_\tsc{gf}}^+=1648$, with sampling every 10 iterations
($\Delta t_{s_\tsc{gf}}^+=10\Delta t^+\approxeq6.47\times10^{-2}\iff f_{s_\tsc{gf}}^+\approxeq15.4$).

%-----------------------------------------------------------------------------------------------------------------------------------
%
\subsection{Fluctuating pressure field}\label{WEPFTCF_s_APC_ss_FPF}
%
%-----------------------------------------------------------------------------------------------------------------------------------

The instantaneous fluctuating pressure $p'$-field reconstructed by the Green's function approach \eqrefsab{Eq_WEPFTCF_s_DNSCpS_ss_GFSPEq_sss_Kim_001}{Eq_WEPFTCF_s_DNSCpS_ss_GFSPEq_sss_Kim_002}
according to the splitting \eqref{Eq_WEPFTCF_s_DNSCpS_ss_GFSPEq_sss_ODEsFT_001} agrees quite well \figref{Fig_WEPFTCF_s_APC_ss_FPF_001} with the instantaneous $p'$-field directly computed 
(without making use of Green's functions) by the compressible \tsc{dns} solver \citep{Gerolymos_Senechal_Vallet_2010a}.

Both the directly computed $p'_\mathrm{rms}$ and the $\mathrm{rms}$ values ($[p_{(\mathrm{r})}']_\mathrm{rms}$, $[p_{(\mathrm{s})}']_\mathrm{rms}$, $[p_{(\tau)}']_\mathrm{rms}$)
of the $p'$-splitting \eqref{Eq_WEPFTCF_s_DNSCpS_ss_GFSPEq_sss_ODEsFT_001} compare globally satisfactorily \figref{Fig_WEPFTCF_s_APC_ss_FPF_002} with standard results from
incompressible \tsc{dns} computations \citep{Kim_Moin_Moser_1987a,
                                             Hoyas_Jimenez_2008a},
both in the wall and the outer regions \figref{Fig_WEPFTCF_s_APC_ss_FPF_002}.
Computed $p'_\mathrm{rms}$ corresponds, for the present \tsc{dns} results \figref{Fig_WEPFTCF_s_APC_ss_FPF_002},
to the fluctuation of the actual thermodynamic pressure, obtained from the equation-of-state \citep{Gerolymos_Senechal_Vallet_2010a}.
There is quite good agreement of present results for $p'_\mathrm{rms}$ with the incompressible \tsc{dns} of \citet{Kim_Moin_Moser_1987a},
obtained with similar resolution on the same computational box \figref{Fig_WEPFTCF_s_APC_ss_FPF_002},
the present data being marginally lower, because of the small mean-density-variation effect ($\lessapprox1.5\%$; \parrefnp{WEPFTCF_s_DNSCpS_ss_PCFCCM}).
The incompressible \tsc{dns} data of \citet{Hoyas_Jimenez_2008a}, with similar resolution but a larger computational box indicate slightly higher $[p']^+_\mathrm{rms}$ \figref{Fig_WEPFTCF_s_APC_ss_FPF_002}.
This is attributed to the higher $Re_{\tau_w}=186$ in the simulations of \citet{Hoyas_Jimenez_2008a}, compared to $Re_{\tau_w}=178$ \citet{Kim_Moin_Moser_1987a} and $Re_{\tau_w}=179$ for the present simulations.
The variation of $[p_w']^+_\mathrm{rms}$ with $Re_{\tau_w}$ \citep{Hu_Morfey_Sandham_2006a,
                                                                   Tsuji_Fransson_Alfredsson_Johansson_2007a},
indicates a $\sim2\%$ increase in $[p_w']^+_\mathrm{rms}$ from $Re_{\tau_w}=178$ to $Re_{\tau_w}=186$.
Incompressible \tsc{dns} data of the different fields ($[p_{(\mathrm{r})}']_\mathrm{rms}$, $[p_{(\mathrm{s})}']_\mathrm{rms}$, $[p_{(\tau)}']_\mathrm{rms}$)
were available only for the simulation of \citet{Hoyas_Jimenez_2008a}, and are expectedly a little higher
than those of the present computations \figref{Fig_WEPFTCF_s_APC_ss_FPF_002},
consistently with the difference in $[p']^+_\mathrm{rms}$.\footnote{\label{ff_WEPFTCF_s_APC_ss_FPF_001}Notice also that $\overline{p'^2}=\overline{(p_{(\mathrm{r})}'+p_{(\mathrm{s})}'+p_{(\tau)}')^2}
                                                                                                                                     \neq\overline{p_{(\mathrm{r})}'^2}
                                                                                                                                        +\overline{p_{(\mathrm{s})}'^2}
                                                                                                                                        +\overline{p_{(\tau)}'^2}$
                                                                   }
There is, nonetheless, a difference in the level of Stokes pressure $[p_{(\tau)}']_\mathrm{rms}$
near the wall \figref{Fig_WEPFTCF_s_APC_ss_FPF_002}, which should be further investigated, but this term is quite small compared to the others.

The new results \figref{Fig_WEPFTCF_s_APC_ss_FPF_003} in the present work concern the further splitting
($p'=p_{(\mathrm{r;\mathfrak{V}})}'+p_{(\mathrm{r};w)}'+p_{(\mathrm{s};\mathfrak{V})}'+p_{(\mathrm{s};w)}'+p_{(\tau)}'$)
into volume and wall-echo terms \eqref{Eq_WEPFTCF_s_DNSCpS_ss_GFSPEq_sss_VWET_001}.
The $\mathrm{rms}$-levels of wall-echo ($[p_{(\mathrm{r};w)}']_\mathrm{rms}$, $[p_{(\mathrm{s};w)}']_\mathrm{rms}$), at the wall ($y=0$; \figrefnp{Fig_WEPFTCF_s_APC_ss_FPF_003}),
are approximately equal to the corresponding volume terms ($[p_{(\mathrm{r};\mathfrak{V})}']_\mathrm{rms}$, $[p_{(\mathrm{s};\mathfrak{V})}']_\mathrm{rms}$).
This implies (\S\ref{WEPFTCF_s_DNSCpS_ss_GFSPEq_sss_IWEbW}) that the dominant contributions to the spectra \eqref{Eq_WEPFTCF_s_DNSCpS_ss_GFSPEq_sss_ODEsFT_004b}
of the source-terms $\hat Q'_{(\mathrm{r})}(\kappa_x,y,\kappa_z,t)$ and $\hat Q'_{(\mathrm{s})}(\kappa_x,y,\kappa_z,t)$ \eqref{Eq_WEPFTCF_s_DNSCpS_ss_GFSPEq_sss_ODEsFT_005a}
occur at sufficiently high nondimensional wavenumbers $\kappa L_y=2\kappa\delta$ \figref{Fig_WEPFTCF_s_DNSCpS_ss_GFSPEq_sss_IWEbW_001},
for the interaction between the wall-echo from the upper wall and the wall-echo from the lower wall to be negligibly small. This approximate equality of
wall-echo and volume terms holds up to $y^+\lessapprox2$ \figref{Fig_WEPFTCF_s_APC_ss_FPF_003}. Further away from the wall, both wall-echo terms ($[p_{(\mathrm{r};w)}']_\mathrm{rms}$, $[p_{(\mathrm{s};w)}']_\mathrm{rms}$)
decay with increasing distance from the wall, always remaining much higher than the Stokes pressure $[p_{(\tau)}']_\mathrm{rms}$ \figref{Fig_WEPFTCF_s_APC_ss_FPF_003}.
In the buffer and outer regions ($y-y_w\gtrapprox\tfrac{5}{100}\delta\stackrel{Re_{\tau_w}\approxeq180}{\iff}y^+\gtrapprox 10$) the well known predominance
of the slow volume term $[p_{(\mathrm{s};\mathfrak{V})}']_\mathrm{rms}$ over the rapid volume term $[p_{(\mathrm{r};\mathfrak{V})}']_\mathrm{rms}$ \citep{Kim_1989a,
                                                                                                                                                          Chang_Piomelli_Blake_1999a}
is evident \figref{Fig_WEPFTCF_s_APC_ss_FPF_003}. Nonetheless, it is essential, from the point-of-view of near-wall modelling, to notice that for $y^+\lessapprox10$,
$[p_{(\mathrm{r};\mathfrak{V})}']_\mathrm{rms}\approxeq[p_{(\mathrm{s};\mathfrak{V})}']_\mathrm{rms}\approxeq[p_{(\mathrm{r};w)}']_\mathrm{rms}\approxeq[p_{(\mathrm{s};w)}']_\mathrm{rms}>[p_{(\tau)}']_\mathrm{rms}$
\figref{Fig_WEPFTCF_s_APC_ss_FPF_003}. Since all of the terms are of the same order-of-magnitude near the wall, but may have different phases,
and hence different correlation-coefficients with the fluctuating velocity field, they all require accurate modelling. Despite the fact that further away from the wall ($y^+\gtrapprox50$)
homogeneous models are reasonably accurate \citep[Fig. 2, p. 9]{Gerolymos_Lo_Vallet_2012a}, global predictive performance of models in actual
flows is dominated by the quality of near-wall modelling \citep{Durbin_1993a,
                                                                Hanjalic_1994a}.

The above quantitative results \figref{Fig_WEPFTCF_s_APC_ss_FPF_003} are also observed qualitatively in the instantaneous levels \figref{Fig_WEPFTCF_s_APC_ss_FPF_004}
of the $5$ terms in the $p'$-splitting \eqref{Eq_WEPFTCF_s_DNSCpS_ss_GFSPEq_sss_VWET_001}. Comparison of instantaneous levels of the slow ($p'_{(\mathrm{s})}$) and rapid ($p'_{(\mathrm{r})}$) terms
\figref{Fig_WEPFTCF_s_APC_ss_FPF_004} clearly shows that both mechanisms of generation of $p'$ \eqref{Eq_WEPFTCF_s_DNSCpS_ss_GFSPEq_sss_ODEsFT_002a} are of the same magnitude at the wall (lower wall; \figrefnp{Fig_WEPFTCF_s_APC_ss_FPF_004}),
while $p'_{(\mathrm{s})}$ is the main mechanism \citep{Chang_Piomelli_Blake_1999a} further away from the wall ($z=\const$ plane; \figrefnp{Fig_WEPFTCF_s_APC_ss_FPF_004}). The same observations apply to the corresponding volume terms,
$p_{(\mathrm{r;\mathfrak{V}})}'$ and $p_{(\mathrm{s};\mathfrak{V})}'$ \figref{Fig_WEPFTCF_s_APC_ss_FPF_004}.
The wall-echo terms, $p_{(\mathrm{r};w)}'$ and $p_{(\mathrm{s};w)}'$, at the wall (lower wall; \figrefnp{Fig_WEPFTCF_s_APC_ss_FPF_004}), are approximately equal to the corresponding volume terms,
$p_{(\mathrm{r;\mathfrak{V}})}'$ and $p_{(\mathrm{s};\mathfrak{V})}'$, but rapidly decay away from the wall ($z=\const$ plane; \figrefnp{Fig_WEPFTCF_s_APC_ss_FPF_004}).
Finally, the Stokes pressure $p_{(\tau)}'$ is substantially lower than the other terms and rapidly decays away from the wall \figref{Fig_WEPFTCF_s_APC_ss_FPF_004}.
                     
The results \figrefsatob{Fig_WEPFTCF_s_APC_ss_FPF_001}{Fig_WEPFTCF_s_APC_ss_FPF_004} on the $p'$-fields \eqref{Eq_WEPFTCF_s_DNSCpS_ss_GFSPEq_sss_VWET_001} 
provide guidance on the relative importance of each of the 5 $p'$-fields. Nonetheless, the transposition of these results to the decomposition of correlations
containing $p'$ is not always straightforward, especially in the near-wall region ($y^+\lessapprox10$). For this reason we study in detail
(\S\ref{WEPFTCF_s_APC_ss_PTPD}, \S\ref{WEPFTCF_s_APC_ss_PSRphiij}, \S\ref{WEPFTCF_s_APC_ss_VPGPiij}) the correlations containing $p'$ which appear in the transport equations
for the Reynolds-stresses (\S\ref{WEPFTCF_s_APC_ss_RST}).

%-----------------------------------------------------------------------------------------------------------------------------------
%
\subsection{Reynolds-stress transport}\label{WEPFTCF_s_APC_ss_RST}
%
%-----------------------------------------------------------------------------------------------------------------------------------

The equations governing the Reynolds-stress tensor are central in single-point closure turbulence modelling~\citep{Lumley_1978a,
                                                                                                                   Hanjalic_1994a}.
The exact transport equations for the Reynolds-stresses in incompressible flow read \citep[pp. 315--320]{Pope_2000a}
\begin{eqnarray}
\underbrace{\frac{\partial\rho\overline{u_i'u_j'}}{\partial t}
           +\frac{\partial}{\partial x_\ell}(\rho\overline{u_i'u_j'}\bar u_\ell)}_{\textstyle\text{convection $C_{ij}$}}=
\underbrace{\frac{\partial}{\partial x_\ell}(-\rho\overline{u_i'u_j'u_\ell'}
                                             -\overline{p'u_j'}\delta_{i\ell}
                                             -\overline{p'u_i'}\delta_{j\ell}
                                             +\mu\dfrac{\partial\overline{u'_iu'_j}}{\partial x_\ell})}_{\textstyle\text{diffusion $d_{ij}^{(u)}+d_{ij}^{(p)}+d_{ij}^{(\mu)}$}}
                                                                                                                                                          \nonumber\\
+\underbrace{\overline{p'\left(\frac{\partial u_i'}{\partial x_j}
                              +\frac{\partial u_j'}{\partial x_i}\right)}}_{\textstyle\text{redistribution $\phi_{ij}$}}
+\underbrace{\left(-\rho\overline{u_i'u_\ell'}\frac{\partial\tilde u_j}{\partial x_\ell}
                   -\rho\overline{u_j'u_\ell'}\frac{\partial\tilde u_i}{\partial x_\ell}\right)}_{\textstyle\text{production $P_{ij}$}}
                                                                                                                                                          \nonumber\\
-\underbrace{\left(\frac{\partial}{\partial x_{\ell}}(\overline{\mu\dfrac{\partial u'_iu'_j}{\partial x_\ell}})
            -(\overline{u_i'\frac{\partial\tau_{j{\ell}}'}{\partial x_{\ell}}
                       +u_j'\frac{\partial\tau_{i{\ell}}'}{\partial x_{\ell}}})\right)}_{\textstyle\text{dissipation $\bar{\rho}\varepsilon^{(\mu)}_{ij}:=2\mu\overline{\partial_{x_\ell}u_i'\partial_{x_\ell}u_j'}$}}
                                                                                                                              \label{Eq_WEPFTCF_s_APC_ss_RST_001}
\end{eqnarray}
Convection $C_{ij}$, production $P_{ij}$ and viscous diffusion $d_{ij}^{(\mu)}$ are exact terms, while all the other terms
($d_{ij}^{(u)}$, $d_{ij}^{(p)}$, $\phi_{ij}$, and $\varepsilon^{(\mu)}_{ij}$) require modelling.
Pressure-diffusion $d_{ij}^{(p)}$ and redistribution $\phi_{ij}$ can be grouped together into the velocity/pressure-gradient correlation tensor $\Pi_{ij}$
\begin{eqnarray}
\Pi_{ij}:=-\overline{u_i'\dfrac{\partial  p'}
                               {\partial x_j}}
          -\overline{u_j'\dfrac{\partial  p'}
                               {\partial x_i}}
         = \underbrace{\overline{p'\left(\frac{\partial u_i'}
                                              {\partial  x_j}
                                        +\frac{\partial u_j'}
                                              {\partial  x_i}\right)}}_{\displaystyle \overline{2p'S_{ij}'}}
           \underbrace{-\left(\overline{\frac{\partial p' u_i'}{\partial x_j}
                                       +\frac{\partial p' u_j'}{\partial x_i}}\right)}_{\displaystyle d_{ij}^{(p)}}
\stackrel{\eqref{Eq_WEPFTCF_s_APC_ss_RST_001}}{=}\phi_{ij}+d_{ij}^{(p)}
                                                                                                                              \label{Eq_WEPFTCF_s_APC_ss_RST_002}
\end{eqnarray}
The velocity/pressure-gradient tensor $\Pi_{ij}$, which is the term appearing in most of the original early developments \citep{Chou_1945a},
tends to $0$ as $y\to0$ (in the viscous sublayer, where $d_{ij}^{(p)}$ and $\phi_{ij}$ cancel one another).
In homogeneous turbulence (\tsc{ht}) $d_{ij}^{(p)}=0\stackrel{\tsc{ht}}{\iff}\phi_{ij}=\Pi_{ij}$.
%-----------------------------------------------------------------------------------------------------------------------------------
\begin{figure}
\begin{center}
\begin{picture}(500,370)
\put(-40,-190){\includegraphics[angle=0,width=470pt]{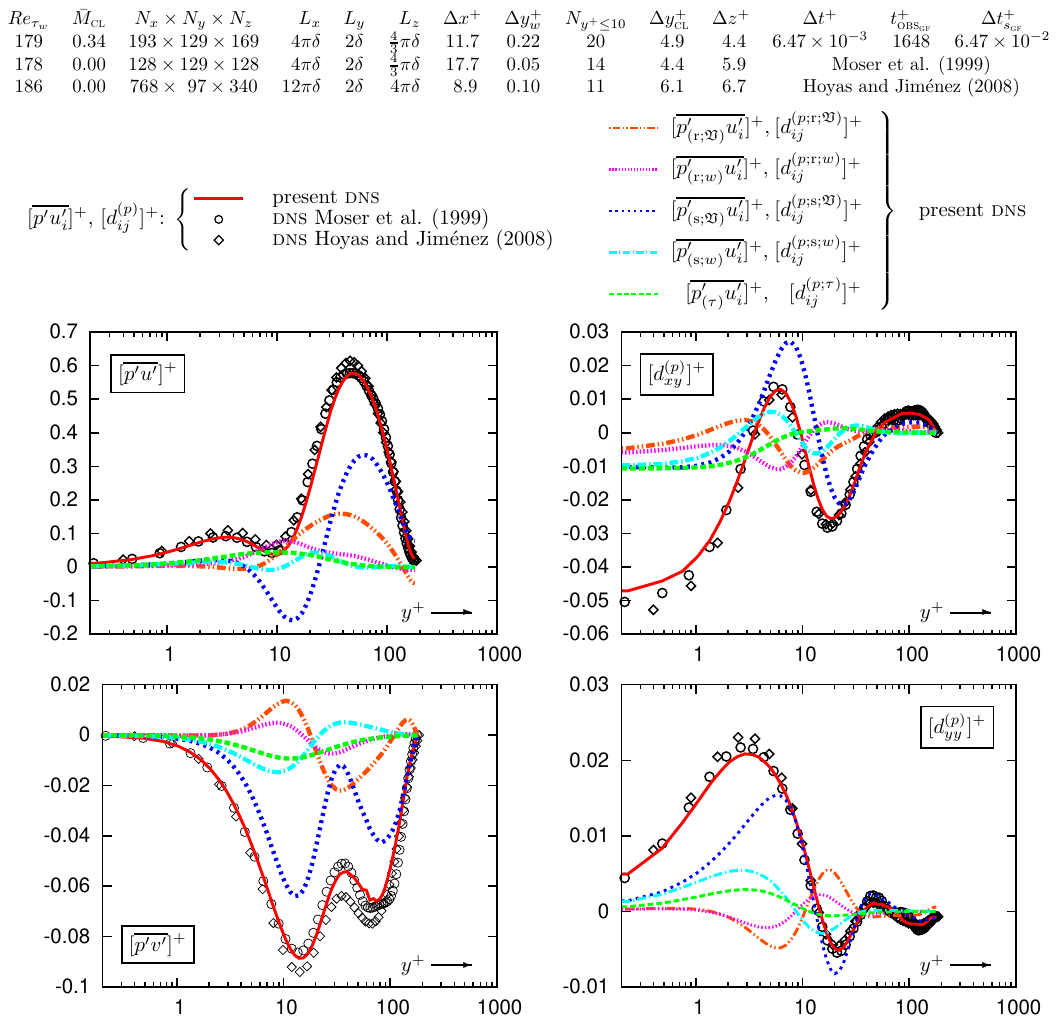}}
\end{picture}
\end{center}
\caption{Distributions of the 5 terms in the incompressible $p'$-splitting \eqref{Eq_WEPFTCF_s_DNSCpS_ss_GFSPEq_sss_VWET_001} of pressure transport,
$\overline{p'_{({\rm r};{\mathfrak V})}u_i'}$, $\overline{p'_{({\rm r};w)}u_i'}$, $\overline{p'_{({\rm s};{\mathfrak V})}u_i'}$, $\overline{p'_{({\rm s};w)}u_i'}$, and $\overline{p'_{({\tau})}u_i'}$,
from the present \tsc{dns} computations ($Re_{\tau_w}=179$; $\bar M_\tsc{cl}=0.34$; $193\times129\times169$ grid; \tabrefnp{Tab_WEPFTCF_s_DNSCpS_ss_PCFCCM_001}),
distributions of pressure transport $\overline{p'u_i'}$ from various \tsc{dns} databases \citep{Moser_Kim_Mansour_1999a,
                                                                                                Hoyas_Jimenez_2008a,
                                                                                                Gerolymos_Senechal_Vallet_2010a},
and corresponding contributions to the pressure-diffusion tensor $d_{ij}^{(p)}$, in wall units, plotted against the nondimensional distance from the wall $y^+$
(in fully developed incompressible plane channel flow $d_{xx}^{(p)}=d_{yz}^{(p)}=d_{zz}^{(p)}=d_{zx}^{(p)}=0$).}
\label{Fig_WEPFTCF_s_APC_ss_PTPD_001}
\end{figure}
%-----------------------------------------------------------------------------------------------------------------------------------
%
%-----------------------------------------------------------------------------------------------------------------------------------
\begin{figure}
\begin{center}
\begin{picture}(500,370)
\put(-40,-190){\includegraphics[angle=0,width=470pt]{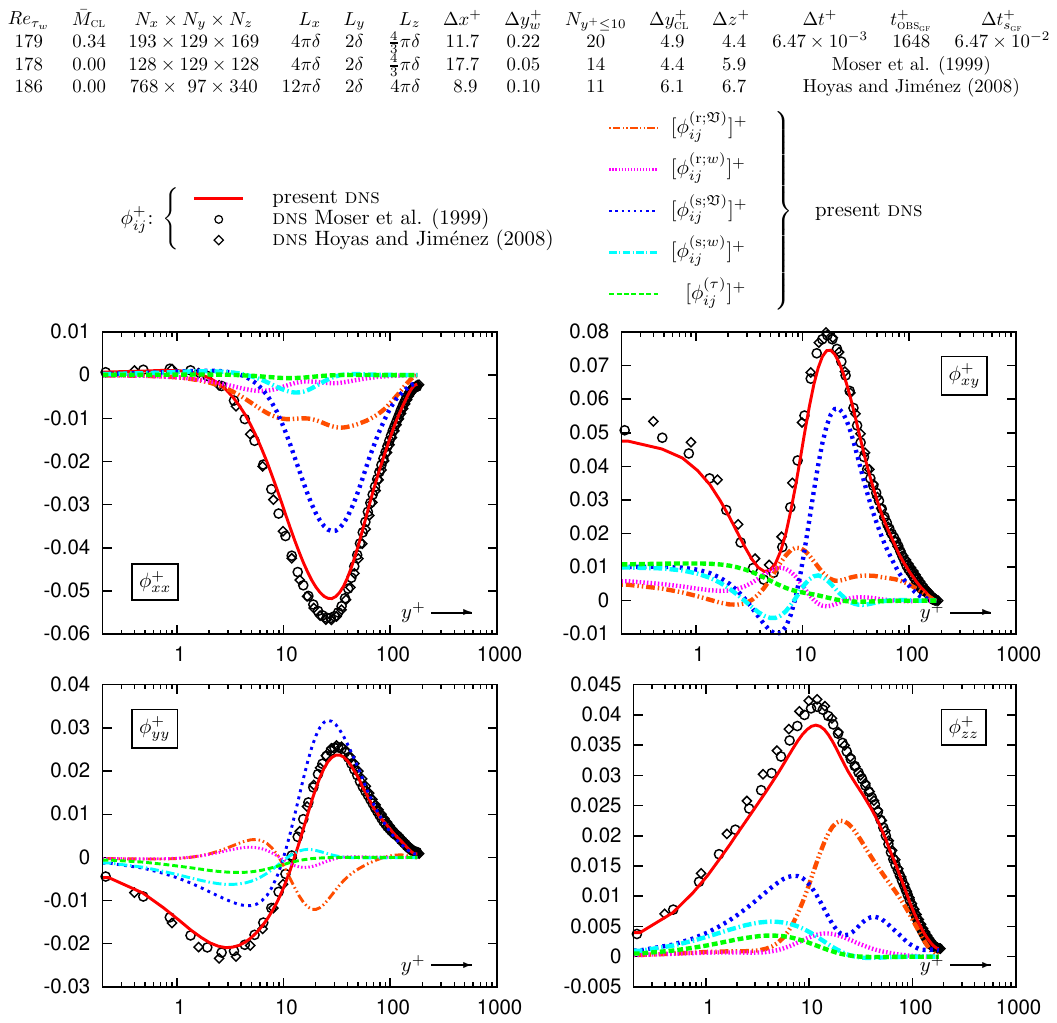}}
\end{picture}
\end{center}
\caption{Distributions of the 5 terms in the incompressible $p'$-splitting \eqref{Eq_WEPFTCF_s_DNSCpS_ss_GFSPEq_sss_VWET_001} of pressure-strain redistribution,
$\phi_{ij}^{(\mathrm{r};\mathfrak{V})}$, $\phi_{ij}^{(\mathrm{r};w)}$, $\phi_{ij}^{(\mathrm{s};\mathfrak{V})}$, $\phi_{ij}^{(\mathrm{s};w)}$, and $\phi_{ij}^{(\tau)}$,
from the present \tsc{dns} computations ($Re_{\tau_w}=179$; $\bar M_\tsc{cl}=0.34$; $193\times129\times169$ grid; \tabrefnp{Tab_WEPFTCF_s_DNSCpS_ss_PCFCCM_001}), and
distributions of pressure-strain redistribution $\phi_{ij}$ from various \tsc{dns} databases \citep{Moser_Kim_Mansour_1999a,
                                                                                                    Hoyas_Jimenez_2008a,
                                                                                                    Gerolymos_Senechal_Vallet_2010a},
in wall units, plotted against the nondimensional distance from the wall $y^+$
(in statistically 2-D plane flow $\phi_{zx}=\phi_{yz}=0$).}
\label{Fig_WEPFTCF_s_APC_ss_PSRphiij_001}
\end{figure}
%-----------------------------------------------------------------------------------------------------------------------------------

%-----------------------------------------------------------------------------------------------------------------------------------
%
\subsection{Pressure transport $\overline{p'u_i'}$ and pressure diffusion $d_{ij}^{(p)}$}\label{WEPFTCF_s_APC_ss_PTPD}
%
%-----------------------------------------------------------------------------------------------------------------------------------

The $p'$-splitting \eqref{Eq_WEPFTCF_s_DNSCpS_ss_GFSPEq_sss_VWET_001} applied to pressure transport $\overline{p'u_i'}$ and pressure diffusion $d_{ij}^{(p)}$ \eqrefsab{Eq_WEPFTCF_s_APC_ss_RST_001}{Eq_WEPFTCF_s_APC_ss_RST_002}
indicates \figref{Fig_WEPFTCF_s_APC_ss_PTPD_001} that the slow volume term $\overline{p'_{({\rm s};{\mathfrak V})}v'}$ is the principal contribution to the normal-to-the-wall transport $\overline{p'v'}$,
and evenmore to $d_{yy}^{(p)}$ (for fully developed plane channel flow $d_{yy}^{(p)}\stackrel{\eqrefsab{Eq_WEPFTCF_s_APC_ss_RST_001}{Eq_WEPFTCF_s_APC_ss_RST_002}}{=}-2\partial_y\overline{p'v'}$),
except in the near-wall region ($y^+\lessapprox5$) where all terms are of comparable importance \figref{Fig_WEPFTCF_s_APC_ss_PTPD_001}. Concerning $d_{xy}^{(p)}$
(for fully developed plane channel flow $d_{xy}^{(p)}\stackrel{\eqrefsab{Eq_WEPFTCF_s_APC_ss_RST_001}{Eq_WEPFTCF_s_APC_ss_RST_002}}{=}-\partial_y\overline{p'u'}$), again the slow volume term $d_{xy}^{(p;\mathrm{s};\mathfrak{V})}$
is the most important contribution in the buffer and outer regions, whereas in the near-wall region ($y^+\lessapprox5$)  all terms are of comparable importance \figref{Fig_WEPFTCF_s_APC_ss_PTPD_001}.

For modelling purposes it is important to notice that, although the slow volume term $d_{ij}^{(p;\mathrm{s};\mathfrak{V})}$ is the main contribution (and a reasonable approximation)
to $d_{ij}^{(p)}$ \figref{Fig_WEPFTCF_s_APC_ss_PTPD_001}, in the buffer and outer regions ($y^+\gtrapprox5$),
this is not the case for pressure transport $\overline{p'u_i'}$, especially for $\overline{p'u'}$ \figref{Fig_WEPFTCF_s_APC_ss_PTPD_001},
where, for $5\lessapprox y^+\lessapprox30$, $\overline{p_{(\mathrm{s};\mathfrak{V})}'u'}$ is of the opposite sign with respect to $\overline{p'u'}$.

These remarks imply that a consistent model for pressure diffusion $d_{ij}^{(p)}$, in the buffer and outer regions ($y^+\gtrapprox5$), can be built using only slow volume terms.
However, such a model for $d_{ij}^{(p)}$ ($y^+\gtrapprox5$), based on $d_{ij}^{(p;\mathrm{s};\mathfrak{V})}$ only, cannot correspond to (or be built from) a satisfactory model
for pressure transport $\overline{p'u_i'}$ \figref{Fig_WEPFTCF_s_APC_ss_PTPD_001}. The near-wall modelling of pressure diffusion $d_{ij}^{(p)}$ ($y^+\lessapprox5$) is more complex,
since \figref{Fig_WEPFTCF_s_APC_ss_PTPD_001} it must contain rapid terms
and wall-echo terms \figref{Fig_WEPFTCF_s_APC_ss_PTPD_001}. To the authors' knowledge \citep{Gerolymos_Lo_Vallet_Younis_2012a} no single-point closure for $d_{ij}^{(p)}$
takes into account the wall-echo terms \citep{Donaldson_1969a,
                                              Hirt_1969a,
                                              Daly_Harlow_1970a,
                                              Lumley_1978a,
                                              Fu_1993a,
                                              Sauret_Vallet_2007a,
                                              Vallet_2007a}.

%-----------------------------------------------------------------------------------------------------------------------------------
%
\subsection{Pressure-strain redistribution $\phi_{ij}$}\label{WEPFTCF_s_APC_ss_PSRphiij}
%
%-----------------------------------------------------------------------------------------------------------------------------------

The application of the $p'$-splitting \eqref{Eq_WEPFTCF_s_DNSCpS_ss_GFSPEq_sss_VWET_001} to pressure-strain redistribution $\phi_{ij}$ \eqrefsab{Eq_WEPFTCF_s_APC_ss_RST_001}{Eq_WEPFTCF_s_APC_ss_RST_002}
shows again that the volume terms ($\phi_{ij}^{(\mathrm{r};\mathfrak{V})}$ and $\phi_{ij}^{(\mathrm{s};\mathfrak{V})}$) are the dominant contribution to $\phi_{ij}$
in the buffer and outer regions ($y^+\gtrapprox10$; \figrefnp{Fig_WEPFTCF_s_APC_ss_PSRphiij_001}). Notice that $\phi_{zz}^{(\mathrm{r};\mathfrak{V})}>\phi_{zz}^{(\mathrm{s};\mathfrak{V})}$
for $y^+\gtrapprox10$ \figref{Fig_WEPFTCF_s_APC_ss_PSRphiij_001},
contrary to the other components for which the slow volume terms are the more important contributions to $\phi_{ij}$ in the buffer and outer regions ($y^+\gtrapprox15$; \figrefnp{Fig_WEPFTCF_s_APC_ss_PSRphiij_001}).

In the near-wall region ($y^+\lessapprox10$) the wall-echo terms ($\phi_{ij}^{(\mathrm{r};w)}$ and $\phi_{ij}^{(\mathrm{s};w)}$) are of the same order-of-magnitude as the corresponding
volume terms ($\phi_{ij}^{(\mathrm{r};\mathfrak{V})}$ and $\phi_{ij}^{(\mathrm{s};\mathfrak{V})}$, respectively), satisfying an approximate equality at the wall,
which ({\em cf} \S\ref{WEPFTCF_s_DNSCpS_ss_GFSPEq_sss_IWEbW}) implies that the energy-containing nondimensional wavenumbers of the source-terms \eqrefsabc{Eq_WEPFTCF_s_DNSCpS_ss_GFSPEq_sss_ODEsFT_002a}
                                                                                                                                                        {Eq_WEPFTCF_s_DNSCpS_ss_GFSPEq_sss_ODEsFT_004b}
                                                                                                                                                        {Eq_WEPFTCF_s_DNSCpS_ss_GFSPEq_sss_ODEsFT_005a}
are sufficiently high for the interaction between upper and lower wall to be negligible \figref{Fig_WEPFTCF_s_DNSCpS_ss_GFSPEq_sss_IWEbW_001}.
Furthermore, for the low-Reynolds-number case studied in the present work ($Re_{\tau_w}\approxeq180$) the Stokes pressure term $\phi_{ij}^{(\tau)}$ is of the same order-of-magnitude,
in the near-wall region ($y^+\lessapprox10$; \figrefnp{Fig_WEPFTCF_s_APC_ss_PSRphiij_001}), as the other terms of the $p'$-splitting \eqref{Eq_WEPFTCF_s_DNSCpS_ss_GFSPEq_sss_VWET_001},
and is particularly important for the shear component $\phi_{xy}$ \figref{Fig_WEPFTCF_s_APC_ss_PSRphiij_001}.
At the wall ($y^+=0$; \figrefnp{Fig_WEPFTCF_s_APC_ss_PSRphiij_001}) $[\phi_{xy}^{(\tau)}]_{y^+=0}$ accounts for $\sim25\%$ of $[\phi_{xy}]_{y^+=0}$.
For a plane wall $\perp\vec{e}_y$, at the wall ($y^+=0$),
$[\phi_{xy}]_{y^+=0}\stackrel{\eqrefsab{Eq_WEPFTCF_s_APC_ss_RST_001}{Eq_WEPFTCF_s_APC_ss_RST_002}}{=}[\overline{p'\partial_y u'}]_{y^+=0}$ (because $v'_{y^+=0}=0\Longrightarrow[\partial_x v']_{y^+=0}=0$).
Therefore, the high level of $[\phi_{xy}^{(\tau)}]_{y^+=0}$ relative to the other terms of the $p'$-splitting \eqref{Eq_WEPFTCF_s_DNSCpS_ss_GFSPEq_sss_VWET_001} implies
that, at the wall, the fluctuating Stokes pressure $[p'_{(\tau)}]_{y^+=0}$ is well correlated with the fluctuating wall-shear-stress $[\tau'_{xy}]_{y^+=0}=\mu[\partial_y u']_{y^+=0}$.
Notice that none of known models for $\phi_{ij}$~\citep{Craft_Launder_1996a,
                                                        Gerolymos_Vallet_2001a,
                                                        Jakirlic_Hanjalic_2002a,
                                                        Suga_2004a}
gives the correct viscous sublayer behaviour of $\phi_{xy}$ and $\phi_{yy}$ ($y^+\lessapprox5$; \figrefnp{Fig_WEPFTCF_s_APC_ss_PSRphiij_001}).
In fact, all known second-moment closures, actually model $\Pi_{ij}=\phi_{ij}+d_{ij}^{(p)}$ in the viscous sublayer \citep{Mansour_Kim_Moin_1988a},
where $\phi_{ij}$ \figref{Fig_WEPFTCF_s_APC_ss_PSRphiij_001} and $d_{ij}^{(p)}$ \figref{Fig_WEPFTCF_s_APC_ss_PTPD_001} cancel one another, since $[\Pi_{ij}]_w=0$,
even when they separately model $\phi_{ij}$ and $d_{ij}^{(p)}$ \citep{Gerolymos_Lo_Vallet_Younis_2012a} further away from the wall ($y^+\lessapprox5$).

%-----------------------------------------------------------------------------------------------------------------------------------
\begin{figure}
\begin{center}
\begin{picture}(500,360)
\put(-40,-185){\includegraphics[angle=0,width=460pt]{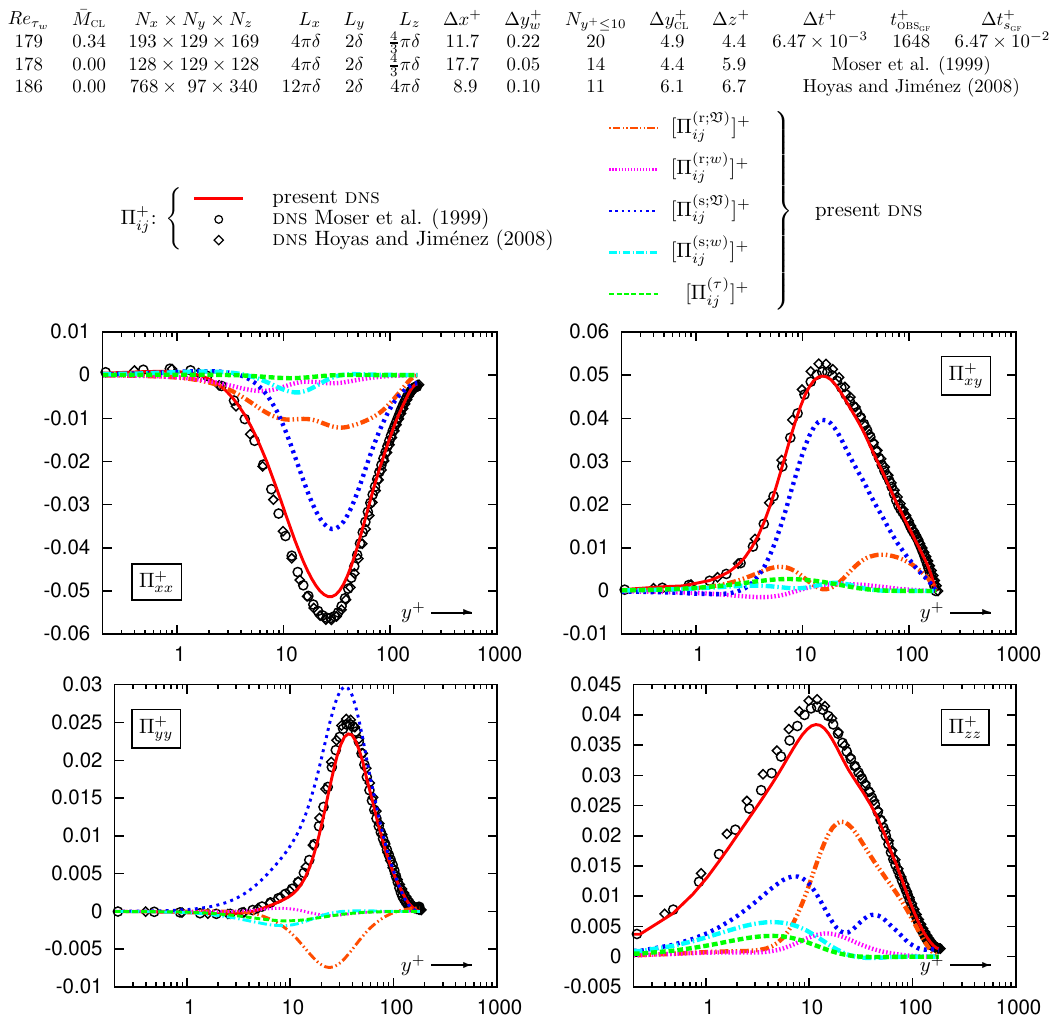}}
\end{picture}
\end{center}
\caption{Distributions of the 5 terms in the incompressible $p'$-splitting \eqref{Eq_WEPFTCF_s_DNSCpS_ss_GFSPEq_sss_VWET_001} of velocity/pressure-gradient correlation,
$\Pi_{ij}^{(\mathrm{r};\mathfrak{V})}$, $\Pi_{ij}^{(\mathrm{r};w)}$, $\Pi_{ij}^{(\mathrm{s};\mathfrak{V})}$, $\Pi_{ij}^{(\mathrm{s};w)}$, and $\Pi_{ij}^{(\tau)}$,
from the present \tsc{dns} computations ($Re_{\tau_w}=179$; $\bar M_\tsc{cl}=0.34$; $193\times129\times169$ grid; \tabrefnp{Tab_WEPFTCF_s_DNSCpS_ss_PCFCCM_001}), and
distributions of velocity/pressure-gradient correlation $\Pi_{ij}$ from various \tsc{dns} databases \citep{Moser_Kim_Mansour_1999a,
                                                                                                           Hoyas_Jimenez_2008a,
                                                                                                           Gerolymos_Senechal_Vallet_2010a},
in wall units, plotted against the nondimensional distance from the wall $y^+$
(in fully developed incompressible plane channel flow $\Pi_{zx}=\Pi_{yz}=0$, $\Pi_{xx}=\phi_{xx}$, and $\Pi_{zz}=\phi_{zz}$).}
\label{Fig_WEPFTCF_s_APC_ss_VPGPiij_001}
\end{figure}
%-----------------------------------------------------------------------------------------------------------------------------------
\vspace*{-0.2in}
%-----------------------------------------------------------------------------------------------------------------------------------
%
\subsection{Velocity/pressure-gradient $\Pi_{ij}$}\label{WEPFTCF_s_APC_ss_VPGPiij}
%
%-----------------------------------------------------------------------------------------------------------------------------------

The velocity/pressure-gradient correlation $\Pi_{ij}$ \eqref{Eq_WEPFTCF_s_APC_ss_RST_002} is exactly the sum of pressure diffusion $d_{ij}^{(p)}$ (\S\ref{WEPFTCF_s_APC_ss_PTPD})
and pressure-strain redistribution $\phi_{ij}$ (\S\ref{WEPFTCF_s_APC_ss_PSRphiij}). By definition \eqref{Eq_WEPFTCF_s_APC_ss_RST_002} $[\Pi_{ij}]_{y^+=0}$ because
of the no-slip wall boundary-condition for the fluctuating velocity $[u_i']_{y^+=0}=0$, implying $[\phi_{ij}]_{y^+=0}=-[d_{ij}^{(p)}]_{y^+=0}$, so that very near the wall
we expect that $\Pi_{ij}:=\phi_{ij}+d_{ij}^{(p)}$ will be asymptotically approaching $0$. Observation of \tsc{dns} data ($y^+\lessapprox5$; \figrefnp{Fig_WEPFTCF_s_APC_ss_VPGPiij_001})
indicates that the rate at which $\Pi_{zz}$ approaches $0$ as $y^+\to0$ is much slower compared to the other components ($\Pi_{xx}$, $\Pi_{xy}$, and $\Pi_{yy}$).

In fully developed incompressible plane channel flow $d_{xx}^{(p)}=d_{zz}^{(p)}=0$ by \eqref{Eq_WEPFTCF_s_APC_ss_RST_002}, implying that  $\Pi_{xx}=\phi_{xx}$ and $\Pi_{zz}=\phi_{zz}$,
but also that $\Pi_{\ell\ell}=d_{yy}^{(p)}$ (since $\phi_{\ell\ell}\stackrel{\eqref{Eq_WEPFTCF_s_APC_ss_RST_002}}{=}0$ by the incompressible fluctuating continuity equation $\partial_{x_\ell} u_\ell'=0$).
For the three components, $\Pi_{xx}$, $\Pi_{xy}$ and $\Pi_{yy}$, the dominant contribution from the mechanisms of the $p'$-splitting \eqref{Eq_WEPFTCF_s_DNSCpS_ss_GFSPEq_sss_VWET_001} comes from
the slow volume terms ($\Pi^{(\mathrm{s};\mathfrak{V})}_{xx}$, $\Pi^{(\mathrm{s};\mathfrak{V})}_{xy}$, and $\Pi^{(\mathrm{s};\mathfrak{V})}_{yy}$),
the rapid volume terms ($\Pi^{(\mathrm{r};\mathfrak{V})}_{xx}$, $\Pi^{(\mathrm{r};\mathfrak{V})}_{xy}$, and $\Pi^{(\mathrm{r};\mathfrak{V})}_{yy}$)
being the main remaining contribution \figref{Fig_WEPFTCF_s_APC_ss_VPGPiij_001}.
On the contrary, all of the 5 mechanisms of $p'$-generation in \eqref{Eq_WEPFTCF_s_DNSCpS_ss_GFSPEq_sss_VWET_001} contribute, with the same order-of-magnitude in the near-wall region
($y^+\lessapprox10$; \figrefnp{Fig_WEPFTCF_s_APC_ss_VPGPiij_001}) to $\Pi_{zz}$ ($=\phi_{zz}$ in fully developed incompressible plane channel flow),
this difference between $\Pi_{zz}$ and the other components being related to the slower rate at which $\Pi_{zz}$ goes to $0$ as $y^+\to0$ ($y^+\lessapprox5$; \figrefnp{Fig_WEPFTCF_s_APC_ss_VPGPiij_001}).
Furthermore, contrary to $\phi_{ij}$ \figref{Fig_WEPFTCF_s_APC_ss_PSRphiij_001}
and $d_{ij}^{(p)}$ \figref{Fig_WEPFTCF_s_APC_ss_PTPD_001}, for which the volume and wall-echo terms of each field (rapid or slow) are of the same sign, almost everywhere, and exhibit the same variation with $y^+$,
this behaviour applies only to $\Pi_{xx}$ and $\Pi_{zz}$. For the normal-to-the-wall $\Pi_{yy}$ and the shear $\Pi_{xy}$ components,
the wall-echo and volume terms of each field (rapid or slow) are of opposite sign.\footnote{\label{ff_WEPFTCF_s_APC_ss_VPGPiij_001}
                                                                                            Actually, in agreement with \cite{Manceau_Wang_Laurence_2001a}, all components of $\phi_{ij}$
                                                                                            \figref{Fig_WEPFTCF_s_APC_ss_PSRphiij_001} are increased in absolute value by wall-echo,
                                                                                            and so are the components of $d_{ij}^{(p)}$ \figref{Fig_WEPFTCF_s_APC_ss_PTPD_001}. This observation is true
                                                                                            both for the rapid and slow fields \figrefsab{Fig_WEPFTCF_s_APC_ss_PTPD_001}
                                                                                                                                         {Fig_WEPFTCF_s_APC_ss_PSRphiij_001},
                                                                                            but with different factors depending on the particular component and $y^+$. Therefore, when these
                                                                                            $\phi_{ij}$ and $d_{ij}^{(p)}$ are combined together, their sum
                                                                                            $\Pi_{ij}\stackrel{\eqref{Eq_WEPFTCF_s_APC_ss_RST_002}}{:=}\phi_{ij}+d_{ij}^{(p)}$
                                                                                            behaves differently.
                                                                                           }
This is not incompatible with their near-equality at the wall, where all components of $\Pi_{ij}$
vanish, because their $y$-gradients can be different. This observation is significant for near-wall modelling, because the term which appears in the Reynolds-stress transport equation is $\Pi_{ij}$, and whether
$\phi_{ij}$ or $d_{ij}^{(p)}$ are modelled separately~\citep{Gerolymos_Lo_Vallet_Younis_2012a} or together~\citep{Mansour_Kim_Moin_1988a}, it is the modelled sum which defines the quality of the model.

In order to explain the different behaviour of the various parts of the
$p'$-splitting
($p'_{(\mathrm{r};\mathfrak{V})}$,
$p'_{(\mathrm{r};w)}$,
$p'_{(\mathrm{s};\mathfrak{V})}$,
$p'_{(\mathrm{s};w)}$)
between $\Pi_{xx}$ and $\Pi_{zz}$ on one hand, and $\Pi_{yy}$ and $\Pi_{xy}$ on the other \parref{WEPFTCF_s_APC_ss_VPGPiij},
consider the analytical expression (hence transfer function) relating different parts of the $p'$-splitting \eqref{Eq_WEPFTCF_s_DNSCpS_ss_GFSPEq_sss_VWET_001}
to the rapid and slow sources \eqref{Eq_WEPFTCF_s_DNSCpS_ss_GFSPEq_sss_ODEsFT_002a}.
Let $m\in\{\mathrm{s},\mathrm{r}\}$ denote the rapid or slow fields, and $n\in\{\mathfrak{V},w\}$ denote the corresponding volume or wall parts.
Then, combining the $xz$-Fourier-transform representation of $p_{(m;n)}'$ \eqref{Eq_WEPFTCF_s_DNSCpS_ss_GFSPEq_sss_ODEsFT_004a}
and $Q_{(m)}'$ \eqref{Eq_WEPFTCF_s_DNSCpS_ss_GFSPEq_sss_ODEsFT_004b}
with the Green's function solution of the corresponding field \eqrefsab{Eq_WEPFTCF_s_DNSCpS_ss_GFSPEq_sss_VWET_003}
                                                                       {Eq_WEPFTCF_s_DNSCpS_ss_GFSPEq_sss_VWET_006},
\begin{subequations}
                                                                                                                 \label{Eq_WEPFTCF_s_APC_ss_SRRSM_001}
\begin{alignat}{6}
&p_{(m;n)}'(x,y,z,t)\stackrel{\eqrefsabc{Eq_WEPFTCF_s_DNSCpS_ss_GFSPEq_sss_ODEsFT_004}
                                        {Eq_WEPFTCF_s_DNSCpS_ss_GFSPEq_sss_VWET_003}
                                        {Eq_WEPFTCF_s_DNSCpS_ss_GFSPEq_sss_VWET_006}}{=}
                                                                                                                 \notag\\
&\int_{-\infty}^{+\infty}\int_{-\infty}^{+\infty}\Bigg(\int_{-\tfrac{1}{2}L_y}^{+\tfrac{1}{2}L_y}G_n(y,Y;\kappa)\hat Q'_{({m})}(\kappa_x,Y,\kappa_z,t)\;dY\Bigg)\;e^{i\kappa_x x+i\kappa_z z}\;d\kappa_x\;d\kappa_z
                                                                                                                 \label{Eq_WEPFTCF_s_APC_ss_VPGPiij_001a}
\end{alignat}
and, by direct differentiation and averaging of \eqref{Eq_WEPFTCF_s_APC_ss_VPGPiij_001a}, we have
\begin{alignat}{6}
&[\Pi_{xx}]_{(m;n)}\stackrel{\eqref{Eq_WEPFTCF_s_APC_ss_RST_002}}{:=}-2\overline{u'\dfrac{\partial p'}
                                                                                         {\partial x }}
\stackrel{\eqref{Eq_WEPFTCF_s_APC_ss_VPGPiij_001a}}{=}
                                                                                                                 \label{Eq_WEPFTCF_s_APC_ss_VPGPiij_001b}\\
&-2
\overline{u'\int_{-\infty}^{+\infty}\int_{-\infty}^{+\infty}\int_{-\tfrac{1}{2}L_y}^{+\tfrac{1}{2}L_y}G_n(y,Y;\kappa)~~\;
                                                                                                      \hat Q'_{({m})}(\kappa_x,Y,\kappa_z,t)i\kappa_xe^{i\kappa_x x+i\kappa_z z}d\kappa_xd\kappa_zdY}
                                                                                                                 \notag\\
&[\Pi_{yy}]_{(m;n)}\stackrel{\eqref{Eq_WEPFTCF_s_APC_ss_RST_002}}{:=}-2\overline{v'\dfrac{\partial p'}
                                                                                         {\partial y }}
\stackrel{\eqref{Eq_WEPFTCF_s_APC_ss_VPGPiij_001a}}{=}
                                                                                                                 \label{Eq_WEPFTCF_s_APC_ss_VPGPiij_001c}\\
&-2
\overline{v'\int_{-\infty}^{+\infty}\int_{-\infty}^{+\infty}\int_{-\tfrac{1}{2}L_y}^{+\tfrac{1}{2}L_y}\dfrac{\partial G_n}
                                                                                                            {\partial y      }(y,Y;\kappa)
                                                                                                      \hat Q'_{({m})}(\kappa_x,Y,\kappa_z,t)~~~~~    e^{i\kappa_x x+i\kappa_z z}d\kappa_xd\kappa_zdY}
                                                                                                                 \notag
\end{alignat}
\begin{alignat}{6}
&[\Pi_{zz}]_{(m;n)}\stackrel{\eqref{Eq_WEPFTCF_s_APC_ss_RST_002}}{:=}-2\overline{w'\dfrac{\partial p'}
                                                                                         {\partial z }}
\stackrel{\eqref{Eq_WEPFTCF_s_APC_ss_VPGPiij_001a}}{=}
                                                                                                                 \label{Eq_WEPFTCF_s_APC_ss_VPGPiij_001d}\\
&-2
\overline{w'\int_{-\infty}^{+\infty}\int_{-\infty}^{+\infty}\int_{-\tfrac{1}{2}L_y}^{+\tfrac{1}{2}L_y}G_n(y,Y;\kappa)~~\;
                                                                                                      \hat Q'_{({m})}(\kappa_x,Y,\kappa_z,t)i\kappa_ze^{i\kappa_x x+i\kappa_z z}d\kappa_xd\kappa_zdY}
                                                                                                                 \notag\\
&[\Pi_{xy}]_{(m;n)}\stackrel{\eqref{Eq_WEPFTCF_s_APC_ss_RST_002}}{:=}-\overline{u'\dfrac{\partial p'}
                                                                                        {\partial y }}
                                                                     -\overline{v'\dfrac{\partial p'}
                                                                                        {\partial x }}
\stackrel{\eqref{Eq_WEPFTCF_s_APC_ss_VPGPiij_001a}}{=}
                                                                                                                 \label{Eq_WEPFTCF_s_APC_ss_VPGPiij_001e}\\
&-~
\overline{v'\int_{-\infty}^{+\infty}\int_{-\infty}^{+\infty}\int_{-\tfrac{1}{2}L_y}^{+\tfrac{1}{2}L_y}G_n(y,Y;\kappa)~~\;
                                                                                                      \hat Q'_{({m})}(\kappa_x,Y,\kappa_z,t)i\kappa_xe^{i\kappa_x x+i\kappa_z z}d\kappa_xd\kappa_zdY}
                                                                                                                 \notag\\
&-~
\overline{u'\int_{-\infty}^{+\infty}\int_{-\infty}^{+\infty}\int_{-\tfrac{1}{2}L_y}^{+\tfrac{1}{2}L_y}\dfrac{\partial G_n}
                                                                                                            {\partial y      }(y,Y;\kappa)
                                                                                                      \hat Q'_{({m})}(\kappa_x,Y,\kappa_z,t)~~~~~    e^{i\kappa_x x+i\kappa_z z}d\kappa_xd\kappa_zdY}
                                                                                                                 \notag
\end{alignat}
\end{subequations}

It is obvious from \eqref{Eq_WEPFTCF_s_APC_ss_SRRSM_001} that there is a fundamental difference in the way the $xz$-Fourier-components of the sources interact with $u_i'$ to built the components of $\Pi_{ij}$.
For $[\Pi_{xx}]_{(m;n)}$ \eqref{Eq_WEPFTCF_s_APC_ss_VPGPiij_001b} and $[\Pi_{zz}]_{(m;n)}$ \eqref{Eq_WEPFTCF_s_APC_ss_VPGPiij_001d},
the contribution of each wavenumber is weighted by $i\kappa_x$ or $i\kappa_z$ (this includes not only modulus weighting but also phase-shift) and by the
appropriate Green's function. This is no longer the case for $[\Pi_{yy}]_{(m;n)}$ \eqref{Eq_WEPFTCF_s_APC_ss_VPGPiij_001c}, where there is no wavenumber-weighting and the $y$-gradient of the Green's function appears.
Therefore, contributions of the sources from different $Y$-locations and different wavenumbers are weighted differently for $[\Pi_{yy}]_{(m;n)}$ compared to the other normal components,
$[\Pi_{xx}]_{(m;n)}$ and $[\Pi_{zz}]_{(m;n)}$. This explains why the behaviour of the different fields in $\Pi_{yy}$ \figref{Fig_WEPFTCF_s_APC_ss_VPGPiij_001}
is so profoundly different compared to $\Pi_{xx}$ or $\Pi_{zz}$. Finally, $\Pi_{xy}$ \eqref{Eq_WEPFTCF_s_APC_ss_VPGPiij_001e} contains both types of integrals, and is arguably more complex to analyze.
Similar analysis, based on \eqrefsab{Eq_WEPFTCF_s_DNSCpS_ss_GFSPEq_sss_ODEsFT_004c}
                                    {Eq_WEPFTCF_s_DNSCpS_ss_GFSPEq_sss_Kim_002},
applies to the Stokes field $p'_{(\tau)}$.

The above analysis hints at the reasons of the different relative behaviour of the contribution of different pressure-fluctuation fields
($p'_{({\rm r};{\mathfrak V})}$, $p'_{({\rm r};w)}$, $p'_{({\rm s};{\mathfrak V})}$, $p'_{({\rm s};w)}$) in the various components of $\Pi_{ij}$.
On the other hand, \eqref{Eq_WEPFTCF_s_APC_ss_SRRSM_001} suggests that further research is required, studying specifically the spectral behaviour of $\hat Q'_{({m})}$  and its convolution
with $u_i'$ $xz$-Fourier-components, continuing the analysis of \citet{Chang_Piomelli_Blake_1999a}, which focused on $p'$ alone.
%}}

%-----------------------------------------------------------------------------------------------------------------------------------
%
%
%
%
%
%
%
%
%
%\section{??????????}\label{WEPFTCF_s_ApESMCsPCs}
%
%
%
%
%
%
%
%
%
%-----------------------------------------------------------------------------------------------------------------------------------

%?????

%-----------------------------------------------------------------------------------------------------------------------------------
%
%
%
%
%
%
%
%
%
\section{Conclusions}\label{WEPFTCF_s_C}
%
%
%
%
%
%
%
%
%
%-----------------------------------------------------------------------------------------------------------------------------------

In the present paper, we consider fully developed ($x$-wise invariant in the mean) turbulent plane channel flow. Examination of the {\sc dns} data of \cite{Hoyas_Jimenez_2008a}, in the range
$Re_{\tau_w}\in[180,2000]$, indicate that, despite the pronounced dependence of $[p'_w]_{\mathrm{rms}}^+$ on $Re_{\tau_w}$, the $y$-distribution of the ratio of slow-to-rapid pressure fluctuations,
$[p'_{(r)}]_{\mathrm{rms}}^{-1}[p'_{(s)}]_{\mathrm{rms}}$, is reasonably independent of $Re_{\tau_w}$, particularly in the near-wall region ($y^+\lessapprox 50$), but also in the outer region
$\left(\tfrac{2}{10}\delta\leq y-y_w\leq \tfrac{7}{10}\delta\right )$ where it takes a nearly constant value ($\sim 1.6$).

The Green's function ($G_{\rm Kim}$) approach of \cite{Kim_1989a}, for the solution of the 1-D modified Helmholtz equation for each parallel-to-the-wall wavenumber,
used in {\sc dns} computations to separate instantaneous pressure
fluctuations $p'$ into rapid, slow and Stokes terms ($p'=p'_{(r)}+p'_({s)}+p'_{(\tau)}$), was extended so as to separate volume ($G_\mathfrak{V}$) and wall-echo ($G_w$) terms 
($p'=p'_{(\mathrm{r};\mathfrak{V})}+p'_{(\mathrm{r};w)}+p'_{(\mathrm{s};\mathfrak{V})}+p'_{(\mathrm{s};w)}+p'_{(\tau)}$), corresponding to volume and surface integrals in the formal solution of the 
the Poisson equation for $p'$ \citep{Chou_1945a}. The algorithm is based on appropriate Green's functions for the volume and wall-echo terms 
($G_{\rm Kim}=G_\mathfrak{V}+G_{w}$; Appendix~\ref{WEPFTCF_s_AppendixGF}), and is directly applicable to existing {\sc dns}
databases containing flowfields sampled at different instants.

The method of images can be represented by an equivalent Green's function ($G_\tsc{mwl}$)
whose expression proves that this method is the superposition (the modified Helmholtz equation for each wavenumber is linear)
of the volume terms ($G_\mathfrak{V}$) and of the wall-echo effect ($G_{w_\pm}$) that each wall would induce in the absence of the other ($G_{\tsc{mwl}}=G_\mathfrak{V}+G_{w_+}+G_{w_-}$). The approximation of the method of images consists
in neglecting the interaction between upper and lower walls, whose effect is shown by the present exact theory to amplify wall-echo, increasingly so as the nondimensional wavenumber $\kappa\delta$ increases.
Theoretical analysis indicates that the method of images is a high-wavenumber approximation, the approximation error growing exponentially with $(\kappa\delta)^{-1}$, but remaining reasonable for the usual wavenumbers of 
energy-containing structures (error $\lessapprox 1\%$ for $\ell_\kappa:=2\pi\kappa^{-1}\leq 4\delta$), consistently with the generally accepted similarity between the near-wall structure of boundary-layer and plane channel flows.
These results are obtained by studying the Green's functions, and as such are independent of the particular distribution of the source-terms 
(hence independent of $Re_{\tau_w}$ or $\bar M_\tsc{cl}$), depending only on the wavenumber.
Nonetheless, although the error estimate qualifies the method of  images as a good engineering approximation, the present exact theory is preferable for the fine analysis of wall-echo effects, to avoid distorsion 
of the large-structures part of the spectrum, especially as it introduces no computational cost overhead.

Analysis of the $p'$-splitting for low $Re_{\tau_w}=180$ flow indicates that:
\begin{enumerate}
\item At the wall, the instantaneous volume ($p'_{(\mathrm{r};\mathfrak{V})}$ and $p'_{(\mathrm{s};\mathfrak{V})}$) and wall-echo ($p'_{(\mathrm{r};w)}$ and $p'_{(\mathrm{s};w)}$) terms of the same field (rapid or slow) are approximately
equal ($[p'_{(\mathrm{r};\mathfrak{V})}]_w\approxeq[p'_{(\mathrm{r};w)}]_w$ and $[p'_{(\mathrm{s};\mathfrak{V})}]_w\approxeq[p'_{(\mathrm{s};w)}]_w$) but this approximate equality only holds very near the wall ($y^+\lessapprox 3$).
\item Although this approximate equality between corresponding volume and wall-echo terms and similar near-wall behaviour ($y^+\lessapprox 20$) holds for $\phi_{ij}$, $\overline{p'u_i'}$ and $d_{ij}^{(p)}$, this does not apply
to the velocity/pressure-gradient correlation $\Pi_{ij}$. For the $\Pi_{yy}$ and $\Pi_{xy}$ components, wall-echo opposes (opposite sign) corresponding volume terms in the range $y^+\lessapprox 20$, contrary to 
$\Pi_{xx}$ and $\Pi_{zz}$. An explanation of this new finding can be sought in the way the $xz$-Fourier components of the sources are weighted by wavenumber and Green's function or its $y$-gradient
in the integrals representing the gradients $\partial_xp'$, $\partial_yp'$ and $\partial_zp'$. 
\end{enumerate}

Theoretical analysis (Appendix~\ref{WEPFTCF_s_AppendixCELMNL}) of compressibility effects in low-Mach-number flow of air shows that the additional compressible terms in the Poisson equation for $p'$
scale with density fluctuations and spatial variations. {\sc dns}-based assessment of the order-of-magnitude of the extra compressible terms indicates that they can be reasonably neglected for centerline 
Mach-number $\bar M_\tsc{cl}\lessapprox0.35$, and that Morkovin's hypothesis stating that the leading-order effect of compressibility on turbulence is related to mean-density variation
($\lessapprox 1.5\%$ for $\bar M_\tsc{cl}\cong 0.35$) applies to the fluctuating-pressure field structure.

The main perspectives of this work are
(a) The application of the algorithm to higher $Re_{\tau_w}$ flows,
(b) The detailed spectral analysis of $\Pi_{ij}$ and the use of the data in improving near-wall modelling of pressure correlations, and
(c) The application of the algorithm to the compressible Poisson equation for $p'$ at higher (supersonic) $\bar M_\tsc{cl}$ flows.

%-----------------------------------------------------------------------------------------------------------------------------------

\begin{acknowledgments}
The authors are listed alphabetically. We are particularly grateful to Prof. B.A. Younis for many enlightening discussions.
Computations were performed using \tsc{hpc} resources from \tsc{genci--idris} (Grant 2010--022139).
Sourcefiles of the code used and computer routines for the splitting are available as part of the
freeware project aerodynamics~\citep[{\tt http://sourceforge.net/projects/aerodynamics}]{Gerolymos_Vallet_aerodynamics_2009}.
Tabulated data are available at {\tt http://www.aerodynamics.fr/DNS\_database/CT\_chnnl}.
The present work was partly supported by the \tsc{eu}-funded research project ProBand,
({\footnotesize STREP--FP6 AST4--CT--2005--012222}).
\end{acknowledgments}
%-----------------------------------------------------------------------------------------------------------------------------------

%-----------------------------------------------------------------------------------------------------------------------------------
%
%
%
%
%
%
%
%
%
\appendix\section{Compressibility effects at the low-Mach-number limit}\label{WEPFTCF_s_AppendixCELMNL}%\oneappendix\section{Green's functions}\label{WEPFTCF_s_AppendixGF}
%
%
%
%
%
%
%
%
%
%-----------------------------------------------------------------------------------------------------------------------------------
\setcounter{figure}{0}\renewcommand{\thefigure}{\Alph{section}\arabic{figure}}

Since the present \tsc{dns} database was obtained using a compressible solver \citep{Gerolymos_Senechal_Vallet_2010a} at $\bar M_\tsc{cl}\approxeq0.34$,
we examine here in more detail the effects of compressibility on fluctuating pressure $p'$, and discuss in particular the parameter that should be used in
the scaling of the compressibility effects. This is also important in assessing how the incompressible flow limit is approached for low-Mach-number aerodynamic flows \citep{Durran_1989a},
as far as correlations containing the fluctuating pressure are concerned. 

%-----------------------------------------------------------------------------------------------------------------------------------
%
%
%
%
%
\subsection{Compressible flow Poisson equation for $p'$}\label{WEPFTCF_s_AppendixCELMNL_ss_CFPEqp'}
%
%
%
%
%
%-----------------------------------------------------------------------------------------------------------------------------------

The flow is modelled by the compressible Navier-Stokes equations \citep[(34--37), pp. 785--786]{Gerolymos_Senechal_Vallet_2010a}.
Taking the divergence of the momentum equation \citep[(34b), p. 785]{Gerolymos_Senechal_Vallet_2010a} readily yields
\begin{subequations}
                                                                                                                                    \label{Eq_WEPFTCF_s_AppendixCELMNL_ss_CFPEqp'_001}
\begin{alignat}{6}
\dfrac{\partial^2}{\partial x_i\partial t  }(\rho u_i)+
\dfrac{\partial^2}{\partial x_i\partial x_j}(\rho u_i u_j)=-\nabla^2 p+\dfrac{\partial^2\tau_{ij}}{\partial x_i\partial x_j}
                                                                      +\dfrac{\partial           }{\partial x_i            }(\rho f_{{\text{\sc v}}_i})
                                                                                                                                    \label{Eq_WEPFTCF_s_AppendixCELMNL_ss_CFPEqp'_001a}
\end{alignat}
where $f_{{\text{\sc v}}_i}$ is the body-acceleration.

\citet{Pantano_Sarkar_2002a} used the inviscid form of \eqref{Eq_WEPFTCF_s_AppendixCELMNL_ss_CFPEqp'_001a}, considering only acoustic pressure ($dp_\mathrm{a}=a^2d\rho$ where $a$ is the speed of sound),
to study compressibility effects in high-speed shear-layers.
\citet{Foysi_Sarkar_Friedrich_2004a} further developed \eqref{Eq_WEPFTCF_s_AppendixCELMNL_ss_CFPEqp'_001a} using the Garrick operator \citep{Garrick_1957a} $[D_{c_t}]^2$
of theoretical compressible unsteady aerodynamics and aeroacoustics~\citep{Miles_1959a_Garrick_Dc_Dt,
                                                                          Bisplinghoff_Ashley_1962a_Garrick_Dc_Dt},
which highlights a wavelike influence of the fluctuating density, and \citet{Mahle_Foysi_Sarkar_Friedrich_2007a} applied it to high-speed compressible mixing-layers
All these high-Mach-number studies used Favre decomposition \citep{Favre_1965a,
                                                  Favre_1965b},
and the form $\partial^2_{x_ix_j}(u_i''u_j''-\overline{u_i''u_j''})$ for the slow terms.
Since we are interested here in establishing the order-of-magnitude of compressibility effects in comparison with the incompressible flow equation \eqref{Eq_WEPFTCF_s_I_001},
we recast the fluctuating part of \eqref{Eq_WEPFTCF_s_AppendixCELMNL_ss_CFPEqp'_001a} in a form containing \eqref{Eq_WEPFTCF_s_I_001} plus compressible terms.
Furthermore, the point can be made that the most general scaling of compressibility effects are density-fluctuations $\rho'_\mathrm{rms}:=\sqrt{\overline{\rho'^2}}$;
therefore, all compressible terms are rewritten in terms of $\rho$ gradients and fluctuations.
This is in particular the case for correlations related to the dilatation \citep{Ristorcelli_1997a}, which can be expressed using the continuity equation
\begin{align}
\dfrac{\partial\rho}{\partial t}+\dfrac{\partial\rho u_\ell}{\partial x_\ell}=0
\qquad\stackrel{\text{$\Theta:=\mathrm{div}\vec{V}=\partial_{x_\ell}u_\ell$}}{\Longrightarrow}\qquad
\Theta=-\dfrac{1}{\rho}\dfrac{D\rho}{Dt}
                                                                                                                                    \label{Eq_WEPFTCF_s_AppendixCELMNL_ss_CFPEqp'_001b}
\end{align}
where $D_t(\cdot):=\partial_t(.)+u_\ell\partial_{x_\ell}(.)$ is the substantial derivative \citep[p.~13]{Pope_2000a}.
Then \eqref{Eq_WEPFTCF_s_AppendixCELMNL_ss_CFPEqp'_001a} reads
\begin{align}
\nabla^2 p\stackrel{\eqrefsab{Eq_WEPFTCF_s_AppendixCELMNL_ss_CFPEqp'_001a}{Eq_WEPFTCF_s_AppendixCELMNL_ss_CFPEqp'_001b}}{=}&
                  \dfrac{\partial^2\tau_{ij}}{\partial x_i\partial x_j}
             +    \dfrac{\partial}{\partial x_i}(\rho f_{{\text{\sc v}}_i})
             -\rho\dfrac{\partial u_i}{\partial x_j}\dfrac{\partial u_j}{\partial x_i}
                                                                                                                              \notag\\
            +&\underbrace{\rho\dfrac{D}{Dt}\Bigg(\dfrac{1}{\rho}\dfrac{D\rho}{Dt}\Bigg)}_{\displaystyle\stackrel{\eqref{Eq_WEPFTCF_s_AppendixCELMNL_ss_CFPEqp'_001b}}{=}-\rho\dfrac{D\Theta}{Dt}}
             -\dfrac{D\vec{V}}{Dt}\cdot{\rm grad}{\rho}
                                                                                                                              \label{Eq_WEPFTCF_s_AppendixCELMNL_ss_CFPEqp'_001c}
\end{align}
by straightforward computation,\footnote{\label{ff_WEPFTCF_s_AppendixCELMNL_ss_CFPEqp'_001}
                                             $\dfrac{\partial^2}{\partial x_i\partial t  }(\rho u_i)+
                                             \dfrac{\partial^2}{\partial x_i\partial x_j}(\rho u_i u_j)=\dfrac{\partial}{\partial t  }\Big(\rho\Theta+u_i\dfrac{\partial\rho}{\partial x_i}\Big)
                                                                                                       +\dfrac{\partial}{\partial x_i}\Big(u_i\dfrac{\partial\rho u_j}{\partial x_j}+\rho u_j\dfrac{\partial u_i}{\partial x_j}\Big)\\
                                             \stackrel{\eqref{Eq_WEPFTCF_s_AppendixCELMNL_ss_CFPEqp'_001b}}{=}
                                              \dfrac{\partial}{\partial t}\Big(\rho\Theta+u_i\dfrac{\partial\rho}{\partial x_i}\Big)
                                             +\Bigg(\Theta\dfrac{\partial\rho u_j}{\partial x_j}+u_i\dfrac{\partial}{\partial x_i}\Big(\dfrac{\partial\rho u_j}{\partial x_j}\Big)
                                             +\dfrac{\partial\rho u_j}{\partial x_i}\dfrac{\partial u_i}{\partial x_j}+\rho u_j\dfrac{\partial\Theta}{\partial x_j}\Bigg)\\
                                             =\Bigg(\overset{\circled{\tsc{i}}}{\rho\dfrac{\partial\Theta}{\partial t}}
                                                   +\overset{\circled{\tsc{ii}}}{\Theta\dfrac{\partial\rho}{\partial t}}
                                                   +\overset{\circled{\tsc{iii}}}{\dfrac{\partial u_i}{\partial t}\dfrac{\partial\rho}{\partial x_i}}
                                                   +\overset{\circled{\tsc{iv}}}{u_i\dfrac{\partial}{\partial x_i}\Big(\dfrac{\partial\rho}{\partial t}\Big)}\Bigg)
                                             +\Bigg(\overset{\circled{\tsc{v}}}{\Theta\dfrac{\partial\rho u_j}{\partial x_j}}
                                                   +\overset{\circled{\tsc{vi}}}{u_i\dfrac{\partial}{\partial x_i}\Big(\dfrac{\partial\rho u_j}{\partial x_j}\Big)}
                                                   +\overset{\circled{\tsc{vii}}}{\rho\dfrac{\partial u_i}{\partial x_j}\dfrac{\partial u_j}{\partial x_i}}
                                                   +\overset{\circled{\tsc{viii}}}{u_j\dfrac{\partial u_i}{\partial x_j}\dfrac{\partial\rho}{\partial x_i}}
                                                   +\overset{\circled{\tsc{ix}}}{\rho u_j\dfrac{\partial\Theta}{\partial x_j}}\Bigg)\\
                                             =\rho\bigg(\underbrace{
                                                        \overset{\circled{\tsc{i}}}{\dfrac{\partial\Theta}{\partial t}}
                                                       +\overset{\circled{\tsc{ix}}}{u_j\dfrac{\partial\Theta}{\partial x_j}}}_{=:\frac{D\Theta}{Dt}}\bigg)
                                              +\Theta\bigg(\underbrace{
                                                           \overset{\circled{\tsc{ii}}}{\dfrac{\partial\rho}{\partial t}}
                                                          +\overset{\circled{\tsc{v}}}{\dfrac{\partial\rho u_j}{\partial x_j}}}_{\stackrel{\eqref{Eq_WEPFTCF_s_AppendixCELMNL_ss_CFPEqp'_001b}}{=}0}\bigg)
                                              +\bigg(\underbrace{\overset{\circled{\tsc{iii}}}{\dfrac{\partial u_i}{\partial t}}
                                                                +\overset{\circled{\tsc{viii}}}{u_j\dfrac{\partial u_i}{\partial x_j}}}_{=:\frac{Du_i}{Dt}}\bigg)
                                               \dfrac{\partial\rho}{\partial x_i}
                                              +u_i\dfrac{\partial}{\partial x_i}\bigg(\underbrace{
                                                                                      \overset{\circled{\tsc{iv}}}{\dfrac{\partial\rho}{\partial t}}
                                                                                     +\overset{\circled{\tsc{vi}}}{\dfrac{\partial\rho u_j}{\partial x_j}}}_{\stackrel{\eqref{Eq_WEPFTCF_s_AppendixCELMNL_ss_CFPEqp'_001b}}{=}0}\bigg)
                                              +\overset{\circled{\tsc{vii}}}{\rho\dfrac{\partial u_i}{\partial x_j}\dfrac{\partial u_j}{\partial x_i}}
                                             $
                                             }
using the continuity equation \eqref{Eq_WEPFTCF_s_AppendixCELMNL_ss_CFPEqp'_001b}.
Substracting from \eqref{Eq_WEPFTCF_s_AppendixCELMNL_ss_CFPEqp'_001c} its Reynolds-average readily yields the working form of the compressible flow Poisson equation for the fluctuating static pressure $p'$
\begin{align}
\nabla^2 p'&=
             \underbrace{\left[-\bar\rho\left(\dfrac{\partial u_i'}{\partial x_j}\dfrac{\partial u_j'}{\partial x_i}
                                   -\overline{\dfrac{\partial u_i'}{\partial x_j}\dfrac{\partial u_j'}{\partial x_i}}\right)\right]}_{\textstyle Q_{(\rm s)}'}
            +\underbrace{\left[-\left(\rho'\dfrac{\partial u_i'}{\partial x_j}\dfrac{\partial u_j'}{\partial x_i}
                           -\overline{\rho'\dfrac{\partial u_i'}{\partial x_j}\dfrac{\partial u_j'}{\partial x_i}}\right)\right]}_{\textstyle Q_{(\rho';{\rm s})}'}
                                                                                                                              \notag\\
           &+\underbrace{\left[-2\bar\rho\left(\dfrac{\partial u_i'}{\partial x_j}\dfrac{\partial \bar u_j}{\partial x_i}\right)\right]}_{\textstyle Q_{(\rm r)}'}
            +\underbrace{\left[-2\left(\rho'\dfrac{\partial u_i'}{\partial x_j}-\overline{\rho'\dfrac{\partial u_i'}{\partial x_j}}\right)
                                            \dfrac{\partial \bar u_j}{\partial x_i}\right]}_{\textstyle Q_{(\rho';{\rm r})}'}
            +\underbrace{\left[-\rho'\dfrac{\partial \bar u_i}{\partial x_j}\dfrac{\partial \bar u_j}{\partial x_i}\right]}_{\textstyle Q_{(\rho')}'}
                                                                                                                              \notag\\
           &+\underbrace{\dfrac{\partial^2\tau_{ij'}}{\partial x_i\partial x_j}}_{\textstyle Q_{(\tau)}'}
            +\underbrace{\dfrac{\partial}{\partial x_i}(\rho f_{{\text{\sc v}}_i}-\overline{\rho f_{{\text{\sc v}}_i}})}_{\textstyle Q_{(\text{\sc bf})}'}
            +\underbrace{\left[\rho\dfrac{D}{Dt}\Bigg(\dfrac{1}{\rho}\dfrac{D\rho}{Dt}\Bigg)\right]'}_{\textstyle Q_{(\Theta)}'}
            +\underbrace{\left[-\dfrac{D\vec{V}}{Dt}\cdot{\rm grad}{\rho}\right]'}_{\textstyle Q_{(\dot V\nabla\rho)}'}
                                                                                                                              \label{Eq_WEPFTCF_s_AppendixCELMNL_ss_CFPEqp'_001d}
\end{align}
where
\begin{align}
\overline{\dfrac{D(\cdot)}{Dt}}&:=\overline{\dfrac{\partial(\cdot)}{\partial t}+u_j\dfrac{\partial(\cdot)}{\partial x_j}}
                                 =\dfrac{\partial\overline{(\cdot)}}{\partial t}+\bar u_j\dfrac{\partial\overline{(\cdot)}}{\partial x_j}
                                                                                    +\overline{u_j'\dfrac{\partial(\cdot)'}{\partial x_j}}
                                                                                                                              \label{Eq_WEPFTCF_s_AppendixCELMNL_ss_CFPEqp'_001e}\\
\left[\dfrac{D(\cdot)}{Dt}\right]'&:=\dfrac{D(\cdot)}{Dt}-\overline{\dfrac{D(\cdot)}{Dt}}
                                     =\dfrac{\partial(\cdot)'}{\partial t}+\bar u_j \dfrac{\partial          (\cdot)'}{\partial x_j}
                                                                          +     u_j'\dfrac{\partial\overline{(\cdot)}}{\partial x_j}
                                                                          +     u_j'\dfrac{\partial          (\cdot)'}{\partial x_j}
                                                                          -\overline{u_j'\frac{\partial      (\cdot)'}{\partial x_j}}
                                                                                                                              \label{Eq_WEPFTCF_s_AppendixCELMNL_ss_CFPEqp'_001f}
\end{align}

Obviously,
\begin{align}
Q_{(\rho';\mathrm{s})}'\stackrel{\eqref{Eq_WEPFTCF_s_AppendixCELMNL_ss_CFPEqp'_001d}}{=}&\dfrac{\rho'}{\bar\rho}Q'_{(\mathrm{s})}
                                                                                                                              \label{Eq_WEPFTCF_s_AppendixCELMNL_ss_CFPEqp'_001g}\\
Q_{(\rho';\mathrm{r})}'\stackrel{\eqref{Eq_WEPFTCF_s_AppendixCELMNL_ss_CFPEqp'_001d}}{=}&\dfrac{\rho'}{\bar\rho}Q'_{(\mathrm{r})}+2\overline{\rho'\dfrac{\partial u_i'}{\partial x_j}}\dfrac{\partial \bar u_j}{\partial x_i}
                                                                                                                              \label{Eq_WEPFTCF_s_AppendixCELMNL_ss_CFPEqp'_001h}
\end{align}
\end{subequations}

%-----------------------------------------------------------------------------------------------------------------------------------
%
%
%
%
%
\subsection{Quasi-incompressible flow approximations}\label{WEPFTCF_s_AppendixCELMNL_ss_QIFA}
%
%
%
%
%
%-----------------------------------------------------------------------------------------------------------------------------------

We start by summarizing some basic approximations for flows with very small mean-density gradients and density fluctuations, which can be termed quasi-incompressible,
and include most low-Mach-number flows without important heat-transfer effects. We assume that for the class of flows under consideration
\begin{subequations}
                                                                                                                                    \label{Eq_WEPFTCF_s_AppendixCELMNL_ss_QIFA_001}
\begin{align}
\dfrac{\rho'}{\bar\rho}\ll&1
                                                                                                                                    \label{Eq_WEPFTCF_s_AppendixCELMNL_ss_QIFA_001a}\\
\abs{\overline{\dfrac{D\rho}{Dt}}}\ll&\left(\dfrac{D\rho}{Dt}\right)'_\mathrm{rms}=:\dfrac{\rho_\mathrm{rms}'}{\mathscr{T}_{(D_t\rho)'}}\stackrel{\eqref{Eq_WEPFTCF_s_AppendixCELMNL_ss_QIFA_001a}}{\ll}\dfrac{\bar\rho}{\mathscr{T}_{(D_t\rho)'}}
                                                                                                                                    \label{Eq_WEPFTCF_s_AppendixCELMNL_ss_QIFA_001b}
\end{align}
\end{subequations}
where $\mathscr{T}_{(D_t\rho)'}:=\rho_\mathrm{rms}'[(D_t\rho)'_\mathrm{rms}]^{-1}$ is an appropriate timescale,\footnote{\label{ff_WEPFTCF_s_AppendixCELMNL_ss_QIFA_001}It will be shown in \parrefnp{WEPFTCF_s_AppendixCELMNL_ss_LMNFDPCF} that,
                                                                                                                                                                        for plane channel flow, this timescale made nondimensional in wall-units,
                                                                                                                                                                        satisfies $[\mathscr{T}_{(D_t\rho)'}]^+=O(1)$
                                                                                                                        }
\ie~density fluctuations are very small compared to mean density \eqref{Eq_WEPFTCF_s_AppendixCELMNL_ss_QIFA_001a},
and furthermore the Reynolds-averaged substantial derivative \eqref{Eq_WEPFTCF_s_AppendixCELMNL_ss_CFPEqp'_001e} of $\rho$ is negligible compared to its fluctuating part \eqref{Eq_WEPFTCF_s_AppendixCELMNL_ss_CFPEqp'_001f},
which is itself very small \eqref{Eq_WEPFTCF_s_AppendixCELMNL_ss_QIFA_001b}.
From \eqref{Eq_WEPFTCF_s_AppendixCELMNL_ss_QIFA_001a} we readily have the usual weakly compressible \citep{Taulbee_vanOsdol_1991a}
approximation\footnote{\label{ff_WEPFTCF_s_AppendixCELMNL_ss_QIFA_002}$x\to0\;\Longrightarrow\;\dfrac{1}{1+x}=1-x+x^2-x^3+x^4+\cdots$}
\begin{subequations}
                                                                                                                                    \label{Eq_WEPFTCF_s_AppendixCELMNL_ss_QIFA_002}
\begin{align}
\dfrac{1}{\rho}\stackrel{\eqref{Eq_WEPFTCF_s_AppendixCELMNL_ss_QIFA_001a}}{    =    }\dfrac{1}{\bar\rho}-\dfrac{\rho'}{\bar\rho^2}+O\left(\dfrac{\rho'^2}{\bar\rho^2}\right)
               \stackrel{\eqref{Eq_WEPFTCF_s_AppendixCELMNL_ss_QIFA_001a}}{\approxeq}\dfrac{1}{\bar\rho}-\dfrac{\rho'}{\bar\rho^2}
                                                                                                                                    \label{Eq_WEPFTCF_s_AppendixCELMNL_ss_QIFA_002a}
\end{align}
Furthermore, \eqrefsab{Eq_WEPFTCF_s_AppendixCELMNL_ss_QIFA_001a}
                      {Eq_WEPFTCF_s_AppendixCELMNL_ss_QIFA_001b} imply
\begin{align}
-\Theta\stackrel{\eqref   {Eq_WEPFTCF_s_AppendixCELMNL_ss_CFPEqp'_001b}}{    =    }&\dfrac{1}{\rho}\dfrac{D\rho}{Dt}
       \stackrel{\eqref   {Eq_WEPFTCF_s_AppendixCELMNL_ss_QIFA_002a}   }{\approxeq}\dfrac{1}{\bar\rho}\dfrac{D\rho}{Dt}-\dfrac{\rho'}{\bar\rho^2}\dfrac{D\rho}{Dt}
       \stackrel{\eqref   {Eq_WEPFTCF_s_AppendixCELMNL_ss_QIFA_001b}   }{\approxeq}\dfrac{1}{\bar\rho}\dfrac{D\rho}{Dt}
                                                                                  +O\left(\dfrac{\rho'^2}{\bar\rho^2}\dfrac{1}{\mathscr{T}_{(D_t\rho)'}}\right)
       \stackrel{\eqref{Eq_WEPFTCF_s_AppendixCELMNL_ss_QIFA_001}}{\approxeq}&\dfrac{1}{\bar\rho}\dfrac{D\rho}{Dt}
                                                                                                                                    \label{Eq_WEPFTCF_s_AppendixCELMNL_ss_QIFA_002b}
\end{align}
and
\begin{align}
-\left (\rho\dfrac{D\Theta}{Dt}\right )'\stackrel{\eqref   {Eq_WEPFTCF_s_AppendixCELMNL_ss_CFPEqp'_001b}}{    =    }&\left(\rho\dfrac{D}{Dt}\left(\dfrac{1}{\rho}\dfrac{D\rho}{Dt}\right)\right )'
                                                                                                              =      \left(\dfrac{D^2\rho}{Dt^2}-\dfrac{1}{\rho}\left (\dfrac{D\rho}{Dt}\right )^2\right )'
                                                                                                                                    \notag\\
                                        \stackrel{\eqrefsab{Eq_WEPFTCF_s_AppendixCELMNL_ss_QIFA_001}
                                                           {Eq_WEPFTCF_s_AppendixCELMNL_ss_QIFA_002a}   }{    =    }&\left(\dfrac{D^2\rho}{Dt^2}\right)'+O\left(\dfrac{\rho'^2}{\bar\rho}\dfrac{1}{\mathscr{T}^2_{(D_t\rho)'}}\right)
                                        \stackrel{\eqref   {Eq_WEPFTCF_s_AppendixCELMNL_ss_QIFA_001}    }{\approxeq} \left(\dfrac{D^2\rho}{Dt^2}\right)'
                                                                                                                                    \label{Eq_WEPFTCF_s_AppendixCELMNL_ss_QIFA_002c}
\end{align}
where \eqref{Eq_WEPFTCF_s_AppendixCELMNL_ss_QIFA_001b} was used to establish the leading term of the error.
\end{subequations}

It was preferred to use density fluctuations and variations to establish order-of-magnitude relations, and to calculate correlations and orders-of-magnitude related to dilatation $\Theta$
via \eqref{Eq_WEPFTCF_s_AppendixCELMNL_ss_CFPEqp'_001b}, both because they are easier to understand physically than
dilation,\footnote{\label{ff_WEPFTCF_s_AppendixCELMNL_ss_QIFA_003}indeed, the transport equation for the dilatation variance $D_t\overline{\Theta'^2}$
                                                                  is actually obtained by multiplying \eqref{Eq_WEPFTCF_s_AppendixCELMNL_ss_CFPEqp'_001d} by $\Theta'$
                  }
but also because they are more reliable numerically (in terms of required grid resolution and associated numerical error),
the more so as $\bar M_\tsc{cl}$ decreases.

%-----------------------------------------------------------------------------------------------------------------------------------
%
%
%
%
%
\subsection{Low-Mach-number fully developed plane channel flow}\label{WEPFTCF_s_AppendixCELMNL_ss_LMNFDPCF}
%
%
%
%
%
%-----------------------------------------------------------------------------------------------------------------------------------

\begin{figure}
\begin{center}
\begin{picture}(380,250)
\put(-50,-185){\includegraphics[angle=0,width=470pt]{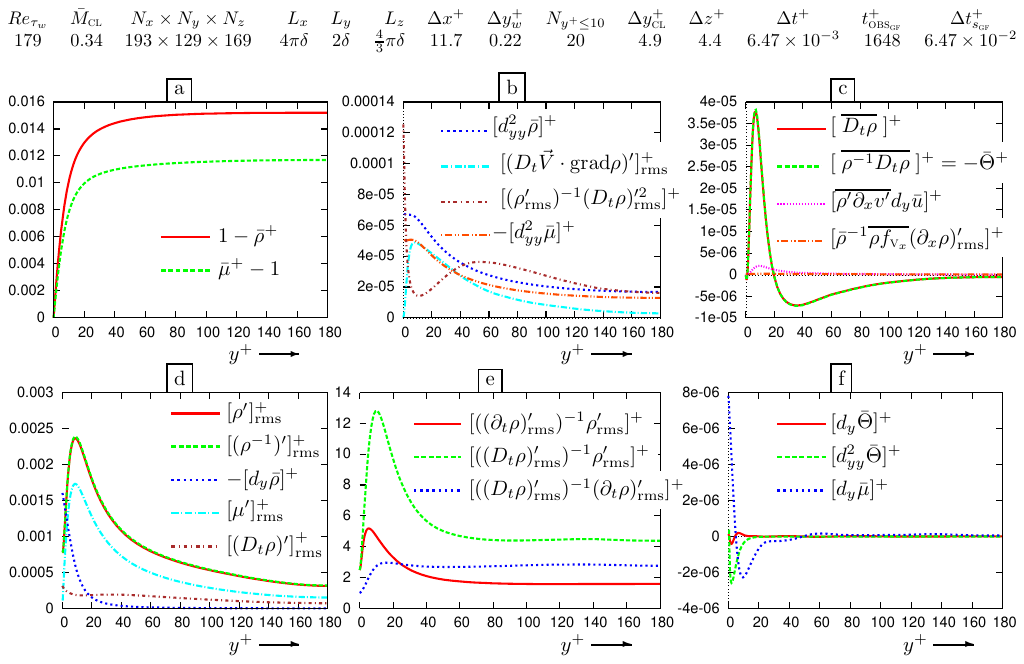}}
\end{picture}
\end{center}
\caption{Present \tsc{dns}-computed data ($Re_{\tau_w}=180$; $\bar M_\tsc{cl}=0.34$; $257\times129\times385$ grid; \tabrefnp{Tab_WEPFTCF_s_DNSCpS_ss_PCFCCM_001})
related to density variation and fluctuation (a) $\bar\rho$, $\bar\mu$,
(b) $d^2_{yy}\bar\rho$, $[(D_t\vec{V}\cdot\mathrm{grad}\rho)']_\mathrm{rms}$, $((\partial_t\rho)_\mathrm{rms}')^{-1}((D_t\rho)'_\mathrm{rms})^2$, -$d^2_{yy}\bar\mu$,
(c) $\overline{D_t\rho}$, $\overline{\rho^{-1}D_t\rho}=-\bar\Theta$, $\overline{\rho'\partial_x v'} d_y \bar u$, $\bar\rho^{-1}\overline{\rho f_{\tsc{v}_x}}(\partial_x \rho)_\mathrm{rms}'$,
(d) $\rho'_\mathrm{rms}$, $(\rho^{-1})'_\mathrm{rms}$, $-d_y\bar\rho$, $\mu'_\mathrm{rms}$, $(D_t\rho)'_\mathrm{rms}$,
(e) $((\partial_t\rho)_\mathrm{rms}')^{-1}\rho_\mathrm{rms}'$, $((D_t\rho)_\mathrm{rms}')^{-1}\rho_\mathrm{rms}'$, $((D_t\rho)_\mathrm{rms}')^{-1}(\partial_t\rho)_\mathrm{rms}'$,
(f) $d_y\bar\Theta$, $d^2_{yy}\bar\Theta$, $d_y\bar\mu$,
in wall units,
relevant to the order-of-magnitude analysis of compressibility effects in \eqref{Eq_WEPFTCF_s_AppendixCELMNL_ss_CFPEqp'_001d}.}
\label{Fig_WEPFTCF_s_AppendixCELMNL_ss_LMNFDPCF_001}
\end{figure}

For the flow studied here (plane channel flow; $Re_{\tau_w}=180$; $\bar M_\tsc{cl}=0.34$),
conditions \eqref{Eq_WEPFTCF_s_AppendixCELMNL_ss_QIFA_001} leading to the approximations \eqref{Eq_WEPFTCF_s_AppendixCELMNL_ss_QIFA_002} are satisfied.
Making all quantities nondimensional in wall units ($\bar\rho_w$,$\bar\nu_w$,$\bar u_\tau$),
density variance $[\rho']_\mathrm{rms}^+\lessapprox 2\tfrac{1}{2}\times10^{-3}$ \subfigref{Fig_WEPFTCF_s_AppendixCELMNL_ss_LMNFDPCF_001}{d}, 
variance of the substantial derivative of density $\left[D_t\rho'\right]_\mathrm{rms}^+\lessapprox3\times10^{-4}$ \subfigref{Fig_WEPFTCF_s_AppendixCELMNL_ss_LMNFDPCF_001}{d},
with mean $\left[\abs{\overline{D_t\rho}}\right]^+\lessapprox 4\times10^{-5}$ \subfigref{Fig_WEPFTCF_s_AppendixCELMNL_ss_LMNFDPCF_001}{c}.
Taking into account that $\bar\rho^+\approxeq 1$ with an accuracy of $1.5\%$ \subfigref{Fig_WEPFTCF_s_AppendixCELMNL_ss_LMNFDPCF_001}{a},
condition \eqref{Eq_WEPFTCF_s_AppendixCELMNL_ss_QIFA_001a} is verified as $\rho'\sim\tfrac{3}{1000}\bar\rho$,
and the separation in \eqref{Eq_WEPFTCF_s_AppendixCELMNL_ss_QIFA_001b} is of 1 order-of-magnitude. Therefore \eqref{Eq_WEPFTCF_s_AppendixCELMNL_ss_QIFA_002} are expected to hold.
Verification of \eqref{Eq_WEPFTCF_s_AppendixCELMNL_ss_QIFA_002a} comes from comparing
$\left [ \rho'\right ]_\mathrm{rms}^+$ and $\left[(\rho^{-1})'\right]_\mathrm{rms}^+$, which are almost equal with an excellent accuracy
(the 2 curves are indistinguishable; \subfigrefnp{Fig_WEPFTCF_s_AppendixCELMNL_ss_LMNFDPCF_001}{d}).\footnote{\label{ff_WEPFTCF_s_AppendixCELMNL_ss_LMNFDPCF_001}
                                                                                                              although not plotted here, the same applies to skewness (opposite) and flatness (equal)
                                                                                                              of $\rho'$ and $(\rho^{-1})'$
                                                                                                              }
Furthermore, because of the very small variation of $\bar\rho(y)$, $\left[\abs{d_y\bar\rho}\right]^+\lessapprox2\times10^{-3}=O([\rho']_\mathrm{rms}^+)$
\subfigref{Fig_WEPFTCF_s_AppendixCELMNL_ss_LMNFDPCF_001}{d}.\footnote{\label{ff_WEPFTCF_s_AppendixCELMNL_ss_LMNFDPCF_002}
                                                                      the order-of-magnitude relation between $\left[d_y\bar\rho\right]^+$ and $[\rho']_\mathrm{rms}^+$
                                                                      can also be justified by considering the production term
                                                                      in the transport equation for density variance $\overline{\rho'^2}$ \citep[(17), p. 4]{Taulbee_vanOsdol_1991a},
                                                                      which, for the flow under consideration, is $-\overline{\rho'v'}d_y\bar\rho$
                                                                     }
The approximation \eqref{Eq_WEPFTCF_s_AppendixCELMNL_ss_QIFA_002b} is corroborated by the excellent superposition
of the \tsc{dns} data for $[\;\overline{D_t\rho}\;]^+$ and $[\;\overline{\rho^{-1}D_t\rho}\;]^+\stackrel{\eqref{Eq_WEPFTCF_s_AppendixCELMNL_ss_CFPEqp'_001b}}{=}-\bar\Theta^+$
\subfigref{Fig_WEPFTCF_s_AppendixCELMNL_ss_LMNFDPCF_001}{c}.

In the case of 2-D in the mean fully developed ($x$-wise invariant in the mean) compressible channel flow with $\vec{f}_{\tsc{v}}=f_{\tsc{v}_x}(t)\vec{e}_x$~\citep{Coleman_Kim_Moser_1995a,
                                                                                                                                                                    Huang_Coleman_Bradshaw_1995a,
                                                                                                                                                                    Gerolymos_Senechal_Vallet_2010a}
and using \eqrefsab{Eq_WEPFTCF_s_AppendixCELMNL_ss_CFPEqp'_001g}{Eq_WEPFTCF_s_AppendixCELMNL_ss_CFPEqp'_001h}, \eqref{Eq_WEPFTCF_s_AppendixCELMNL_ss_CFPEqp'_001d} made nondimensional in wall units reads
\begin{eqnarray}
\left[\nabla^2 p'\right ]^+=\underbrace{\left (1+\dfrac{[\rho']^+}{\bar{\rho}^+}\right )
                                        \left [ \textstyle Q_{(\rm s)}' + \textstyle Q_{(\rm r)}' \right ]^+
       + 2\left [\overline{\rho'\dfrac{\partial v'}{\partial x}}\dfrac{d\bar{u}}{dy}\right ]^+}_{\textstyle\left[Q_{(\rm s)}'+Q_{(\rho';{\rm s})}'+Q_{(\rm r)}'+Q_{(\rho';{\rm r})}' \right ]^+}
                           +\underbrace{0}_{\left [ \textstyle Q_{(\rho')}'\right ]^+=0}
                                                                                                                                                             \notag\\
                           +\underbrace{f_{\tsc{v}_x}^+\left [\dfrac{\partial \rho'}{\partial x}\right ]^+}_{\left [\textstyle Q_{(\text{\sc bf})}'\right ]^+}
                           +\left [ Q_{(\Theta)}'\right ]^+ + \left [ Q_{(\dot V\nabla\rho)}'\right ]^+ +  \left [ Q_{(\tau)}'\right ]^+
                                                                                                                                                              \label{Eq_WEPFTCF_s_AppendixCELMNL_ss_LMNFDPCF_001}
\end{eqnarray}
exactly. Statistics of the individual source-terms in \eqref{Eq_WEPFTCF_s_AppendixCELMNL_ss_LMNFDPCF_001} were not available, but it is possible to make order-of-magnitude estimates,
by relating them to available \tsc{dns} data.
%-----------------------------------------------------------------------------------------------------------------------------------
%-----------------------------------------------------------------------------------------------------------------------------------
\begin{itemize}
\item $\left [ Q_{(\rm s)}' + Q_{(\rm r)}' \right ]^+$: Observation of instantaneous values indicates that $\left [ Q_{(\rm s)}' + Q_{(\rm r)}' \right ]^+\in[-1,1]$,
implying that the terms which were retained in the incompressible analysis \eqref{Eq_WEPFTCF_s_I_001} are $\left [ Q_{(\rm s)}' + Q_{(\rm r)}' \right ]^+\sim O(\tfrac{1}{10})$
for the present flow conditions (plane channel flow; $Re_{\tau_w}=180$; $\bar M_\tsc{cl}=0.34$).
\item $\left[Q_{(\rho';{\rm s})}'\right]^+$, $\left[Q_{(\rho';{\rm r})}'\right]^+$, $\left[Q_{(\rho')}'\right]^+$:
The term $Q_{(\rho')}'$ \eqref{Eq_WEPFTCF_s_AppendixCELMNL_ss_CFPEqp'_001d} is identically $=0$ for this flow.
Obviously, the term $2\overline{\rho'\partial_x v'}d_y\bar u=O(2c_{(\rho',\partial_x v')}\rho'_\mathrm{rms}(\partial_x v')_\mathrm{rms}d_y\bar u)$
and using the definition or $Q'_{(\mathrm{r})}$ \eqref{Eq_WEPFTCF_s_AppendixCELMNL_ss_CFPEqp'_001d},
$2\overline{\rho'\partial_x v'}d_y\bar u=O(c_{(\rho',\partial_x v')}\bar\rho^{-1} \rho'_\mathrm{rms} (Q'_{(\mathrm{r})})_\mathrm{rms})$.
In the present flow $\left[\abs{\overline{\rho'\partial_x v'}}d_y\bar u\right]^+\lessapprox3\times10^{-6}$
is negligibly small \subfigref{Fig_WEPFTCF_s_AppendixCELMNL_ss_LMNFDPCF_001}{b}.\footnote{\label{ff_WEPFTCF_s_AppendixCELMNL_ss_LMNFDPCF_003}
                                                                                         the present \tsc{dns} data indicate that the correlation coefficient $c_{(\rho',\partial_x v')}\approxeq0.3\pm0.05$
                                                                                         in the range $y^+\in[1,100]$, falling to $0$ at the wall and at channel centerline
                                                                                         }
\item $\left[Q_{(\text{\sc bf})}'\right]^+$: 
Concerning the body-force term $[Q_{(\text{\sc bf})}']^+$ \eqref{Eq_WEPFTCF_s_AppendixCELMNL_ss_LMNFDPCF_001}, it obviously scales with $[\rho']_\mathrm{rms}^+$.
It also scales with $Re_{\tau_w}^{-1}$ (decreases with increasing $Re_{\tau_w}$), because volume-integration and subsequent averaging of the momentum equation \citep[(34b), p. 785]{Gerolymos_Senechal_Vallet_2010a},
yields the exact relation $\overline{\rho_\tsc{b}f_{\tsc{v}_x}}=Re_{\tau_w}^{-1}$, where, because of the small variation and fluctuation of density,
the bulk density \citep[(46a), p. 791]{Gerolymos_Senechal_Vallet_2010a} $\rho^+_\tsc{b}\approxeq\bar\rho^+\approxeq1$. Present \tsc{dns} data
indicate that $[\bar\rho^{-1}\overline{\rho f_{\tsc{v}_x}}(\partial_x \rho)_\mathrm{rms}']^+\lessapprox2\times10^{-7}$ \subfigref{Fig_WEPFTCF_s_AppendixCELMNL_ss_LMNFDPCF_001}{c}.
\item $\left[Q_{(\dot V\nabla\rho)}'\right]^+$: It is easy to show that the acceleration/density-gradient source-term $[(D_t\vec{V}\cdot\mathrm{grad}\rho)']^+$ \eqref{Eq_WEPFTCF_s_AppendixCELMNL_ss_CFPEqp'_001d} should scale
with $[\rho']_\mathrm{rms}+$ \subfigref{Fig_WEPFTCF_s_AppendixCELMNL_ss_LMNFDPCF_001}{d} and $\left[\abs{d_y\bar\rho}\right]^+$ \subfigref{Fig_WEPFTCF_s_AppendixCELMNL_ss_LMNFDPCF_001}{d},
weighted by acceleration statistics. The \tsc{dns} data indicate $[(D_t\vec{V}\cdot\mathrm{grad}\rho)']^+_\mathrm{rms}\lessapprox5\times10^{-5}$ \subfigref{Fig_WEPFTCF_s_AppendixCELMNL_ss_LMNFDPCF_001}{b}.
\item $\left[Q_{(\Theta)}'\right]^+$:
By \eqref{Eq_WEPFTCF_s_AppendixCELMNL_ss_QIFA_002c} $Q_{(\Theta)}'\stackrel{\eqref{Eq_WEPFTCF_s_AppendixCELMNL_ss_CFPEqp'_001d}}{:=}-\left(\rho D_t\Theta\right)'
                                                                  \stackrel{\eqref{Eq_WEPFTCF_s_AppendixCELMNL_ss_CFPEqp'_001b}}{ =}\rho(D_t(\rho^{-1} D_t\rho)'
                                                                  \stackrel{\eqref{Eq_WEPFTCF_s_AppendixCELMNL_ss_QIFA_002c}}{\approxeq}(D_{tt}^2\rho)'$.
Statistics for $\rho D_t(\rho^{-1} D_t\rho)$ or $D_{tt}^2\rho$ were not available. We can make a rough estimate of its order-of-magnitude.
 From the available statistics for $\left(D_t\rho\right)_\mathrm{rms}'$ \subfigref{Fig_WEPFTCF_s_AppendixCELMNL_ss_LMNFDPCF_001}{d} and
$\left(\partial_t\rho\right)_\mathrm{rms}'$ (not plotted) we can define associated timescales, $\mathscr{T}_{(D_t\rho)'}:=\rho_\mathrm{rms}'[(D_t\rho)'_\mathrm{rms}]^{-1}$
and $\mathscr{T}_{(\partial_t\rho)'}:=\rho_\mathrm{rms}'[(\partial_t\rho)'_\mathrm{rms}]^{-1}$. These specific timescales \subfigref{Fig_WEPFTCF_s_AppendixCELMNL_ss_LMNFDPCF_001}{e}, associated with the fluctuation of the
Eulerian and Lagrangian time-derivatives are practically constant in the major part of the channel ($y^+\gtrapprox15$),
and their ratio \subfigref{Fig_WEPFTCF_s_AppendixCELMNL_ss_LMNFDPCF_001}{e} is $\sim1$ near the wall where convection is small,
and grows to $\sim3$ for $y^+\gtrapprox15$, which is also a typical value of the ratio of Lagrangian to Eulerian timescales \citep{Dosio_VilaGueraudeArellano_Holtslag_Builtjes_2005a},
although these are not the timescales defined here. Then, by analogy, we assume that the same timescale $\mathscr{T}_{(D_t\rho)'}$
relates the fluctuation of the time-derivative of $D_t\rho$ to $\left(D_t\rho\right)_\mathrm{rms}'$, \viz
\begin{align}
\left [\rho\dfrac{D}{Dt}\left (\rho^{-1}\dfrac{D\rho}{Dt}\right )\right ]_\mathrm{rms}'
\stackrel{\eqref{Eq_WEPFTCF_s_AppendixCELMNL_ss_QIFA_002c}}{\approxeq}
=\left(\dfrac{D(D_t\rho)}{Dt}\right)_\mathrm{rms}'\sim\dfrac{\left(D_t\rho\right)_\mathrm{rms}'}{\mathscr{T}_{(D_t\rho)'}}
=O\left (\dfrac{\left[\left(D_t\rho\right )'\right]_\mathrm{rms}^2}{\rho'_\mathrm{rms}}\right )
                                                                                                                                    \label{Eq_WEPFTCF_s_AppendixCELMNL_ss_LMNFDPCF_002}
\end{align}
which suggests that this term is not expected to be important in the present flow, where $\left[(\rho'_\mathrm{rms})^{-1}\left[\left ( D_t\rho\right )'\right]_\mathrm{rms}^{2}\right]^+\lessapprox4. 10^{-5}$
\subfigref{Fig_WEPFTCF_s_AppendixCELMNL_ss_LMNFDPCF_001}{b}. Notice that the term associated with $\bar\rho$ in $D^2_{tt}$ is of the same order-of-magnitude as the estimate \eqref{Eq_WEPFTCF_s_AppendixCELMNL_ss_LMNFDPCF_002},
\viz~$\left[\abs{d_{yy}^2\bar{\rho}}\right ]^+\lessapprox6\times10^{-5}$ \subfigref{Fig_WEPFTCF_s_AppendixCELMNL_ss_LMNFDPCF_001}{b}.
\item $\left[Q_{(\tau)}'\right]^+$: 
The present calculations used a Newtonian constitutive relation \citep[(36a), p. 786]{Gerolymos_Senechal_Vallet_2010a},
with $\mu_\mathrm{b}=0$ and $\mu=\mu(T)$ \citep[(37), p. 786]{Gerolymos_Senechal_Vallet_2010a}.  The exact expression of
$Q_{(\tau)}'$\footnote{\label{ff_WEPFTCF_s_AppendixCELMNL_ss_LMNFDPCF_004}
                       $Q_{(\tau)}'\stackrel{\eqref{Eq_WEPFTCF_s_AppendixCELMNL_ss_CFPEqp'_001d}}{:=}\dfrac{\partial^2\tau_{ij'}}{\partial x_i\partial x_j}\\
~~~~~~~~~~~~~~~~~~~~~~~~ =\left(         (\tfrac{4}{3}\mu+\mu_\mathrm{b})\nabla^2\Theta
                                         +4\dfrac{\partial\mu}{\partial x_i}\dfrac{\partial S_{ij}}{\partial x_j}
                                         +2S_{ij}\dfrac{\partial^2\mu}{\partial x_i\partial x_j}
                                         +\Theta\nabla^2(\mu_\mathrm{b}-\tfrac{2}{3}\mu)
                                         +2\dfrac{\partial(\mu_\mathrm{b}-\tfrac{2}{3}\mu)}{\partial x_j}\dfrac{\partial\Theta}{\partial x_j}\right)'\\
~~~~~~~~~~~~~~~~~~~~~~~~ =\left(         (\tfrac{4}{3}\mu'+\mu'_\mathrm{b})\nabla^2\bar\Theta
                                         +4\dfrac{\partial\mu'}{\partial x_i}\dfrac{\partial\bar S_{ij}}{\partial x_j}
                                         +2S'_{ij}\dfrac{\partial^2\bar\mu}{\partial x_i\partial x_j}
                                         +\Theta'\nabla^2(\bar\mu_\mathrm{b}-\tfrac{2}{3}\bar\mu)
                                         +2\dfrac{\partial(\mu'_\mathrm{b}-\tfrac{2}{3}\mu')}{\partial x_j}\dfrac{\partial\bar\Theta}{\partial x_j}\right)\\
~~~~~~~~~~~~~~~~~~~~~~~~~+\left(         (\tfrac{4}{3}\bar\mu+\bar\mu_\mathrm{b})\nabla^2\Theta'
                                         +4\dfrac{\partial\bar\mu}{\partial x_i}\dfrac{\partial S'_{ij}}{\partial x_j}
                                         +2\bar S_{ij}\dfrac{\partial^2\mu'}{\partial x_i\partial x_j}
                                         +\bar \Theta\nabla^2(\mu'_\mathrm{b}-\tfrac{2}{3}\mu')
                                         +2\dfrac{\partial(\bar\mu_\mathrm{b}-\tfrac{2}{3}\bar\mu)}{\partial x_j}\dfrac{\partial\Theta'}{\partial x_j}\right)\\
~~~~~~~~~~~~~~~~~~~~~~~~~+\left(          (\tfrac{4}{3}\mu'+\mu'_\mathrm{b})\nabla^2\Theta'
                                         +4\dfrac{\partial\mu'}{\partial x_i}\dfrac{\partial S'_{ij}}{\partial x_j}
                                         +2S'_{ij}\dfrac{\partial^2\mu'}{\partial x_i\partial x_j}
                                         +\Theta'\nabla^2(\mu'_\mathrm{b}-\tfrac{2}{3}\mu')
                                         +2\dfrac{\partial(\mu'_\mathrm{b}-\tfrac{2}{3}\mu')}{\partial x_j}\dfrac{\partial\Theta'}{\partial x_j}\right)\\
~~~~~~~~~~~~~~~~~~~~~~~~~-\left(\overline{(\tfrac{4}{3}\mu'+\mu'_\mathrm{b})\nabla^2\Theta'
                                         +4\dfrac{\partial\mu'}{\partial x_i}\dfrac{\partial S'_{ij}}{\partial x_j}
                                         +2S'_{ij}\dfrac{\partial^2\mu'}{\partial x_i\partial x_j}
                                         +\Theta'\nabla^2(\mu'_\mathrm{b}-\tfrac{2}{3}\mu')
                                         +2\dfrac{\partial(\mu'_\mathrm{b}-\tfrac{2}{3}\mu')}{\partial x_j}\dfrac{\partial\Theta'}{\partial x_j}}\right)
                       $
                      }
involves terms depending on the variation and fluctuation of $\mu$ and $\Theta$. For this reason, in strictly incompressible flow, $Q_{(\tau)}'=0$.
In the present flow (plane channel flow; $Re_{\tau_w}=180$; $\bar M_\tsc{cl}=0.34$), mean viscosity varies in a way very similar to $\rho$,
$\bar\mu^+\approxeq 1$ with an accuracy of $1.2\%$ \subfigref{Fig_WEPFTCF_s_AppendixCELMNL_ss_LMNFDPCF_001}{a}, and the same similarity with density applies
to spatial variations, $\left[\abs{d_y\bar\mu}\right]^+\lessapprox8\times10^{-6}$ \subfigref{Fig_WEPFTCF_s_AppendixCELMNL_ss_LMNFDPCF_001}{f} and
$\left[\abs{d^2_{yy}\bar\mu}\right]^+\lessapprox5\times10^{-5}$ \subfigref{Fig_WEPFTCF_s_AppendixCELMNL_ss_LMNFDPCF_001}{b},
and fluctuation, $[\mu']_\mathrm{rms}^+\lessapprox 2\times10^{-3}$ \subfigref{Fig_WEPFTCF_s_AppendixCELMNL_ss_LMNFDPCF_001}{d}.
As already stated, mean dilatation $\abs{\bar\Theta^+}\stackrel{\eqref{Eq_WEPFTCF_s_AppendixCELMNL_ss_CFPEqp'_001b}}{=}\left[\abs{\;\overline{\rho^{-1}D_t\rho}\;}\right]^+\lessapprox4\times10^{-5}$,
with gradients $\left[\abs{d_y\bar\Theta}\right]^+\lessapprox4\times10^{-7}$ \subfigref{Fig_WEPFTCF_s_AppendixCELMNL_ss_LMNFDPCF_001}{f} and
$\left[\abs{d^2_{yy}\bar\Theta}\right]^+\lessapprox2\times10^{-6}$ \subfigref{Fig_WEPFTCF_s_AppendixCELMNL_ss_LMNFDPCF_001}{f},
and fluctuation $[\Theta']_\mathrm{rms}^+\stackrel{\eqref{Eq_WEPFTCF_s_AppendixCELMNL_ss_QIFA_002b}}{\approxeq}\left[D_t\rho'\right]_\mathrm{rms}^+\lessapprox3\times10^{-4}$ \subfigref{Fig_WEPFTCF_s_AppendixCELMNL_ss_LMNFDPCF_001}{d}.
Therefore, terms containing products of the small (in wall units) quantities $d_y\bar\mu$, $d^2_{yy}\bar\mu$, $\bar\Theta$, $d_y\bar\Theta$, $d^2_{yy}\bar\Theta$, $\mu'$ (and its space-derivatives),
and $\Theta'$ (and its space-derivatives), can be considered negligible compared to other terms, and we have ($\mu_\mathrm{b}=0$)
\begin{align}
Q_{(\tau)}'=\tfrac{4}{3}\bar\mu\nabla^2\Theta'
           +4\dfrac{\partial S_{yj}'}{\partial x_j}\dfrac{d\bar\mu}{dy}
           +2\dfrac{\partial v'}{\partial y}\dfrac{d^2\bar\mu}{dy^2}
           +2\dfrac{\partial\mu'}{\partial x}\dfrac{d^2\bar u}{dy^2}
           +2\dfrac{d\bar u}{dy}\dfrac{\partial^2\mu'}{\partial x\partial y}+o\left(Q_{(\tau)}'\right)
                                                                                                                                    \label{Eq_WEPFTCF_s_AppendixCELMNL_ss_LMNFDPCF_003}
\end{align}
%\vspace*{-0.1in}
where $S_{ij}:=\tfrac{1}{2}(\partial_{x_j}u_i+\partial_{x_i}u_j)$ is the rate-of-strain tensor.
\end{itemize}
Using the above estimates in \eqrefsab{Eq_WEPFTCF_s_AppendixCELMNL_ss_CFPEqp'_001d}{Eq_WEPFTCF_s_AppendixCELMNL_ss_LMNFDPCF_001} we may write
\begin{align}
\left[\nabla^2 p'\right ]^+=&\left[Q_{(\rm s)}' +Q_{(\rm r)}'\right]^+
                           + \overbrace{\left[Q_{(\rho';{\rm s})}'+Q_{(\rho';{\rm r})}'\right]^+}^{\textstyle \dfrac{[\rho']^+}{\bar\rho^+}\left[Q_{(\rm s)}' +Q_{(\rm r)}'\right]^+
                                                                                                            +O\left(([\rho']^+_\mathrm{rms}\left[Q'_{(\mathrm{r})}\right]^+_\mathrm{rms}\right)}
                           + \overbrace{\left[Q_{(\rho')}'\right ]^+}^{\textstyle=0} \notag\\
%~                      \notag\\
                           +&\underbrace{\left[Q_{(\tsc{bf})}'\right ]^+}_{\textstyle O\left(\dfrac{[\rho']^+_\mathrm{rms}}{Re_{\tau_w}}\right)}
                           + \underbrace{\left [ Q_{(\Theta)}'\right ]^+}_{\textstyle O\left (\dfrac{{\left[\left(D_t\rho\right )'\right]_\mathrm{rms}^+}^2}{[\rho']^+_\mathrm{rms}}\right )}
                           + \underbrace{\left [ Q_{(\dot V\nabla\rho)}'\right ]^+}_{\textstyle O\left([\rho']^+_\mathrm{rms},\left[d_y\bar\rho\right]^+\right)}
                                                                                                                                    \notag\\
%~            \notag\\
                           +&\underbrace{\left [ Q_{(\tau)}'\right ]^+}_{\textstyle O\left(\left[(D_t\rho)'\right]_\mathrm{rms}^+, [\mu']_\mathrm{rms}^+,\left[d_y\bar\mu\right]^+,\left[d^2_{yy}\bar\mu\right]^+\right)}
                                                                                                                                    \label{Eq_WEPFTCF_s_AppendixCELMNL_ss_LMNFDPCF_004}
\end{align}
By \eqref{Eq_WEPFTCF_s_AppendixCELMNL_ss_LMNFDPCF_004} it is clear that all extra compressible terms in \eqref{Eq_WEPFTCF_s_AppendixCELMNL_ss_CFPEqp'_001d}, compared to \eqref{Eq_WEPFTCF_s_I_001},
scale with $[\rho']^+_\mathrm{rms}$ or $\left[(D_t\rho)'\right]_\mathrm{rms}^+$ and $\left[d_y\bar\rho\right]^+$ ($\mu$ gradients and fluctuations, in wall units, being of the same order-of-magnitude as corresponding
gradients and fluctuations of $\rho$; \figrefnp{Fig_WEPFTCF_s_AppendixCELMNL_ss_LMNFDPCF_001}). The order-of-magnitude analysis of different terms indicates that they can be reasonably neglected for
the flow studied in the present work (plane channel flow; $Re_{\tau_w}=180$; $\bar M_\tsc{cl}=0.34$). As a consequence, the main influence of compressibility appears in the retained quasi-incompressible terms
$Q_{(\rm s)}' +Q_{(\rm r)}'$ \eqrefsab{Eq_WEPFTCF_s_AppendixCELMNL_ss_CFPEqp'_001d}
                                      {Eq_WEPFTCF_s_I_001},
through the variation of $\bar\rho(y)$, indicating the validity of Morkovin's hypothesis \citep{So_Gatski_Sommer_1998a}, which states that, for low-Mach-number flow,
the effects of compressibility on turbulence are due to the mean-density-gradient, the influence of $\rho'$ being a higher-order effect.
%}}

%-----------------------------------------------------------------------------------------------------------------------------------
%
%
%
%
%
%
%
%
%
\appendix\setcounter{section}{1}\section{Green's functions}\label{WEPFTCF_s_AppendixGF}%\oneappendix\section{Green's functions}\label{WEPFTCF_s_AppendixGF}
%
%
%
%
%
%
%
%
%
%-----------------------------------------------------------------------------------------------------------------------------------

We analyze the mathematical tools for the solution of the modified Helmholtz equation \eqref{Eq_WEPFTCF_s_DNSCpS_ss_GFSPEq_sss_ODEsFT_005}
which provides the terms of the $p'$-splitting \eqrefsab{Eq_WEPFTCF_s_DNSCpS_ss_GFSPEq_sss_ODEsFT_001}
                                                        {Eq_WEPFTCF_s_DNSCpS_ss_GFSPEq_sss_VWET_001}
for each Fourier $xz$-Fourier-component of $p'$ \eqref{Eq_WEPFTCF_s_DNSCpS_ss_GFSPEq_sss_ODEsFT_004a}.
The Green's functions approach for the solution of \eqref{Eq_WEPFTCF_s_DNSCpS_ss_GFSPEq_sss_ODEsFT_005} is briefly summarized
in \parrefnp{WEPFTCF_s_AppendixGF_ss_GFSMHEq}.
In \parrefnp{WEPFTCF_s_AppendixGF_ss_kappaneq0} we study the case of spatially $xz$-varying Fourier components ($\kappa:=\sqrt{\kappa_x^2+\kappa_z^2}\neq0$),
for which the Green's function has the general form \eqref{Eq_WEPFTCF_s_AppendixGF_ss_kappaneq0_002},
parametrized by 2 functions, $A_1(Y;\kappa)$ and $A_2(Y;\kappa)$, which are determined by the boundary-conditions.
Homogeneous Neumann boundary-conditions \eqref{Eq_WEPFTCF_s_DNSCpS_ss_GFSPEq_sss_ODEsFT_005b} yield \parref{WEPFTCF_s_AppendixGF_ss_kappaneq0_sss_Kim}
the well known solution of Kim \citep{Kim_1989a}, $G_{\rm Kim}(y,Y;\kappa\neq0)$ \eqref{Eq_WEPFTCF_s_AppendixGF_ss_kappaneq0_sss_Kim_002c},
providing the solution \eqref{Eq_WEPFTCF_s_DNSCpS_ss_GFSPEq_sss_Kim_001} for the rapid $p_{(\mathrm{r})}'$ and slow $p_{(\mathrm{s})}'$ parts \eqref{Eq_WEPFTCF_s_DNSCpS_ss_GFSPEq_sss_ODEsFT_001},
while the solution \citep{Moser_Kim_Mansour_1999a,
                          Chang_Piomelli_Blake_1999a}
for Stokes part $p_{(\tau)}'$ \eqref{Eq_WEPFTCF_s_DNSCpS_ss_GFSPEq_sss_ODEsFT_001}, with $Q_{(\tau)}'=0$ in strictly incompressible flow \eqrefsab{Eq_WEPFTCF_s_DNSCpS_ss_GFSPEq_sss_ODEsFT_002a}
                                                                                                                                                       {Eq_WEPFTCF_s_DNSCpS_ss_GFSPEq_sss_ODEsFT_005a},
which is driven by the inhomogeneous Neumann boundary-conditions \eqrefsabc{Eq_WEPFTCF_s_DNSCpS_ss_GFSPEq_sss_ODEsFT_002b}
                                                                           {Eq_WEPFTCF_s_DNSCpS_ss_GFSPEq_sss_ODEsFT_003}
                                                                           {Eq_WEPFTCF_s_DNSCpS_ss_GFSPEq_sss_ODEsFT_005b},
given by $q_\tsc{bc}(y;\kappa\neq0,B_\pm)$ \eqref{Eq_WEPFTCF_s_AppendixGF_ss_kappaneq0_sss_ABCs_002}, is calculated in \parrefnp{WEPFTCF_s_AppendixGF_ss_kappaneq0_sss_ABCs}.
The free-space Green's function $G_\mathfrak{V}(y,Y;\kappa\neq0)$ \eqref{Eq_WEPFTCF_s_AppendixGF_ss_kappaneq0_sss_FSGF_002},
which provides the solution \eqref{Eq_WEPFTCF_s_DNSCpS_ss_GFSPEq_sss_VWET_003} for the rapid $p'_{(\mathrm{r};\mathfrak{V})}$ and slow $p'_{(\mathrm{s};\mathfrak{V})}$ volume-parts of
the $p'$-splitting \eqref{Eq_WEPFTCF_s_DNSCpS_ss_GFSPEq_sss_VWET_001} is calculated in \parrefnp{WEPFTCF_s_AppendixGF_ss_kappaneq0_sss_FSGF}.
Finally, for use in the analysis \parref{WEPFTCF_s_DNSCpS_ss_GFSPEq_sss_AMI} of the method-of-images approach \citep{Manceau_Wang_Laurence_2001a},
we calculate in \parrefnp{WEPFTCF_s_AppendixGF_ss_kappaneq0_sss_ULWGF} the Green's functions $G_\pm(y,Y;\kappa\neq0)$ \eqref{Eq_WEPFTCF_s_AppendixGF_ss_kappaneq0_sss_ULWGF_004}
and in \parrefnp{WEPFTCF_s_AppendixGF_ss_kappaneq0_sss_ABCsHP} the boundary-conditions functions $q_{\tsc{bc}_\pm}(y;\kappa\neq0,B_\pm)$ \eqref{Eq_WEPFTCF_s_AppendixGF_ss_kappaneq0_sss_ABCs_003},
for the halfspace problems with only 1 (upper or lower) of the walls present.
In \parrefnp{WEPFTCF_s_AppendixGF_ss_kappaeq0} we discuss the singular case $\kappa:=\sqrt{\kappa_x^2+\kappa_z^2}=0\Longrightarrow\kappa=\kappa_x=\kappa_z=0$,
for which the modified Helmholtz equation \eqref{Eq_WEPFTCF_s_DNSCpS_ss_GFSPEq_sss_ODEsFT_005} can be solved iff the compatibility conditions \eqref{Eq_WEPFTCF_s_AppendixGF_ss_kappaeq0_002}
hold. We revisit \parrefsatob{WEPFTCF_s_AppendixGF_ss_kappaeq0_sss_Kim}
                             {WEPFTCF_s_AppendixGF_ss_kappaeq0_sss_ABCsHP}
the same problems as for the $\kappa\neq0$ case, and calculate the corresponding
Green's functions $G_{\rm Kim}(y,Y;\kappa=0)G_\mathfrak{V}(y,Y;\kappa=0)=G_\pm(y,Y;\kappa=0)$ \eqrefsabc{Eq_WEPFTCF_s_AppendixGF_ss_kappaeq0_sss_Kim_003}
                                                                                                        {Eq_WEPFTCF_s_AppendixGF_ss_kappaeq0_sss_FSGF_001}
                                                                                                        {Eq_WEPFTCF_s_AppendixGF_ss_kappaeq0_sss_ULWGF_001}
boundary-conditions functions $q_\tsc{bc}(y;\kappa=0,B_-,B_+)$ \eqref{Eq_WEPFTCF_s_AppendixGF_ss_kappaeq0_sss_ABCs_003}
and $q_{\tsc{bc}_\pm}(y;\kappa=0,B_\pm)$ \eqref{Eq_WEPFTCF_s_AppendixGF_ss_kappaeq0_sss_ABCsHP_002},
for the $\kappa=0$ case.
Finally, in \parrefnp{WEPFTCF_s_AppendixGF_ss_AEMI} we calculate the Green's function 
$G_\tsc{mwl}(y,Y;\kappa)$ \eqref{Eq_WEPFTCF_s_AppendixGF_ss_AEMI_004} which corresponds to the method-of-images evaluation \eqref{Eq_WEPFTCF_s_DNSCpS_ss_GFSPEq_sss_AMI_001}
of the rapid $p'_{(\mathrm{r};\tsc{mwl})}$ and slow $p'_{(\mathrm{s};\tsc{mwl})}$ parts,
and evaluate the error of the method-of-images approximation as a function of the wavenumber \eqref{Eq_WEPFTCF_s_AppendixGF_ss_AEMI_008}.
The application of these mathematical results to the calculation of the $p'$-splitting is described in the main paper \parref{WEPFTCF_s_DNSCpS_ss_GFSPEq}.
%}}

%-----------------------------------------------------------------------------------------------------------------------------------
%
%
%
%
%
\subsection{Green's function solution of the modified Helmholtz equation}\label{WEPFTCF_s_AppendixGF_ss_GFSMHEq}
%
%
%
%
%
%-----------------------------------------------------------------------------------------------------------------------------------

As discussed in \S\ref{WEPFTCF_s_DNSCpS_ss_GFSPEq}, we have to solve the generic modified Helmholtz~\citep{Cheng_Huang_Leiterman_2006a} equation
\begin{alignat}{6}
\dfrac{d^2 q(y;\kappa)}
      {dy^2           } -\kappa^2 q(y;\kappa) = Q(y;\kappa)\qquad y\in(L_-,L_+)
                                                                                                                              \label{Eq_WEPFTCF_s_AppendixGF_ss_GFSMHEq_001}
\end{alignat}
with associated boundary-conditions at $y\in\{L_-,L_+\}$,
for the complex-valued function $q:\mathbb{R}\longrightarrow\mathbb{C}$, $Q(y)$ being a given complex-valued function $Q:\mathbb{R}\longrightarrow\mathbb{C}$, and $\kappa\in\mathbb{R}_{\geq0}$.
Notice that there is no loss of generality in assuming $\kappa\in\mathbb{R}_{\geq0}$ because \eqref{Eq_WEPFTCF_s_AppendixGF_ss_GFSMHEq_001} depends on $\kappa^2$ and is therefore independent of the sign of $\kappa$.
The general method of solution~\citep{Bender_Orszag_1978a_1D_Greenfunction} of the linear \tsc{ode}
\eqref{Eq_WEPFTCF_s_AppendixGF_ss_GFSMHEq_001} is based on the determination of the appropriate Green's function $G(y,Y;\kappa)$, solution of
\begin{subequations}
                                                                                                                              \label{Eq_WEPFTCF_s_AppendixGF_ss_GFSMHEq_002}
\begin{alignat}{6}
\dfrac{\partial^2 G(y,Y;\kappa)}
      {\partial y^2            } -\kappa^2 G(y,Y;\kappa) = \delta(y-Y)\qquad\forall y,Y\in[L_-,L_+]
                                                                                                                              \label{Eq_WEPFTCF_s_AppendixGF_ss_GFSMHEq_002a}\\
\lim_{\epsilon\to0^+}\left[\dfrac{\partial G}
                                 {\partial y}(y=Y+\epsilon,Y;\kappa)
                          -\dfrac{\partial G}
                                 {\partial y}(y=Y-\epsilon,Y;\kappa)\right]=1\qquad\forall Y\in[L_-,L_+]
                                                                                                                              \label{Eq_WEPFTCF_s_AppendixGF_ss_GFSMHEq_002b}
\end{alignat}
\begin{alignat}{6}
&G(y,Y;\kappa)\in C^0(L_-,L_+)
                                                                                                                              \label{Eq_WEPFTCF_s_AppendixGF_ss_GFSMHEq_002c}\\
&G(y,Y;\kappa)\in C^1(L_-,L_+)\setminus\{Y\}
                                                                                                                              \label{Eq_WEPFTCF_s_AppendixGF_ss_GFSMHEq_002d}
\end{alignat}
\end{subequations}
so that
\begin{alignat}{6}
q(y;\kappa):=\int_{L_-}^{L_+} G(y,Y;\kappa)\;Q(Y;\kappa)\;dY
                                                                                                                              \label{Eq_WEPFTCF_s_AppendixGF_ss_GFSMHEq_003}
\end{alignat}
satisfies \eqref{Eq_WEPFTCF_s_AppendixGF_ss_GFSMHEq_001} because of \eqref{Eq_WEPFTCF_s_AppendixGF_ss_GFSMHEq_002},
as can be verified by substituting \eqref{Eq_WEPFTCF_s_AppendixGF_ss_GFSMHEq_003} in \eqref{Eq_WEPFTCF_s_AppendixGF_ss_GFSMHEq_001}.
The Green's function satisfies the symmetry (reciprocity) condition
\begin{alignat}{6}
G(y,Y;\kappa)=G(Y,y;\kappa) \qquad \forall y,Y\in(L_-,L_+) \qquad \forall \kappa\in\mathbb{R}_{\geq0}
                                                                                                                              \label{Eq_WEPFTCF_s_AppendixGF_ss_GFSMHEq_004}
\end{alignat}
because the modified Helmholz operator $[d^2_{yy}(\cdot)-\kappa^2(\cdot)]$ is a self-adjoint linear differential operator~\citep{Ince_1926a_1D_Greenfunction_and_compatibility_relation,
                                                                                                                                 Courant_Hilbert_1953a_1D_Greenfunction}.
As shown in \cite{Bender_Orszag_1978a_1D_Greenfunction},
the general solution of \eqref{Eq_WEPFTCF_s_AppendixGF_ss_GFSMHEq_002} is
\begin{alignat}{6}
G(y,Y;\kappa)=\left\{\begin{array}{rcr} A_1(Y;\kappa)\;q_1(y;\kappa)
                                       +A_2(Y;\kappa)\;q_2(y;\kappa)                                           &\quad&y\leq Y\\
                                                                                                              ~&\quad&~  \\
                                        \left(A_1(Y;\kappa)-\dfrac{q_2(Y;\kappa)         }
                                                                  {[W(q_1,q_2)](Y;\kappa)}\right) q_1(y;\kappa)&\quad&   \\
                                       +\left(A_2(Y;\kappa)+\dfrac{q_1(Y;\kappa)         }
                                                                  {[W(q_1,q_2)](Y;\kappa)}\right) q_2(y;\kappa)&\quad&y\geq Y\\\end{array}\right.
                                                                                                                              \label{Eq_WEPFTCF_s_AppendixGF_ss_GFSMHEq_005}
\end{alignat}
where $q_1(y;\kappa)$ and $q_2(y;\kappa)$ are 2 linearly independent solutions of the corresponding homogeneous equation $q''(y;\kappa)-\kappa^2q(y;\kappa)=0$,
\begin{alignat}{6}
[W(q_1,q_2)](y;\kappa):={\rm det}\left[\begin{array}{rcr}q_1 (y;\kappa)&\quad&q_2 (y;\kappa)\\
                                                         q_1'(y;\kappa)&\quad&q_2'(y;\kappa)\\\end{array}\right]\neq0
                                                                                                                              \label{Eq_WEPFTCF_s_AppendixGF_ss_GFSMHEq_006}
\end{alignat}
is the Wronskian ($\cdot'$ in \eqref{Eq_WEPFTCF_s_AppendixGF_ss_GFSMHEq_006} denotes differentiation by $y$, $\kappa$ being a parameter), which is $\neq0$
iff the 2 solutions are linearly independent. The functions $A_1(Y;\kappa)$ and $A_2(Y;\kappa)$ are determined by the boundary-conditions.
It is straightforward to verify that $G(y,Y;\kappa)$ \eqref{Eq_WEPFTCF_s_AppendixGF_ss_GFSMHEq_005} is continuous at $y=Y$.
Notice that by straightforward differentiation of \eqref{Eq_WEPFTCF_s_AppendixGF_ss_GFSMHEq_003}
\begin{alignat}{6}
q'(y;\kappa)=\int_{L_-}^{L_+} \dfrac{\partial G}{\partial y}(y,Y;\kappa)\;Q(Y;\kappa)\;dY
                                                                                                                              \label{Eq_WEPFTCF_s_AppendixGF_ss_GFSMHEq_007}
\end{alignat}
The cases $\kappa\neq0$ and $\kappa=0$ are fundamentally different, not only because the 2 linearly independent solutions of the homogeneous equation $q''(y;\kappa)-\kappa^2 q(y;\kappa)=0$
differ, but principally because $\kappa=0$ is a characteristic number (eigenvalue) of \eqref{Eq_WEPFTCF_s_AppendixGF_ss_GFSMHEq_001}, {\em viz} $q''(y;\kappa)-\kappa^2 q(y;\kappa)\stackrel{\kappa=0}{=}q''(y;\kappa=0)=0$ with
homogeneous Neumann boundary-conditions admits the constant function as an eigensolution~\citep{Ince_1926a_1D_Greenfunction_and_compatibility_relation,
                                                                                                    Courant_Hilbert_1953a_1D_Greenfunction}.

%-----------------------------------------------------------------------------------------------------------------------------------
%
%
%
%
%
\subsection{$\kappa\neq0$ ($xz$-varying components)}\label{WEPFTCF_s_AppendixGF_ss_kappaneq0}
%
%
%
%
%
%-----------------------------------------------------------------------------------------------------------------------------------

When $\kappa\neq0$ the 2 linearly independent solutions of the homogeneous equation $q''(y;\kappa\neq0)-\kappa^2q(y;\kappa\neq0)=0$ are
\begin{subequations}
                                                                                                                              \label{Eq_WEPFTCF_s_AppendixGF_ss_kappaneq0_001}
\begin{alignat}{6}
q_1(y;\kappa\neq0)         =&{\rm e}^{+\kappa y}
                                                                                                                              \label{Eq_WEPFTCF_s_AppendixGF_ss_kappaneq0_001a}\\
q_2(y;\kappa\neq0)         =&{\rm e}^{-\kappa y}
                                                                                                                              \label{Eq_WEPFTCF_s_AppendixGF_ss_kappaneq0_001b}
\end{alignat}
with Wronskian
\begin{alignat}{6}
[W(q_1,q_2)](y;\kappa\neq0)=&{\rm det}\left[\begin{array}{rcr}       {\rm e}^{+\kappa y}&\quad&       {\rm e}^{-\kappa y}\\
                                                              +\kappa{\rm e}^{+\kappa y}&\quad&-\kappa{\rm e}^{-\kappa y}\end{array}\right]=-2\kappa\neq0
                                                                                                                              \label{Eq_WEPFTCF_s_AppendixGF_ss_kappaneq0_001c}
\end{alignat}
\end{subequations}
so that solution \eqref{Eq_WEPFTCF_s_AppendixGF_ss_GFSMHEq_005} reads
\begin{alignat}{6}
G(y,Y;\kappa\neq0)=\left\{\begin{array}{rcr} A_1(Y;\kappa)\;{\rm e}^{+\kappa y}
                                            +A_2(Y;\kappa)\;{\rm e}^{-\kappa y}                                &\quad&y\leq Y\\
                                                                                                              ~&\quad&~      \\
                                             A_1(Y;\kappa)\;{\rm e}^{+\kappa y}
                                            +A_2(Y;\kappa)\;{\rm e}^{-\kappa y}                                &\quad&       \\
                                            +\dfrac{1}{\kappa}\sinh[\kappa(y-Y)]                               &\quad&y\geq Y\\\end{array}\right.
                                                                                                                              \label{Eq_WEPFTCF_s_AppendixGF_ss_kappaneq0_002}
\end{alignat}

The functions $A_1(Y;\kappa)$ and $A_2(Y;\kappa)$ are determined so as to satisfy the boundary-conditions.

%-----------------------------------------------------------------------
%
\subsubsection{$G_{\rm Kim}(y,Y;\kappa\neq0)$}\label{WEPFTCF_s_AppendixGF_ss_kappaneq0_sss_Kim}
%
%-----------------------------------------------------------------------

\cite{Kim_1989a} solves for the homogeneous Neumann boundary-conditions \eqref{Eq_WEPFTCF_s_DNSCpS_ss_GFSPEq_sss_ODEsFT_005b}
\begin{subequations}
                                                                                                                              \label{Eq_WEPFTCF_s_AppendixGF_ss_kappaneq0_sss_Kim_002}
\begin{alignat}{6}
\dfrac{\partial G_{\rm Kim}}
      {\partial y          }(y=\pm\dfrac{L_y}{2},Y;\kappa)=0\stackrel{\eqref{Eq_WEPFTCF_s_AppendixGF_ss_GFSMHEq_003}}{\iff}\dfrac{\partial q_{\rm Kim}}
                                                                                                                                 {\partial y          }(y=\pm\dfrac{L_y}{2};\kappa)=0
                                                                                                                              \label{Eq_WEPFTCF_s_AppendixGF_ss_kappaneq0_sss_Kim_002a}
\end{alignat}
Straightforward calculations yield the functions $A_1(Y;\kappa)$ and $A_2(Y;\kappa)$ in \eqref{Eq_WEPFTCF_s_AppendixGF_ss_kappaneq0_002} which satisfy
\eqref{Eq_WEPFTCF_s_AppendixGF_ss_kappaneq0_sss_Kim_002a}, giving
\begin{alignat}{6}
G_{\rm Kim}(y,Y;\kappa\neq0)=-\dfrac{\cosh[\kappa(L_y-\abs{y-Y})]+\cosh[\kappa(y+Y)]}
                                    {2\kappa\sinh{\kappa L_y}                       }
                                                                                                                              \label{Eq_WEPFTCF_s_AppendixGF_ss_kappaneq0_sss_Kim_002b}
\end{alignat}
This if-less expression, which highlights the symmetry (reciprocity) condition \eqref{Eq_WEPFTCF_s_AppendixGF_ss_GFSMHEq_004}, is equivalent to the form in \citet[(8), p. 440]{Kim_1989a},
written here using dimensional variables
\begin{alignat}{6}
G_{\rm Kim}(y,Y;\kappa\neq0)=\left\{\begin{array}{rcr} -\dfrac{\cosh[\kappa(\tfrac{1}{2}L_y-Y)]\cosh[\kappa(\tfrac{1}{2}L_y+y)]}
                                                              {\kappa\sinh{\kappa L_y}                                         }
                                                                                                                              &\quad&y\leq Y\\
                                                                                                                             ~&\quad&~      \\
                                                       -\dfrac{\cosh[\kappa(\tfrac{1}{2}L_y-y)]\cosh[\kappa(\tfrac{1}{2}L_y+Y)]}
                                                              {\kappa\sinh{\kappa L_y}                                         }
                                                                                                                              &\quad&y\geq Y\\\end{array}\right.
                                                                                                                              \label{Eq_WEPFTCF_s_AppendixGF_ss_kappaneq0_sss_Kim_002c}
\end{alignat}
because using well-known identities\footnote{\label{ff_WEPFTCF_s_AppendixGF_ss_kappaneq0_sss_Kim_001}$\cosh(a+b)=\cosh a\cosh b+\sinh a\sinh b$\newline
                                                                                                      $\cosh(a-b)=\cosh a\cosh b-\sinh a\sinh b$
                                            }
\citep[p. 249]{Harris_Stocker_1998a}
\begin{alignat}{6}
2 \cosh[\kappa(\tfrac{1}{2}L_y-Y)]\cosh[\kappa(\tfrac{1}{2}L_y+y)] =& \cosh[\kappa(L_y-Y+y)]+\cosh[\kappa(-Y-y)]
                                                                                                                              \notag\\
                                                                   =& \cosh[\kappa(L_y-\abs{y-Y})]+\cosh[\kappa(y+Y)]
                                                                                                                              \notag\\
                                                                    &\forall y\leq Y
                                                                                                                              \label{Eq_WEPFTCF_s_AppendixGF_ss_kappaneq0_sss_Kim_002d}\\
2 \cosh[\kappa(\tfrac{1}{2}L_y-y)]\cosh[\kappa(\tfrac{1}{2}L_y+Y)] =& \cosh[\kappa(L_y-y+Y)]+\cosh[\kappa(-y-Y)]
                                                                                                                              \notag\\
                                                                   =& \cosh[\kappa(L_y-\abs{y-Y})]+\cosh[\kappa(y+Y)]
                                                                                                                              \notag\\
                                                                    &\forall y\geq Y
                                                                                                                              \label{Eq_WEPFTCF_s_AppendixGF_ss_kappaneq0_sss_Kim_002e}
\end{alignat}
\end{subequations}

%-----------------------------------------------------------------------
%
\subsubsection{Accommodating the boundary-conditions ($\kappa\neq0$)}\label{WEPFTCF_s_AppendixGF_ss_kappaneq0_sss_ABCs}
%
%-----------------------------------------------------------------------

To take into account normal gradients $d_y q(y=\pm\tfrac{1}{2}L_y;\kappa\neq0)=B_\pm(\kappa\neq0)\neq0$ at the channel walls,
corresponding to the Stokes pressure \eqrefsab{Eq_WEPFTCF_s_DNSCpS_ss_GFSPEq_sss_ODEsFT_002}{Eq_WEPFTCF_s_DNSCpS_ss_GFSPEq_sss_ODEsFT_005},
\cite{Chang_Piomelli_Blake_1999a} superpose a boundary-condition $q_\tsc{bc}$ term to the solution of~\cite{Kim_1989a}
\begin{subequations}
                                                                                                                              \label{Eq_WEPFTCF_s_AppendixGF_ss_kappaneq0_sss_ABCs_001}
\begin{alignat}{6}
q(y;\kappa\neq0,B_\pm)=\int_{-\tfrac{L_y}{2}}^{+\tfrac{L_y}{2}} G_{\rm Kim}(y,Y;\kappa\neq0)Q(Y;\kappa\neq0)dY+q_\tsc{bc}(y;\kappa\neq0,B_\pm)
                                                                                                                              \label{Eq_WEPFTCF_s_AppendixGF_ss_kappaneq0_sss_ABCs_001a}\\
\dfrac{d^2 q         (y;\kappa\neq0,B_\pm)}
      {dy^2                               } -\kappa^2 q         (y;\kappa\neq0,B_\pm) = Q(y;\kappa\neq0)\qquad y\in(-\tfrac{1}{2}L_y,+\tfrac{1}{2}L_y)
                                                                                                                              \label{Eq_WEPFTCF_s_AppendixGF_ss_kappaneq0_sss_ABCs_001b}\\
\dfrac{d^2 q_\tsc{bc}(y;\kappa\neq0,B_\pm)}
      {dy^2                               } -\kappa^2 q_\tsc{bc}(y;\kappa\neq0,B_\pm) = 0\qquad y\in(-\tfrac{1}{2}L_y,+\tfrac{1}{2}L_y)
                                                                                                                              \label{Eq_WEPFTCF_s_AppendixGF_ss_kappaneq0_sss_ABCs_001c}\\
\dfrac{dq}
      {dy}(y=\pm\tfrac{1}{2}L_y;\kappa\neq0,B_\pm)\stackrel{\eqref{Eq_WEPFTCF_s_AppendixGF_ss_kappaneq0_sss_Kim_002a}}{=}
\dfrac{dq_\tsc{bc}}
      {dy         }(y=\pm\tfrac{1}{2}L_y;\kappa\neq0,B_\pm)=B_\pm(\kappa\neq0)
                                                                                                                              \label{Eq_WEPFTCF_s_AppendixGF_ss_kappaneq0_sss_ABCs_001d}
\end{alignat}
\end{subequations}
By \eqrefsab{Eq_WEPFTCF_s_AppendixGF_ss_GFSMHEq_003}{Eq_WEPFTCF_s_AppendixGF_ss_kappaneq0_sss_Kim_002},
the superposition \eqref{Eq_WEPFTCF_s_AppendixGF_ss_kappaneq0_sss_ABCs_001a} is the solution to \eqref{Eq_WEPFTCF_s_AppendixGF_ss_kappaneq0_sss_ABCs_001b},
with boundary-conditions \eqref{Eq_WEPFTCF_s_AppendixGF_ss_kappaneq0_sss_ABCs_001d}. The solution of \eqref{Eq_WEPFTCF_s_AppendixGF_ss_kappaneq0_sss_ABCs_001c} with boundary-conditions
\eqref{Eq_WEPFTCF_s_AppendixGF_ss_kappaneq0_sss_ABCs_001d} is readily obtained by direct integration \citep{Chang_Piomelli_Blake_1999a,
                                                                                                            Foysi_Sarkar_Friedrich_2004a},
and reads
\begin{alignat}{6}
q_\tsc{bc}(y;\kappa\neq0,B_\pm)=\dfrac{B_+\cosh[\kappa(\tfrac{1}{2}L_y+y)]-B_-\cosh[\kappa(\tfrac{1}{2}L_y-y)]}
                                      {\kappa\sinh{\kappa L_y}                                                }
                                                                                                                              \label{Eq_WEPFTCF_s_AppendixGF_ss_kappaneq0_sss_ABCs_002}
\end{alignat}

%-----------------------------------------------------------------------
%
\subsubsection{Freespace Green's function $G_\mathfrak{V}(y,Y;\kappa\neq0)$}\label{WEPFTCF_s_AppendixGF_ss_kappaneq0_sss_FSGF}
%
%-----------------------------------------------------------------------

The freespace Green's function is directly obtained from the general solution \eqref{Eq_WEPFTCF_s_AppendixGF_ss_kappaneq0_002}
by requiring that the solution should tend to $0$ as $y\to\pm\infty$, for any finite distribution of sources $Q(y; \kappa\neq0)$ with compact support, which is equivalent to
\begin{alignat}{6}
\lim_{y\to\pm\infty} q_\mathfrak{V}(y;\kappa\neq0) = 0 \stackrel{\eqref{Eq_WEPFTCF_s_AppendixGF_ss_GFSMHEq_003}}{\iff}\lim_{y\to\pm\infty} G_\mathfrak{V}(y,Y;\kappa\neq0)=0 \qquad\forall Y\in\mathbb{R}\setminus\{y\}
                                                                                                                              \label{Eq_WEPFTCF_s_AppendixGF_ss_kappaneq0_sss_FSGF_001}
\end{alignat}
Using \eqref{Eq_WEPFTCF_s_AppendixGF_ss_kappaneq0_sss_FSGF_001} to determine the functions $A_1(Y;\kappa)$ and $A_2(Y;\kappa)$ in \eqref{Eq_WEPFTCF_s_AppendixGF_ss_kappaneq0_002} readily gives
\begin{alignat}{6}
G_\mathfrak{V}(y,Y;\kappa\neq0)&=-\dfrac{\mathrm{e}^{-\kappa|y-Y|}}{2\kappa}
                                                                                                                              \label{Eq_WEPFTCF_s_AppendixGF_ss_kappaneq0_sss_FSGF_002}
\end{alignat}

%-----------------------------------------------------------------------
%
\subsubsection{Upper/lower wall Green's functions $G_\pm(y,Y;\kappa\neq0)$}\label{WEPFTCF_s_AppendixGF_ss_kappaneq0_sss_ULWGF}
%
%-----------------------------------------------------------------------

These Green's functions correspond to the problems where the domain of interest is $(-\infty,\tfrac{1}{2}L_y)$ (isolated upper-wall influence; $G_+$) or $(-\tfrac{1}{2}L_y,+\infty)$ (isolated lower-wall influence; $G_-$),
{\em ie} they are halfspace problems ($\pm y\in(-\infty,\tfrac{1}{2}L_y)$). For each of these problems only one of the walls is present.
The associated boundary-conditions are $0$-gradient at the wall of the problem ($y=\pm\tfrac{1}{2}L_y$), and bounded influence at the infinity of the problem ($y\to\mp\infty$),
{\em ie} in terms of Green's functions
\begin{subequations}
                                                                                                                              \label{Eq_WEPFTCF_s_AppendixGF_ss_kappaneq0_sss_ULWGF_001}
\begin{alignat}{6}
G_+(y,Y;\kappa\neq0)~\text{satisfies \eqref{Eq_WEPFTCF_s_AppendixGF_ss_GFSMHEq_002}$\Longrightarrow$\eqrefsab{Eq_WEPFTCF_s_AppendixGF_ss_GFSMHEq_003}
                                                                                                             {Eq_WEPFTCF_s_AppendixGF_ss_GFSMHEq_007}}
                                                                                                                              \label{Eq_WEPFTCF_s_AppendixGF_ss_kappaneq0_sss_ULWGF_001a}\\
\dfrac{\partial G_+}
      {\partial y  }(y=+\tfrac{1}{2}L_y,Y;\kappa\neq0)=0
                                                                                                                              \label{Eq_WEPFTCF_s_AppendixGF_ss_kappaneq0_sss_ULWGF_001b}\\
\lim_{y\to-\infty}G_+(y,Y;\kappa\neq0) = 0 \quad\forall Y\in\mathbb{R}\setminus\{y\}
                                                                                                                              \label{Eq_WEPFTCF_s_AppendixGF_ss_kappaneq0_sss_ULWGF_001c}
\end{alignat}
\end{subequations}
to study the virtual influence of the upper wall, and
\begin{subequations}
                                                                                                                              \label{Eq_WEPFTCF_s_AppendixGF_ss_kappaneq0_sss_ULWGF_002}
\begin{alignat}{6}
G_-(y,Y;\kappa\neq0)~\text{satisfies \eqref{Eq_WEPFTCF_s_AppendixGF_ss_GFSMHEq_002}$\Longrightarrow$\eqrefsab{Eq_WEPFTCF_s_AppendixGF_ss_GFSMHEq_003}
                                                                                                             {Eq_WEPFTCF_s_AppendixGF_ss_GFSMHEq_007}}
                                                                                                                              \label{Eq_WEPFTCF_s_AppendixGF_ss_kappaneq0_sss_ULWGF_002a}\\
\dfrac{\partial G_-}
      {\partial y  }(y=-\tfrac{1}{2}L_y,Y;\kappa\neq0)=0
                                                                                                                              \label{Eq_WEPFTCF_s_AppendixGF_ss_kappaneq0_sss_ULWGF_002b}\\
\lim_{y\to+\infty}G_-(y,Y;\kappa\neq0) = 0 \quad\forall Y\in\mathbb{R}\setminus\{y\}
                                                                                                                              \label{Eq_WEPFTCF_s_AppendixGF_ss_kappaneq0_sss_ULWGF_002c}
\end{alignat}
\end{subequations}
to study the virtual influence of the lower wall.
By \eqref{Eq_WEPFTCF_s_AppendixGF_ss_GFSMHEq_003}, boundary-conditions \eqrefsab{Eq_WEPFTCF_s_AppendixGF_ss_kappaneq0_sss_ULWGF_001b}{Eq_WEPFTCF_s_AppendixGF_ss_kappaneq0_sss_ULWGF_002b} imply
$0$-gradient at the wall of the problem (upper for $G_+$ and lower for $G_-$). Boundary-conditions \eqrefsab{Eq_WEPFTCF_s_AppendixGF_ss_kappaneq0_sss_ULWGF_001c}{Eq_WEPFTCF_s_AppendixGF_ss_kappaneq0_sss_ULWGF_002c}
ensure boundedness at infinity for the halfspace corresponding to each problem. Straightforward computation of the functions $A_1(Y;\kappa)$ and $A_2(Y;\kappa)$ in \eqref{Eq_WEPFTCF_s_AppendixGF_ss_kappaneq0_002}
yields
\begin{subequations}
                                                                                                                              \label{Eq_WEPFTCF_s_AppendixGF_ss_kappaneq0_sss_ULWGF_003}
\begin{alignat}{6}
G_-(y,Y;\kappa\neq0)=-\dfrac{\mathrm{e}^{-\kappa\abs{y-Y}}}
                            {2\kappa                      }
                     -\dfrac{\mathrm{e}^{-\kappa(L_y+y+Y)}}
                           {2\kappa                       }\qquad y,Y\in[-\tfrac{1}{2}L_y,+\infty)
                                                                                                                              \label{Eq_WEPFTCF_s_AppendixGF_ss_kappaneq0_sss_ULWGF_003a}\\
G_+(y,Y;\kappa\neq0)=-\dfrac{\mathrm{e}^{-\kappa\abs{y-Y}}}
                            {2\kappa                      }
                     -\dfrac{\mathrm{e}^{-\kappa(L_y-y-Y)}}
                            {2\kappa                      }\qquad y,Y\in(-\infty,+\tfrac{1}{2}L_y]
                                                                                                                              \label{Eq_WEPFTCF_s_AppendixGF_ss_kappaneq0_sss_ULWGF_003b}
\end{alignat}
\end{subequations}
which by \eqref{Eq_WEPFTCF_s_AppendixGF_ss_kappaneq0_sss_FSGF_002} reads
\begin{subequations}
                                                                                                                              \label{Eq_WEPFTCF_s_AppendixGF_ss_kappaneq0_sss_ULWGF_004}
\begin{alignat}{6}
G_-(y,Y;\kappa\neq0)=G_\mathfrak{V}(y,Y;\kappa\neq0)
                     -\dfrac{\mathrm{e}^{-\kappa(L_y+y+Y)}}
                           {2\kappa                       }\qquad y,Y\in[-\tfrac{1}{2}L_y,+\infty)
                                                                                                                              \label{Eq_WEPFTCF_s_AppendixGF_ss_kappaneq0_sss_ULWGF_004a}\\
G_+(y,Y;\kappa\neq0)=G_\mathfrak{V}(y,Y;\kappa\neq0)
                     -\dfrac{\mathrm{e}^{-\kappa(L_y-y-Y)}}
                            {2\kappa                      }\qquad y,Y\in(-\infty,+\tfrac{1}{2}L_y]
                                                                                                                              \label{Eq_WEPFTCF_s_AppendixGF_ss_kappaneq0_sss_ULWGF_004b}
\end{alignat}
\end{subequations}
{\em ie} the presence of a single wall in any of the 2 halfspace problems induces an additive correction to the freespace Green's function,
the wall-echo $G_{w_\pm}(y,Y;\kappa\neq0)$
\begin{alignat}{6}
G_{w_\pm}(y,Y;\kappa\neq0):=G_\pm(y,Y;\kappa\neq0)-G_\mathfrak{V}(y,Y;\kappa\neq0)
                                                                                                                              \label{Eq_WEPFTCF_s_AppendixGF_ss_kappaneq0_sss_ULWGF_005}
\end{alignat}

%-----------------------------------------------------------------------
%
\subsubsection{Accommodating the boundary-conditions for the halfspace problems ($\kappa\neq0$)}\label{WEPFTCF_s_AppendixGF_ss_kappaneq0_sss_ABCsHP}
%
%-----------------------------------------------------------------------

To take into account normal gradients $d_y q(y=\pm\tfrac{1}{2}L_y;\kappa\neq0)=B_\pm(\kappa\neq0)\neq0$ at the wall for each of the halfspace problems of \S\ref{WEPFTCF_s_AppendixGF_ss_kappaneq0_sss_ULWGF}
we may proceed exactly as in \S\ref{WEPFTCF_s_AppendixGF_ss_kappaneq0_sss_ABCs}, by
superposing a boundary-condition $q_{\tsc{bc}_\pm}$ term to the corresponding solution \eqref{Eq_WEPFTCF_s_AppendixGF_ss_kappaneq0_sss_ULWGF_002} with $0$-gradient condition, {\em viz}
\begin{subequations}
                                                                                                                              \label{Eq_WEPFTCF_s_AppendixGF_ss_kappaneq0_sss_ABCsHP_001}
\begin{alignat}{6}
q(y;\kappa\neq0,B_-)=\int_{-\tfrac{L_y}{2}}^{+\tfrac{L_y}{2}} G_-(y,Y;\kappa\neq0)Q(Y;\kappa\neq0)dY+q_{\tsc{bc}_-}(y;\kappa\neq0,B_-)
                                                                                                                              \label{Eq_WEPFTCF_s_AppendixGF_ss_kappaneq0_sss_ABCsHP_001a}\\
\dfrac{d^2 q         (y;\kappa\neq0,B_-)}
      {dy^2                             } -\kappa^2 q         (y;\kappa\neq0,B_-) = Q(y;\kappa\neq0)\qquad y\in(-\tfrac{1}{2}L_y,+\infty)
                                                                                                                              \label{Eq_WEPFTCF_s_AppendixGF_ss_kappaneq0_sss_ABCsHP_001b}\\
\dfrac{d^2 q_{\tsc{bc}_-}(y;\kappa\neq0,B_-)}
      {dy^2                                 } -\kappa^2 q_{\tsc{bc}_-}(y;\kappa,B_-) = 0\qquad y\in(-\tfrac{1}{2}L_y,+\infty)
                                                                                                                              \label{Eq_WEPFTCF_s_AppendixGF_ss_kappaneq0_sss_ABCsHP_001c}\\
\dfrac{dq}
      {dy}(y=-\tfrac{1}{2}L_y;\kappa\neq0,B_-)\stackrel{\eqref{Eq_WEPFTCF_s_AppendixGF_ss_kappaneq0_sss_ULWGF_002b}}{=}
\dfrac{dq_{\tsc{bc}_-}}
      {dy             }(y=-\tfrac{1}{2}L_y;\kappa\neq0,B_-)=B_-(\kappa\neq0)
                                                                                                                              \label{Eq_WEPFTCF_s_AppendixGF_ss_kappaneq0_sss_ABCsHP_001d}\\
\lim_{y\to+\infty}q_{\tsc{bc}_-}(y;\kappa\neq0,B_-) = 0
                                                                                                                              \label{Eq_WEPFTCF_s_AppendixGF_ss_kappaneq0_sss_ABCsHP_001e}
\end{alignat}
\end{subequations}
to study the virtual influence of the lower wall boundary-condition on the solution, and
\begin{subequations}
                                                                                                                              \label{Eq_WEPFTCF_s_AppendixGF_ss_kappaneq0_sss_ABCsHP_002}
\begin{alignat}{6}
q(y;\kappa\neq0,B_+)=\int_{-\tfrac{L_y}{2}}^{+\tfrac{L_y}{2}} G_+(y,Y;\kappa\neq0)Q(Y;\kappa\neq0)dY+q_{\tsc{bc}_+}(y;\kappa\neq0,B_+)
                                                                                                                              \label{Eq_WEPFTCF_s_AppendixGF_ss_kappaneq0_sss_ABCsHP_002a}\\
\dfrac{d^2 q         (y;\kappa\neq0,B_+)}
      {dy^2                             } -\kappa^2 q         (y;\kappa\neq0,B_+) = Q(y;\kappa\neq0)\qquad y\in(-\infty,+\tfrac{1}{2}L_y)
                                                                                                                              \label{Eq_WEPFTCF_s_AppendixGF_ss_kappaneq0_sss_ABCsHP_002b}\\
\dfrac{d^2 q_{\tsc{bc}_+}(y;\kappa\neq0,B_+)}
      {dy^2                                 } -\kappa^2 q_{\tsc{bc}_+}(y;\kappa\neq0,B_+) = 0\qquad y\in(-\infty,+\tfrac{1}{2}L_y)
                                                                                                                              \label{Eq_WEPFTCF_s_AppendixGF_ss_kappaneq0_sss_ABCsHP_002c}\\
\dfrac{dq}
      {dy}(y=+\tfrac{1}{2}L_y;\kappa\neq0,B_+)\stackrel{\eqref{Eq_WEPFTCF_s_AppendixGF_ss_kappaneq0_sss_ULWGF_001b}}{=}
\dfrac{dq_{\tsc{bc}_+}}
      {dy             }(y=+\tfrac{1}{2}L_y;\kappa\neq0,B_+)=B_+(\kappa\neq0)
                                                                                                                              \label{Eq_WEPFTCF_s_AppendixGF_ss_kappaneq0_sss_ABCsHP_002d}\\
\lim_{y\to-\infty}q_{\tsc{bc}_+}(y;\kappa\neq0,B_+) = 0
                                                                                                                              \label{Eq_WEPFTCF_s_AppendixGF_ss_kappaneq0_sss_ABCsHP_002e}
\end{alignat}
\end{subequations}
to study the virtual influence of the upper wall boundary-condition on the solution.
By \eqref{Eq_WEPFTCF_s_AppendixGF_ss_kappaneq0_sss_ULWGF_002}, the superposition \eqref{Eq_WEPFTCF_s_AppendixGF_ss_kappaneq0_sss_ABCsHP_001a} is the solution to \eqref{Eq_WEPFTCF_s_AppendixGF_ss_kappaneq0_sss_ABCsHP_001b},
with boundary-conditions \eqrefsab{Eq_WEPFTCF_s_AppendixGF_ss_kappaneq0_sss_ABCsHP_001d}{Eq_WEPFTCF_s_AppendixGF_ss_kappaneq0_sss_ABCsHP_001e}.
By \eqref{Eq_WEPFTCF_s_AppendixGF_ss_kappaneq0_sss_ULWGF_001}, the superposition \eqref{Eq_WEPFTCF_s_AppendixGF_ss_kappaneq0_sss_ABCsHP_002a} is the solution to \eqref{Eq_WEPFTCF_s_AppendixGF_ss_kappaneq0_sss_ABCsHP_002b},
with boundary-conditions \eqrefsab{Eq_WEPFTCF_s_AppendixGF_ss_kappaneq0_sss_ABCsHP_002d}{Eq_WEPFTCF_s_AppendixGF_ss_kappaneq0_sss_ABCsHP_002e}.
Straightforward computation yields
\begin{alignat}{6}
q_{\tsc{bc}_\pm}(y;\kappa\neq0,B_\pm)=\pm\dfrac{B_\pm }
                                               {\kappa}{\rm e}^{-\kappa(\tfrac{1}{2}L_y\mp y)}
                                                                                                                              \label{Eq_WEPFTCF_s_AppendixGF_ss_kappaneq0_sss_ABCs_003}
\end{alignat}

%-----------------------------------------------------------------------------------------------------------------------------------
%
%
%
%
%
\subsection{$\kappa=0$ ($xz$-constant component)}\label{WEPFTCF_s_AppendixGF_ss_kappaeq0}
%
%
%
%
%
%-----------------------------------------------------------------------------------------------------------------------------------

When $\kappa=0$ \eqref{Eq_WEPFTCF_s_AppendixGF_ss_GFSMHEq_001} becomes a Poisson equation~\citep{Katz_Plotkin_1991a_Poisson_eq_freespace_Greenfuncion}
\begin{alignat}{6}
\dfrac{d^2 q(y;\kappa=0)}
      {dy^2             }                       = Q(y;\kappa=0)\qquad y\in(-\tfrac{1}{2}L_y,\tfrac{1}{2}L_y)
                                                                                                                              \label{Eq_WEPFTCF_s_AppendixGF_ss_kappaeq0_001}
\end{alignat}
The limits as $\kappa\to0$ of the solutions obtained in \S\ref{WEPFTCF_s_AppendixGF_ss_kappaneq0} for $\kappa\neq0$ are all singular,
as can be easily verified by straightforward computation, contrary to the case
of the same problem with Dirichlet boundary-conditions~\citep{Zauderer_2006a_1D_Greenfunction_and_compatibility_relation}.
This is related to the existence of solutions of \eqref{Eq_WEPFTCF_s_AppendixGF_ss_kappaeq0_001} with Neumann boundary-conditions at $y=\pm\tfrac{1}{2}L_y$,
because integrating \eqref{Eq_WEPFTCF_s_AppendixGF_ss_kappaeq0_001} yields
\begin{alignat}{6}
\dfrac{d q}
      {dy }(y=+\tfrac{1}{2}L_y;\kappa=0)-
\dfrac{d q}
      {dy }(y=-\tfrac{1}{2}L_y;\kappa=0)                       = \int_{-\frac{1}{2}L_y}^{+\frac{1}{2}L_y}Q(Y;\kappa=0)\;dY
                                                                                                                              \label{Eq_WEPFTCF_s_AppendixGF_ss_kappaeq0_002}
\end{alignat}
implying that \eqref{Eq_WEPFTCF_s_AppendixGF_ss_kappaeq0_001} with gradient boundary-conditions $q'(y=\pm\tfrac{1}{2}L_y;\kappa=0)=B_\pm(\kappa=0)$ can only be solved iff the compatibility condition \eqref{Eq_WEPFTCF_s_AppendixGF_ss_kappaeq0_002}
is satisfied by the integral of the source-term and the boundary-conditions.

In the particular case of homogeneous Neumann boundary-conditions $B_\pm(\kappa=0)=0$,
which is the most important since $B_\pm(\kappa=0)\neq0$ is only concerned with Stokes pressure \eqrefsab{Eq_WEPFTCF_s_DNSCpS_ss_GFSPEq_sss_ODEsFT_002}{Eq_WEPFTCF_s_DNSCpS_ss_GFSPEq_sss_ODEsFT_005},
the homogeneous equivalent of equation \eqref{Eq_WEPFTCF_s_AppendixGF_ss_kappaeq0_001}, $q''(y;\kappa=0)=0$,
admits any constant function as solution, {\em ie} any constant function is an eigenfunction~\citep{Ince_1926a_1D_Greenfunction_and_compatibility_relation,
                                                                                                    Courant_Hilbert_1953a_1D_Greenfunction,
                                                                                                    MyintU_Debnath_2007a_1D_Greenfunction}
of the homogeneous equivalent of \eqref{Eq_WEPFTCF_s_AppendixGF_ss_kappaeq0_001}, $q''(y;\kappa=0)=0$, with boundary-conditions $q'(y=\pm\tfrac{1}{2}L_y;\kappa=0)=0$,
and $\kappa=0$ is a characteristic number of the homogeneous equivalent of \eqref{Eq_WEPFTCF_s_AppendixGF_ss_GFSMHEq_001}, $q''(y;\kappa)-\kappa^2q(y;\kappa)=0$ with boundary-conditions $q'(y=\pm\tfrac{1}{2}L_y;\kappa)=0$.
Therefore a solution only exists provided the corresponding compatibility relation~\citep{Ince_1926a_1D_Greenfunction_and_compatibility_relation,
                                                                                          Courant_Hilbert_1953a_1D_Greenfunction}
holds, {\em ie} the integral in \eqref{Eq_WEPFTCF_s_AppendixGF_ss_kappaeq0_002} is equal to $0$.
This is analogous to the 3-D compatibility relation for the Poisson equation with Neumann boundary-conditions~\citep{Ockendon_Howison_Lacey_Movchan_2003a_Greenfuncion,
                                                                                                                     Zauderer_2006a_1D_Greenfunction_and_compatibility_relation}.
%-----------------------------------------------------------------------
%
\subsubsection{$G_{\rm Kim}(y,Y;\kappa=0)$}\label{WEPFTCF_s_AppendixGF_ss_kappaeq0_sss_Kim}
%
%-----------------------------------------------------------------------

In this case \eqrefsab{Eq_WEPFTCF_s_DNSCpS_ss_GFSPEq_sss_ODEsFT_002b}{Eq_WEPFTCF_s_DNSCpS_ss_GFSPEq_sss_ODEsFT_005b} $0$-gradient boundary-conditions apply on both walls ($y=\pm\tfrac{1}{2}L_y$).
Assuming that the compatibility relation
\begin{alignat}{6}
\int_{-\frac{1}{2}L_y}^{+\frac{1}{2}L_y}Q(Y;\kappa=0)\;dY=0
                                                                                                                              \label{Eq_WEPFTCF_s_AppendixGF_ss_kappaeq0_sss_Kim_001}
\end{alignat}
holds, there are 2 different ways for constructing a modified Green's function~\citep{Courant_Hilbert_1953a_1D_Greenfunction,
                                                                                      Zauderer_2006a_1D_Greenfunction_and_compatibility_relation}
for solving \eqref{Eq_WEPFTCF_s_AppendixGF_ss_kappaeq0_001} with homogeneous Neumann boundary-conditions.
The first method described in~\cite{Courant_Hilbert_1953a_1D_Greenfunction} makes explicit
use of the eigensolution of \eqref{Eq_WEPFTCF_s_AppendixGF_ss_kappaeq0_001} in a modified definition of the problem \eqref{Eq_WEPFTCF_s_AppendixGF_ss_GFSMHEq_002} for calculating
the modified Green's function which satisfies the boundary-conditions, but has also to modify \eqref{Eq_WEPFTCF_s_AppendixGF_ss_GFSMHEq_003}. The second method~\citep{Zauderer_2006a_1D_Greenfunction_and_compatibility_relation}
which leads to the Green's function used by~\cite{Kim_1989a} uses the standard definition of the Green's function \eqref{Eq_WEPFTCF_s_AppendixGF_ss_GFSMHEq_002} along with the standard integral solution \eqref{Eq_WEPFTCF_s_AppendixGF_ss_GFSMHEq_003}.
In this latter case however the boundary-conditions are not explicitly included in the solution but are enforced by the compatibility relation \eqref{Eq_WEPFTCF_s_AppendixGF_ss_kappaeq0_sss_Kim_001}.

Hence, assuming that the compatibility relation \eqref{Eq_WEPFTCF_s_AppendixGF_ss_kappaeq0_sss_Kim_001} holds,
so that \eqref{Eq_WEPFTCF_s_AppendixGF_ss_kappaneq0_sss_Kim_002a} for $\kappa=0$ may be  satisfied, the Green's function corresponding to \eqref{Eq_WEPFTCF_s_AppendixGF_ss_kappaeq0_001}
satisfies
\begin{subequations}
                                                                                                                              \label{Eq_WEPFTCF_s_AppendixGF_ss_kappaeq0_sss_Kim_002}
\begin{alignat}{6}
\dfrac{\partial^2 G_{\rm Kim}(y,Y;\kappa=0)}
      {\partial y^2                        }                         =\delta(y-Y)
                                                                                                                              \label{Eq_WEPFTCF_s_AppendixGF_ss_kappaeq0_sss_Kim_002a}\\
\lim_{\epsilon\to0^+}\left[\dfrac{\partial G_{\rm Kim}}
                                             {\partial y          }(Y+\epsilon,Y;\kappa=0)
                                      -\dfrac{\partial G_{\rm Kim}}
                                             {\partial y          }(Y-\epsilon,Y;\kappa=0)\right]=1
                                                                                                                              \label{Eq_WEPFTCF_s_AppendixGF_ss_kappaeq0_sss_Kim_002b}\\
G_{\rm Kim}(y,Y;\kappa=0)\in C^0(-\tfrac{1}{2}L_y,+\tfrac{1}{2}L_y)
                                                                                                                              \label{Eq_WEPFTCF_s_AppendixGF_ss_kappaeq0_sss_Kim_002c}\\
G_{\rm Kim}(y,Y;\kappa)\in C^1(-\tfrac{1}{2}L_y,+\tfrac{1}{2}L_y)\setminus\{Y\}
                                                                                                                              \label{Eq_WEPFTCF_s_AppendixGF_ss_kappaeq0_sss_Kim_002d}
\end{alignat}
\end{subequations}
It is therefore made up by 2 straight lines joined together at $Y$, since by \eqref{Eq_WEPFTCF_s_AppendixGF_ss_kappaeq0_sss_Kim_002a} $\partial^2_{yy}G(y,Y;\kappa=0)=0\;\forall y\neq Y$.
If we further admit that $G_{\rm Kim}(y,Y;\kappa=0)=G_{\rm Kim}(Y,y;\kappa=0)$ \eqref{Eq_WEPFTCF_s_AppendixGF_ss_GFSMHEq_004}, as expected because of the self-adjointedness of the
the operator $[d^2_{yy}(\cdot)]$~\citep{Ince_1926a_1D_Greenfunction_and_compatibility_relation,
                                         Courant_Hilbert_1953a_1D_Greenfunction},
we must have~\citep{Katz_Plotkin_1991a_Poisson_eq_freespace_Greenfuncion} $G_{\rm Kim}(y,Y;\kappa=0)=G_{\rm Kim}(|y-Y|;\kappa=0)$ so that
\begin{align}
G_{\rm Kim}(y,Y;\kappa=0)=\dfrac{|y-Y|}{2}+c
                                                                                                                 \label{Eq_WEPFTCF_s_AppendixGF_ss_kappaeq0_sss_Kim_003}
\end{align}
The factor $\tfrac{1}{2}$ is required to satisfy \eqref{Eq_WEPFTCF_s_AppendixGF_ss_kappaeq0_sss_Kim_002b}, and $c\in\mathbb{R}$ is a constant. The value of $c$ has no influence whatsoever
on the solution \eqref{Eq_WEPFTCF_s_AppendixGF_ss_GFSMHEq_003} because of the compatibility condition
\eqref{Eq_WEPFTCF_s_AppendixGF_ss_kappaeq0_sss_Kim_001}.\footnote{\label{ff__WEPFTCF_s_AppendixGF_ss_kappaeq0_001}
                                                                   $\displaystyle\int_{-\frac{1}{2}L_y}^{+\frac{1}{2}L_y}\left(\dfrac{|y-Y|}{2}+c\right)Q(Y;\kappa=0)\;dY=
                                                                                 \int_{-\frac{1}{2}L_y}^{+\frac{1}{2}L_y}\dfrac{|y-Y|}{2}\;Q(Y;\kappa=0)\;dY+c\int_{-\frac{1}{2}L_y}^{+\frac{1}{2}L_y}Q(Y;\kappa=0)\;dY$\\
                                                                   $\stackrel{\eqref{Eq_WEPFTCF_s_AppendixGF_ss_kappaeq0_sss_Kim_001}}{=}
                                                                    \displaystyle\int_{-\frac{1}{2}L_y}^{+\frac{1}{2}L_y}\dfrac{|y-Y|}{2}\;Q(Y;\kappa=0)\;dY$
                                                                   }
The simplest choice $c=0$ corresponding to the reasonable situation $G_{\rm Kim}(y,y;\kappa=0)=0$ is made
\begin{align}
G_{\rm Kim}(y,Y;\kappa=0)=\dfrac{|y-Y|}{2}
                                                                                                                 \label{Eq_WEPFTCF_s_AppendixGF_ss_kappaeq0_sss_Kim_004}
\end{align}
and we have
\begin{align}
\dfrac{d q_{\rm Kim}}
      {dy           }(y=\pm\tfrac{1}{2}L_y;\kappa=0)\stackrel{\eqref{Eq_WEPFTCF_s_AppendixGF_ss_GFSMHEq_003}}{=}            &\int_{-\frac{1}{2}L_y}^{+\frac{1}{2}L_y}\dfrac{\partial G_{\rm Kim}}
                                                                                                                                                                           {\partial y          }(y,Y;\kappa=0)\;Q(Y;\kappa=0)\;dY
                                                                                                                 \notag\\
                                                  \stackrel{\eqref{Eq_WEPFTCF_s_AppendixGF_ss_kappaeq0_sss_Kim_004}}{=}&\pm\tfrac{1}{2}\int_{-\frac{1}{2}L_y}^{+\frac{1}{2}L_y}Q(Y;\kappa=0)\;dY
                                                  \stackrel{\eqref{Eq_WEPFTCF_s_AppendixGF_ss_kappaeq0_sss_Kim_001}}{=} 0
                                                                                                                 \label{Eq_WEPFTCF_s_AppendixGF_ss_kappaeq0_sss_Kim_005}
\end{align}
Notice that in this case it is the compatibility relation \eqref{Eq_WEPFTCF_s_AppendixGF_ss_kappaeq0_sss_Kim_001},
and not the Green's function alone, which is responsible for satisfying the boundary-conditions.

%-----------------------------------------------------------------------
%
\subsubsection{Accommodating the boundary-conditions ($\kappa=0$)}\label{WEPFTCF_s_AppendixGF_ss_kappaeq0_sss_ABCs}
%
%-----------------------------------------------------------------------

Finally, the solution of the boundary-conditions problem (required for $p_{(\tau)}'$) can be obtained directly by integrating
\begin{subequations}
                                                                                                                              \label{Eq_WEPFTCF_s_AppendixGF_ss_kappaeq0_sss_ABCs_001}
\begin{alignat}{6}
\dfrac{d^2 q_\tsc{bc}(y;\kappa=0,B_\pm)}
      {dy^2                            }                       = 0\qquad y\in(-\tfrac{1}{2}L_y,+\tfrac{1}{2}L_y)
                                                                                                                              \label{Eq_WEPFTCF_s_AppendixGF_ss_kappaeq0_sss_ABCs_001a}\\
\dfrac{d q_\tsc{bc}}
      {dy          }(y=\pm\tfrac{1}{2}L_y;\kappa=0,B_\pm) = B_\pm(\kappa=0)
                                                                                                                              \label{Eq_WEPFTCF_s_AppendixGF_ss_kappaeq0_sss_ABCs_001b}
\end{alignat}
\end{subequations}
The solution exists iff
\begin{alignat}{6}
B_-(\kappa=0)=B_+(\kappa=0)
                                                                                                                              \label{Eq_WEPFTCF_s_AppendixGF_ss_kappaeq0_sss_ABCs_002}
\end{alignat}
and is obviously a straight line, defined up to an additive constant
\begin{alignat}{6}
q_\tsc{bc}(y;\kappa=0,B_\pm)=\tfrac{1}{2}B_- y + \tfrac{1}{2}B_+ y +q_{\tsc{bc}_0}
                                                                                                                              \label{Eq_WEPFTCF_s_AppendixGF_ss_kappaeq0_sss_ABCs_003}
\end{alignat}
where by \eqref{Eq_WEPFTCF_s_AppendixGF_ss_kappaeq0_sss_ABCs_003} $q_\tsc{bc}(y=0;\kappa=0)=q_{\tsc{bc}_0}$ is the solution at centerline.

%-----------------------------------------------------------------------
%
\subsubsection{Freespace Green's function $G_\mathfrak{V}(y,Y;\kappa=0)$}\label{WEPFTCF_s_AppendixGF_ss_kappaeq0_sss_FSGF}
%
%-----------------------------------------------------------------------

The solution for the freespace Green's function $G_\mathfrak{V}(y,Y;\kappa=0)$ is obtained exactly in the same way as for $G_{\rm Kim}(y,Y;\kappa=0)$ (\S\ref{WEPFTCF_s_AppendixGF_ss_kappaeq0_sss_Kim}),
yielding by \eqref{Eq_WEPFTCF_s_AppendixGF_ss_kappaeq0_sss_Kim_004}
\begin{align}
G_\mathfrak{V}(y,Y;\kappa=0)=G_\mathrm{Kim}(y,Y;\kappa=0)=\dfrac{|y-Y|}{2}
                                                                                                                 \label{Eq_WEPFTCF_s_AppendixGF_ss_kappaeq0_sss_FSGF_001}
\end{align}
Again, the sources $Q(y;\kappa=0)$ must satisfy the compatibility condition \eqref{Eq_WEPFTCF_s_AppendixGF_ss_kappaeq0_sss_Kim_001} to ensure boundedness as $\abs{y}\to\infty$
\begin{alignat}{6}
&\lim_{\abs{y}\to\infty}\int_{-\frac{1}{2}L_y}^{+\frac{1}{2}L_y} G_\mathfrak{V}(y,Y\kappa=0)\;Q(Y;\kappa=0)\;dY
\stackrel{\eqref{Eq_WEPFTCF_s_AppendixGF_ss_kappaeq0_sss_FSGF_001}}{=}
                                                                                                                 \notag\\
&\lim_{\abs{y}\to\infty}\left\{\left(\int_{-\frac{1}{2}L_y}^{+\frac{1}{2}L_y}Q(Y;\kappa=0)\;dY\right)\dfrac{|y|}{2}\right\}
\stackrel{\eqref{Eq_WEPFTCF_s_AppendixGF_ss_kappaeq0_sss_Kim_001}}{=}0 \quad\forall Y\in\mathbb{R}\setminus\{y\}
                                                                                                                 \label{Eq_WEPFTCF_s_AppendixGF_ss_kappaeq0_sss_FSGF_002}
\end{alignat}
which is $0$ iff \eqref{Eq_WEPFTCF_s_AppendixGF_ss_kappaeq0_sss_Kim_001} holds.

%-----------------------------------------------------------------------
%
\subsubsection{Upper/lower wall Green's functions $G_\pm(y,Y;\kappa=0)$}\label{WEPFTCF_s_AppendixGF_ss_kappaeq0_sss_ULWGF}
%
%-----------------------------------------------------------------------

The solution for $G_\pm(y,Y;\kappa=0)$ is again obtained in exactly the same way as for $G_{\rm Kim}(y,Y;\kappa=0)$ (\S\ref{WEPFTCF_s_AppendixGF_ss_kappaeq0_sss_Kim}),
yielding by \eqref{Eq_WEPFTCF_s_AppendixGF_ss_kappaeq0_sss_FSGF_001}
\begin{align}
G_\pm(y,Y;\kappa=0)=G_\mathfrak{V}(y,Y;\kappa=0)=G_\mathrm{Kim}(y,Y;\kappa=0)=\dfrac{|y-Y|}{2}
                                                                                                                 \label{Eq_WEPFTCF_s_AppendixGF_ss_kappaeq0_sss_ULWGF_001}
\end{align}
The boundary-conditions for the gradient at the wall ($y=\mp\tfrac{1}{2}L_y$), and for boundedness at $\pm\infty$,
are again satisfied by relations analogous to \eqrefsab{Eq_WEPFTCF_s_AppendixGF_ss_kappaeq0_sss_Kim_005}{Eq_WEPFTCF_s_AppendixGF_ss_kappaeq0_sss_FSGF_002},
iff the compatibility relation \eqref{Eq_WEPFTCF_s_AppendixGF_ss_kappaeq0_sss_Kim_001} holds.

%-----------------------------------------------------------------------
%
\subsubsection{Accommodating the boundary-conditions for the halfspace problems ($\kappa=0$)}\label{WEPFTCF_s_AppendixGF_ss_kappaeq0_sss_ABCsHP}
%
%-----------------------------------------------------------------------

Finally, the solution of the boundary-conditions (required for $p_{(\tau)}'$), in the case of the halfspace problems, can be obtained directly by integrating
\begin{subequations}
                                                                                                                              \label{Eq_WEPFTCF_s_AppendixGF_ss_kappaeq0_sss_ABCsHP_001}
\begin{alignat}{6}
\dfrac{d^2 q_{\tsc{bc}_\pm}(y;\kappa=0,B_\pm)}
      {dy^2                                  }                       = 0\qquad \mp y\in(-\tfrac{1}{2}L_y,+\infty)
                                                                                                                              \label{Eq_WEPFTCF_s_AppendixGF_ss_kappaeq0_sss_ABCsHP_001a}\\
\dfrac{d q_{\tsc{bc}_\pm}}
      {dy                }(y=\pm\tfrac{1}{2}L_y;\kappa=0,B_\pm) = B_\pm(\kappa=0)
                                                                                                                              \label{Eq_WEPFTCF_s_AppendixGF_ss_kappaeq0_sss_ABCsHP_001b}
\end{alignat}
\end{subequations}
The solution is obviously a straight line of slope $B_\pm(\kappa=0)$, and is bounded as $\abs{y}\to\infty$ iff the compatibility condition
\begin{alignat}{6}
B_\pm(\kappa=0)=0\Longrightarrow q_{\tsc{bc}_\pm}(y;\kappa=0,B_\pm)=0
                                                                                                                              \label{Eq_WEPFTCF_s_AppendixGF_ss_kappaeq0_sss_ABCsHP_002}
\end{alignat}
holds.

%-----------------------------------------------------------------------------------------------------------------------------------
%
%
%
%
%
\subsection{Approximation error in the method of images}\label{WEPFTCF_s_AppendixGF_ss_AEMI}
%
%
%
%
%
%-----------------------------------------------------------------------------------------------------------------------------------

One interesting application of the halfspace problems is that, by superposition, they give the solution obtained by the method of images~\citep{Manceau_Wang_Laurence_2001a}.
Indeed the method of images corresponds to applying the freespace Green's function on the domain $y\in(-\tfrac{1}{2}L_y,+\tfrac{1}{2}L_y)$, and on 2 ghost domains
$y\in(-\tfrac{3}{2}L_y,-\tfrac{1}{2}L_y)\cup(+\tfrac{1}{2}L_y,+\tfrac{3}{2}L_y)$, with appropriately reflected source-terms \figref{Fig_WEPFTCF_s_DNSCpS_ss_GFSPEq_sss_ODEsFT_001}
\begin{subequations}
                                                                                                                              \label{Eq_WEPFTCF_s_AppendixGF_ss_AEMI_001}
\begin{alignat}{6}
Q_\tsc{mwl}(y;\kappa)&=& Q(-L_y-y;\kappa)&\qquad&y\in(-\tfrac{3}{2}L_y,-\tfrac{1}{2}L_y)
                                                                                                                              \label{Eq_WEPFTCF_s_AppendixGF_ss_AEMI_001a}\\
Q_\tsc{mwl}(y;\kappa)&=& Q(     y;\kappa)&\qquad&y\in(-\tfrac{1}{2}L_y,+\tfrac{1}{2}L_y)
                                                                                                                              \label{Eq_WEPFTCF_s_AppendixGF_ss_AEMI_001b}\\
Q_\tsc{mwl}(y;\kappa)&=& Q(+L_y-y;\kappa)&\qquad&y\in(+\tfrac{1}{2}L_y,+\tfrac{3}{2}L_y)
                                                                                                                              \label{Eq_WEPFTCF_s_AppendixGF_ss_AEMI_001c}
\end{alignat}
\end{subequations}
and the approximation to the solution obtained by the method of images used by~\cite{Manceau_Wang_Laurence_2001a}\footnote{\label{ff_WEPFTCF_s_AppendixGF_ss_AEMI_001}
                                                                                                                           Although \cite{Manceau_Wang_Laurence_2001a} neglect $q_\tsc{bc}(y;\kappa,B_\pm)$,
                                                                                                                           this can be added to $q_\tsc{mwl}(y;\kappa)$ \eqref{Eq_WEPFTCF_s_AppendixGF_ss_AEMI_002}
                                                                                                                           to obtain the approximation of the complete solution \eqref{Eq_WEPFTCF_s_AppendixGF_ss_kappaneq0_sss_ABCs_001a}.
                                                                                                                          }
is
\begin{alignat}{6}
q_\tsc{mwl}(y;\kappa):=\int_{-\frac{3}{2}L_y}^{+\frac{3}{2}L_y}\;G_\mathfrak{V}(y,Y;\kappa)\;Q_\tsc{mwl}(Y;\kappa)\;dY
                                                                                                                              \label{Eq_WEPFTCF_s_AppendixGF_ss_AEMI_002}
\end{alignat}
It is straightforward to construct a kernel $G_\tsc{mwl}(y,Y;\kappa)$ which applied on the source-terms in the actual computational domain $y\in(-\tfrac{1}{2}L_y,-\tfrac{1}{2}L_y)$
will return the method of images approximation \eqref{Eq_WEPFTCF_s_AppendixGF_ss_AEMI_002}, because, defining $Y_\pm:=\pm L_y-Y$, we have
\begin{alignat}{6}
q_\tsc{mwl}(y;\kappa)                                                                                    = &\int_{-\frac{3}{2}L_y}^{+\frac{3}{2}L_y}\;G_\mathfrak{V}(y,Y;\kappa)\;Q_\tsc{mwl}(Y;\kappa)\;dY
                                                                                                                              \notag\\
                                             \stackrel{\eqrefsab{Eq_WEPFTCF_s_AppendixGF_ss_AEMI_001}
                                                                {Eq_WEPFTCF_s_AppendixGF_ss_AEMI_002}}{=}&\int_{-\frac{3}{2}L_y}^{-\frac{1}{2}L_y}\;G_\mathfrak{V}(y,Y;\kappa)\;Q(-L_y-Y;\kappa)\;dY
                                                                                                                              \notag\\
                                                                                                         + &\int_{-\frac{1}{2}L_y}^{+\frac{1}{2}L_y}\;G_\mathfrak{V}(y,Y;\kappa)\;Q(Y;\kappa)\;dY
                                                                                                                              \notag\\
                                                                                                         + &\int_{+\frac{1}{2}L_y}^{+\frac{3}{2}L_y}\;G_\mathfrak{V}(y,Y;\kappa)\;Q(+L_y-Y;\kappa)\;dY
                                                                                                                              \notag\\
                                                                                                         =-&\int_{+\frac{1}{2}L_y}^{-\frac{1}{2}L_y}\;G_\mathfrak{V}(y,-L_y-Y_-;\kappa)\;Q(Y_-;\kappa)\;dY_-
                                                                                                                              \notag\\
                                                                                                          +&\int_{-\frac{1}{2}L_y}^{+\frac{1}{2}L_y}\;G_\mathfrak{V}(y,Y;\kappa)\;Q(Y;\kappa)\;dY
                                                                                                                              \notag\\
                                                                                                          -&\int_{+\frac{1}{2}L_y}^{-\frac{1}{2}L_y}\;G_\mathfrak{V}(y,L_y-Y_+;\kappa)\;Q(Y_+;\kappa)\;dY_+
                                                                                                                              \label{Eq_WEPFTCF_s_AppendixGF_ss_AEMI_003}
\end{alignat}
yielding
\begin{subequations}
                                                                                                                              \label{Eq_WEPFTCF_s_AppendixGF_ss_AEMI_004}
\begin{alignat}{6}
q_\tsc{mwl}(y;\kappa)   =&\int_{-\frac{1}{2}L_y}^{+\frac{1}{2}L_y}\;G_\tsc{mwl}(y,Y;\kappa)\;Q(Y;\kappa)\;dY
                                                                                                                              \label{Eq_WEPFTCF_s_AppendixGF_ss_AEMI_004a}\\
G_\tsc{mwl}(y,Y;\kappa):=&G_\mathfrak{V}(y,Y;\kappa)+G_\mathfrak{V}(y,-L_y-Y;\kappa)+G_\mathfrak{V}(y,L_y-Y;\kappa)
                                                                                                                              \label{Eq_WEPFTCF_s_AppendixGF_ss_AEMI_004b}
\end{alignat}
and by \eqref{Eq_WEPFTCF_s_AppendixGF_ss_kappaneq0_sss_FSGF_002}
\begin{alignat}{6}
G_\tsc{mwl}(y,Y;\kappa\neq0)\stackrel{\eqref{Eq_WEPFTCF_s_AppendixGF_ss_kappaneq0_sss_FSGF_002}}{=}&
                          -\dfrac{\mathrm{e}^{-\kappa|y-Y|}}{2\kappa}
                          -\dfrac{\mathrm{e}^{-\kappa\abs{L_y+y+Y}}}
                                 {2\kappa                         }
                          -\dfrac{\mathrm{e}^{-\kappa\abs{L_y-y-Y}}}
                                 {2\kappa                         }
                                                                                                                              \label{Eq_WEPFTCF_s_AppendixGF_ss_AEMI_004c}
\end{alignat}
\end{subequations}
Notice that \eqref{Eq_WEPFTCF_s_AppendixGF_ss_AEMI_004} gives\footnote{\label{ff_WEPFTCF_s_AppendixGF_ss_AEMI_002}
                                                                       \eqref{Eq_WEPFTCF_s_AppendixGF_ss_AEMI_004} is the 1-D equivalent of \citep[(4.2), p. 314]{Manceau_Wang_Laurence_2001a}
                                                                      }
the solution by integrating the sources in the actual domain $Y\in[-\tfrac{1}{2}L_y,+\tfrac{1}{2}L_y]$ with
an approximate Green's functions $G_\tsc{mwl}(y,Y;\kappa)$ \eqref{Eq_WEPFTCF_s_AppendixGF_ss_AEMI_004b}. This representation \eqref{Eq_WEPFTCF_s_AppendixGF_ss_AEMI_004}
readily yields the approximation error of the method of images, by comparison with the exact solution \eqref{Eq_WEPFTCF_s_AppendixGF_ss_kappaneq0_sss_Kim_002b} obtained using $G_\mathrm{Kim}(y,Y;\kappa)$.
If we consider the actual domain $y,Y\in[-\tfrac{1}{2}L_y,+\tfrac{1}{2}L_y]$, we have
\begin{subequations}
                                                                                                                              \label{Eq_WEPFTCF_s_AppendixGF_ss_AEMI_005}
\begin{alignat}{6}
\left. \begin{array}{c}-\tfrac{1}{2}L_y\leq y\leq+\tfrac{1}{2}L_y\\
                       -\tfrac{1}{2}L_y\leq Y\leq+\tfrac{1}{2}L_y\\\end{array}\right\}\Longrightarrow \left\{\begin{array}{c}L_y+(y+Y)\geq0\\
                                                                                                                             L_y-(y+Y)\geq0\\\end{array}\right.
                                                                                                                              \label{Eq_WEPFTCF_s_AppendixGF_ss_AEMI_005a}
\end{alignat}
so that
\begin{alignat}{6}
G_\tsc{mwl}(y,Y;\kappa\neq0)\stackrel{\eqrefsab{Eq_WEPFTCF_s_AppendixGF_ss_AEMI_004c}
                                               {Eq_WEPFTCF_s_AppendixGF_ss_AEMI_005}}{=}& -\dfrac{\mathrm{e}^{-\kappa|y-Y|}}{2\kappa}
                                                                                                                              \notag\\
                                                                                      & -\dfrac{\mathrm{e}^{-\kappa(L_y+y+Y})}
                                                                                               {2\kappa                      }
                                                                                                                              \notag\\
                                                                                      & -\dfrac{\mathrm{e}^{-\kappa(L_y-y-Y})}
                                                                                               {2\kappa                      }\qquad\forall y,Y\in[-\tfrac{1}{2}L_y,+\tfrac{1}{2}L_y]
                                                                                                                              \label{Eq_WEPFTCF_s_AppendixGF_ss_AEMI_005b}\\
                            \stackrel{\eqrefsab{Eq_WEPFTCF_s_AppendixGF_ss_kappaneq0_sss_FSGF_002}
                                               {Eq_WEPFTCF_s_AppendixGF_ss_kappaneq0_sss_ULWGF_004}}{=}& G_\mathfrak{V}(y,Y;\kappa\neq0)
                                                                                                                              \notag\\
                                                                                                   + & \underbrace{G_-           (y,Y;\kappa\neq0)
                                                                                                                  -G_\mathfrak{V}(y,Y;\kappa\neq0)}_{\displaystyle G_{w_-}(y,Y;\kappa\neq0)\;
                                                                                                                                                     \eqref{Eq_WEPFTCF_s_AppendixGF_ss_kappaneq0_sss_ULWGF_005}}
                                                                                                                              \notag\\
                                                                                                   + & \underbrace{G_+           (y,Y;\kappa\neq0)
                                                                                                                  -G_\mathfrak{V}(y,Y;\kappa\neq0)}_{\displaystyle G_{w_+}(y,Y;\kappa\neq0)\;
                                                                                                                                                     \eqref{Eq_WEPFTCF_s_AppendixGF_ss_kappaneq0_sss_ULWGF_005}}
                                                                                                                              \notag\\
                                                                                                     &                        \qquad\forall y,Y\in[-\tfrac{1}{2}L_y,+\tfrac{1}{2}L_y]
                                                                                                                              \label{Eq_WEPFTCF_s_AppendixGF_ss_AEMI_005c}
\end{alignat}
\end{subequations}
{\em ie}, the images problem~\citep{Manceau_Wang_Laurence_2001a} is equivalent to adding to the freespace Green's function an independent wall-echo correction for each wall,
the wall-echo correction for the upper (lower) wall being exactly the wall-echo of the halfspace problem with only the upper (lower)
 wall present (\S\ref{WEPFTCF_s_AppendixGF_ss_kappaneq0_sss_ULWGF}).
Notice that by \eqref{Eq_WEPFTCF_s_AppendixGF_ss_kappaeq0_sss_ULWGF_001} $G_{w_-}(y,Y;\kappa=0)=G_{w_+}(y,Y;\kappa=0)=0$ so that \eqref{Eq_WEPFTCF_s_AppendixGF_ss_AEMI_005c} holds $\forall\kappa\in\mathbb{R}_{\geq0}$.

By \eqrefsab{Eq_WEPFTCF_s_AppendixGF_ss_kappaneq0_sss_Kim_002b}{Eq_WEPFTCF_s_AppendixGF_ss_AEMI_005b} we may directly compute the error made when approximating the Green's function by the method of images
\begin{alignat}{6}
\dfrac{1  }
      {L_y}\Big(G_\tsc{mwl}(y,Y;\kappa\neq0)-G_\mathrm{Kim}(y,Y;\kappa\neq0)\Big)
\stackrel{\eqrefsab{Eq_WEPFTCF_s_AppendixGF_ss_kappaneq0_sss_Kim_002b}
                   {Eq_WEPFTCF_s_AppendixGF_ss_AEMI_005b}}{=}
                                                                                                                              \notag\\
\dfrac{\mathrm{e}^{-\kappa L_y}  }
      {2\kappa L_y\sinh\kappa L_y}\Big(\cosh\kappa\abs{y-Y}+\mathrm{e}^{-\kappa L_y}\cosh\kappa(y+Y)\Big)
                                                                                                                              \label{Eq_WEPFTCF_s_AppendixGF_ss_AEMI_006}\\
\qquad\qquad\qquad\qquad\forall y,Y\in[-\tfrac{1}{2}L_y,+\tfrac{1}{2}L_y]
                                                                                                                              \notag
\end{alignat}
where \eqref{Eq_WEPFTCF_s_AppendixGF_ss_AEMI_006} was made nondimensional by dividing the Green's functions by $L_y$. Obviously
\begin{alignat}{6}
y,Y\in[-\tfrac{1}{2}L_y,+\tfrac{1}{2}L_y]\Longrightarrow
\left. \begin{array}{c}0\leq\abs{y-Y}\leq L_y\\
                       0\leq\abs{y+Y}\leq L_y\\\end{array}\right\}\Longrightarrow \left\{\begin{array}{c}0\leq\cosh\kappa\abs{y-Y}\leq\cosh\kappa L_y\\
                                                                                                         0\leq\cosh\kappa\abs{y+Y}\leq\cosh\kappa L_y\\\end{array}\right.
                                                                                                                              \label{Eq_WEPFTCF_s_AppendixGF_ss_AEMI_007}
\end{alignat}
since $\cosh x=\cosh\abs{x}$ is positive ans strictly increasing $\forall\abs{x}\in\mathbb{R}_{\geq0}$. Hence\footnote{\label{ff_WEPFTCF_s_AppendixGF_ss_AEMI_003}
                                                                                                                       notice also that by \eqref{Eq_WEPFTCF_s_AppendixGF_ss_AEMI_006}
                                                                                                                       $G_\tsc{mwl}(y,Y;\kappa\neq0)-G_\mathrm{Kim}(y,Y;\kappa\neq0)\geq0$
                                                                                                                       $\forall y,Y\in[-\tfrac{1}{2}L_y,+\tfrac{1}{2}L_y]$
                                                                                                                      }
\begin{alignat}{6}
\dfrac{1  }
      {L_y}\Bigg(\max_{y,Y\in[-\frac{1}{2}L_y,+\frac{1}{2}L_y]}{\Big\lvert G_\tsc{mwl}(y,Y;\kappa)-G_\mathrm{Kim}(y,Y;\kappa)\Big\rvert}\Bigg)\leq
                                                                                                                              \notag\\
\underbrace{\dfrac{\mathrm{e}^{-\kappa L_y}(1+\mathrm{e}^{-\kappa L_y})\cosh\kappa L_y}
                  {2\kappa L_y\sinh\kappa L_y                                         }}_{\displaystyle E_\tsc{mwl}(\kappa L_y)}
                                                                                                                              \label{Eq_WEPFTCF_s_AppendixGF_ss_AEMI_008}
\end{alignat}
showing that the upper bound of approximation error made by the method of images is a function of the nondimensional (outer scaling) wavenumber $\kappa L_y$ \figref{Fig_WEPFTCF_s_DNSCpS_ss_GFSPEq_sss_AMI_001}.

%-----------------------------------------------------------------------------------------------------------------------------------
%
%
%
%
%
%
%
%
%
\bibliographystyle{jfm}\footnotesize%\bibliography{Aerodynamics,GV,GV_news}\normalsize
%
%
%
%
%
%
%
%
%
%-----------------------------------------------------------------------------------------------------------------------------------

\normalsize
\end{document}